\documentclass[12pt,a4paper]{article}       

\usepackage{xcolor}
 \usepackage[skins,theorems]{tcolorbox}
\tcbset{highlight math style={enhanced,
  colframe=red,colback=white,arc=0pt,boxrule=1pt}}
  \usepackage[bookmarksopen, bookmarksnumbered, bookmarksopenlevel=2]{hyperref}
  \usepackage{tikz}
  \usepackage{tikz-3dplot}
 \usetikzlibrary{calc}
 \usetikzlibrary{decorations} %
 \usepackage[UKenglish]{babel}
 \usepackage[toc,page]{appendix}
 \usepackage{amsmath}
 \usepackage{amssymb}
 \usepackage{graphicx}
 \usepackage{hhline}
 \usepackage{tikz-feynman}

 \usepackage[bf]{caption}
\usepackage{cite}
\usepackage[vcentermath]{youngtab}
\usepackage{geometry}
\usepackage{slashed}
\usepackage{stackrel}
\usepackage{tikz-cd} 
\usepackage{mathtools}
\usepackage{cancel} 
\usepackage{multirow}
\usepackage[margin=0pt,font=small,labelfont=normalfont,skip=22pt]{subcaption}

\usepackage{empheq}
\usepackage{arydshln}
\newcommand{\bqa}{\begin{eqnarray}}
\newcommand{\eqa}{\end{eqnarray}}

\newenvironment{eqn*}{\begin{equation*}\begin{aligned}}{\end{aligned}\end{equation*}\noindent}
\hypersetup{
    pdftitle={},
    pdfauthor={},
    pdfsubject={}
}
\numberwithin{equation}{section}
\numberwithin{table}{section}

\makeatletter

\definecolor{dgreen}{rgb}{0,0.45,0.2}
\definecolor{dblue}{rgb}{0,0.0,0.5}
\definecolor{mathorange}{rgb}{0.65,0.45,0.}

\newcommand{\diam}{\text{diam}}
\newcommand{\be}{\begin{equation}}
\newcommand{\ee}{\end{equation}}
\newcommand{\beq}{\begin{equation}}
\newcommand{\eeq}{\end{equation}}
\newcommand{\ba}{\begin{aligned}}
\newcommand{\ea}{\end{aligned}}

\newcommand{\bea}{\begin{eqnarray}}
\newcommand{\eea}{\end{eqnarray}}

\newcommand{\cO}{\mathcal{O}}

\newcommand{\cE}{\mathcal{E}}

\newcommand{\cK}{\mathcal{K}}
\newcommand{\cN}{\mathcal{N}}

\newcommand{\cF}{\mathcal{F}}
\newcommand{\cI}{\mathcal{I}}

\newcommand{\cV}{\mathcal{V}}

\newcommand{\cM}{\mathcal M}

\newcommand\bi{\begin{itemize}}
\newcommand\ei{\end{itemize}}

\def\Re{\mathop{\mathrm{Re}}\nolimits}

\def\Tr{\mathop{\mathrm{Tr}}\nolimits}
\def\tr{\mathop{\mathrm{tr}}\nolimits}

\def\unit{{1\kern-.65ex {\rm l}}}
\def\1{{1\kern-.65ex {\rm l}}}

\def\bbR{{\mathbb{R}}}

\def\bbZ{{\mathbb{Z}}}

\def\dd{{\mathrm{d}}}

\newcount\hour \newcount\minute
\hour=\time \divide \hour by 60
\minute=\time
\def\now{%
\ifnum \hour<13
  \ifnum \hour=0 \advance \hour by 12 \number\hour:\else \number\hour:\fi%
     \ifnum \minute<10 0\fi%
     \number\minute%
\ A.M.%
\else \advance \hour by -12 \number\hour:%
  \ifnum \minute<10 0\fi%
  \number\minute%
  \ P.M.%
\fi%
}

\makeatother

\begin{document}

\begin{titlepage}
\begin{center}
\rightline{\small }

\vskip 15 mm

{\large \bf
Species Scale in Diverse Dimensions
} 
\vskip 11 mm
Damian van de Heisteeg,$^{1}$ Cumrun Vafa,$^{2}$ Max Wiesner,$^{1,2}$ David H. Wu$^{2}$

\vskip 11 mm
\small ${}^{1}$ 
{\it Center of Mathematical Sciences and Applications, Harvard University,\\ Cambridge, MA 02138, USA}  \\[3 mm]
\small ${}^{2}$ 
{\it Jefferson Physical Laboratory, Harvard University, Cambridge, MA 02138, USA}

\end{center}
\vskip 17mm

\begin{abstract}
In a quantum theory of gravity, the species scale $\Lambda_s$ can be defined as the scale at which corrections to the Einstein action become important or alternatively as codifying the ``number of light degrees of freedom", due to the fact that $\Lambda_s^{-1}$ is the smallest size black hole described by the EFT involving only the Einstein term.  In this paper, we check the validity of this picture in diverse dimensions and with different amounts of supersymmetry and verify the expected behavior of the species scale at the boundary of moduli space.  This also leads to the evaluation of the species scale in the interior of the moduli space as well as to the computation of the diameter of the moduli space. We also find evidence that the species scale satisfies the bound $\big|{\nabla \Lambda_s \over \Lambda_s} \big|^2\leq {1\over d-2}$ all over moduli space including the interior.
\end{abstract}

\vfill
\end{titlepage}

\newpage

\tableofcontents

\setcounter{page}{1}

\section{Introduction and Summary}
The Swampland program has given detailed insights about the boundaries of moduli spaces in quantum gravitational theories.  In particular the distance conjecture \cite{Ooguri:2006in} combined with the emergent string conjecture \cite{Lee:2019oct} has led to a complete characterization of how the masses of towers of light particles decay as asymptotic boundaries of moduli space are approached (see also \cite{Etheredge:2022opl} and \cite{Agmon:2022thq} for a recent review). Therefore in these regions the effective theory of gravity needs to include a large number of light degrees of freedom. The species scale $\Lambda_s$~\cite{Dvali:2007hz,Dvali:2009ks,Dvali:2010vm,Dvali:2012uq} provides a measure for the number of light species and, as anticipated from the distance conjecture, has to decrease as we traverse large distances in field space \cite{Heidenreich:2018kpg, Grimm:2018ohb}. As explained in \cite{vandeHeisteeg:2023ubh}, the emergent string conjecture fixes the exponential decay rate of this cut-off scale. Further detailed studies of the species scale in asymptotic regimes have been carried out in~\cite{Calderon-Infante:2023ler}. 

However, since the species scale can be regarded as an invariant way to capture towers of light states, it can also be used to study light towers of states away from asymptotic boundaries where in general it might be difficult to identify the exact spectrum of states. For the particular case of Calabi--Yau threefold compactifications of Type II string theory, it was shown in \cite{vandeHeisteeg:2022btw} that the behavior of the species scale as a function of moduli can be effectively computed by considering a certain $R^2$-correction to the effective action. More precisely, due to supersymmetric protection, the dependence of the $R^2$-term on the scalars in the vector multiplets can be calculated explicitly using the one-loop topological string free energy $F_1$~\cite{Bershadsky:1993ta} leading to the evaluation of the species scale $M_{\rm pl}^2/\Lambda_s^2=F_1$ everywhere in vector multiplet moduli space. And indeed, as shown in \cite{vandeHeisteeg:2022btw}, the behavior of the species scale defined via the higher-derivative term matches the results obtained by studying the details of the light tower of states in the asymptotic regimes of moduli space. More importantly, for the first time, the identification between $F_1$ and the species scale allowed one to study the behavior of the species scale in the interior of the moduli space of a theory. The identification between $F_1$ and the species scale has been further supported using black hole entropy arguments in \cite{Cribiori:2022nke}.\footnote{See also \cite{Cribiori:2023ffn} for a detailed analysis of the relation between species scale and black hole thermodynamics and \cite{Cribiori:2023sch} for an extension of the relation between modular invariant functions, such as $F_1$, and the species scale to 4d theories with potentials.}

We argued in \cite{vandeHeisteeg:2023ubh} that in general the coefficients of generic higher-curvature terms in the effective action capture the moduli dependence of the species scale. This in principle can be used to generalize the 4d $\cN=2$ results to any general gravitational theories --- provided the moduli dependence of the higher-derivative terms can be calculated explicitly. Still, even without having access to the explicit coefficients of the higher-derivative terms, we showed in \cite{vandeHeisteeg:2023ubh} that consistency of the perturbative expansion is enough to constrain the slope of the species scale as 
\begin{equation}\label{nablaboundintro}
    \frac{|\nabla \Lambda_s|^2}{\Lambda_s^2} \leq \frac{c}{M_{\rm pl}^{d-2}}\,. 
\end{equation}
This bound is saturated in asymptotic regions of field space where the emergent string conjecture predicts an exponentially decaying species scale. The general EFT argument alone does not, however, fix the $\mathcal{O}(1)$ coefficient $c$ appearing in the above bound though the emergent string conjecture \cite{Lee:2019oct} leads to the prediction \cite{vandeHeisteeg:2023ubh} that, at least asymptotically, the bound is satisfied with $c=\frac{1}{d-2}$. Nevertheless, the examples studied in \cite{vandeHeisteeg:2023ubh} suggested that the bound with this value of $c$ may in fact be violated in the interior of field space. 

Knowledge of the behavior of the species scale in the interior of moduli space allows us to have some global insights into physical aspects of the moduli space.  In particular, we can identify special points where the species scale is maximized, dubbed `desert points' in the spirit of \cite{Long:2021jlv}. These points can be viewed as the center of moduli space where the amount of light states is minimized. In addition, the species scale can be used to identify the effective diameter of the moduli space at a cut-off scale $\mu$. In order for the EFT description to be valid at a cut-off scale $\mu$, we need to require $\mu \leq \Lambda_s$. Having a moduli-dependent expression for the species scale therefore allows us to exactly determine these regions and thus identify the diameter of the moduli space available at any cut-off scale $\mu$. For Type II compactifications on Calabi--Yau threefolds with mildly broken supersymmetry, a similar notion of the diameter of the field space was considered in \cite{vandeHeisteeg:2023uxj} by considering regions available for a slowly varying positive potential due to the condition $V<\Lambda_s^2$. Furthermore, in \cite{Andriot:2023isc} it was proposed that the gradient of the species scale bounds the slope of the scalar potential from below.

The goal of this paper is two-fold: First we extend the study of the species scale via higher-derivative terms to theories in higher dimensions in order to gather further evidence for the proposed relation between species scale and higher-derivative terms. To that end we focus on theories with eight supercharges in five and six dimensions, where we consider the $R^2$-term studied already in 4d $\cN=2$ \cite{vandeHeisteeg:2022btw,vandeHeisteeg:2023ubh} and theories with 16 and 32 supercharges for which $R^4$-couplings are used to compute the species scale. Again, we compare the asymptotic behavior with the expected behavior predicted by the properties of light towers of states. Second we aim to use the respective higher-curvature couplings to infer the properties of the species scale in the interior of field space to identify the desert points and the diameter of field space in these classes of theories. In particular, we also study the slope of the species scale in order to confirm the bound \eqref{nablaboundintro} and find evidence that the $\mathcal{O}(1)$ constant appearing in it is given by the naive expectation $c={1\over d-2}$.  We explain why the apparent counter-examples found in \cite{vandeHeisteeg:2023ubh} may be avoided if we delete the contribution of fields within the EFT to the species scale.

The rest of this paper is structured as follows: In the remainder of this section, we provide a summary of the main results of the analysis presented in this paper. In section~\ref{sec:basics}, we provide a review of how to extract the species scale from the coefficient of higher-derivative terms in general effective theories of gravity and introduce the main quantities that we calculate for the different classes of theories in the following sections. In section~\ref{sec:32Qs}, we consider theories with maximal supersymmetry in $d\geq 8$ and study the properties of the species scale using the coefficient of a certain $R^4$-coupling. In section~\ref{sec:16Qs}, we perform a similar analysis for theories with 16 supercharges focusing on $d\geq 9$. In section~\ref{sec:8Qs}, we then turn to theories with eight supercharges in six and five dimensions and use certain supersymmetrically protected $R^2$-couplings to study the properties of the species scale. Based on the results of these sections, we then revisit the bound~\eqref{nablaboundintro} in section~\ref{sec:refine} and argue how a proper treatment of the massless modes suggests $c={1\over d-2}$. The appendices contain some useful details about Eisenstein series that appear in the coefficients of the $R^4$-term in various dimensions and on the F-theory geometries considered in section~\ref{sec:8Qs}. 

\subsubsection*{Summary of Results}
We study the species scale in theories with 32, 16 and 8 supercharges by considering the moduli dependence of higher-derivative corrections at the eight- and four-derivative level. The location of the desert point and the value of the species scale at the desert point depend crucially on the details of the theory in consideration. The values of the species scale at the desert points of the different theories are summarized in table~\ref{tab:summary}. The diameter, $\diam (\mu)$, of the region of moduli space for which $\Lambda_s\geq \mu$ has the general form\footnote{The logarithmic dependence of the diameter of the moduli space on the cut-off scale has previously been observed in \cite{Hebecker:2017lxm,Scalisi:2018eaz}.} 
\begin{equation}\label{eq:diameter}
    \diam(\mu) = -\alpha \log\left[\frac{\mu}{M_{\rm pl}}\right] + b\,.
\end{equation}
This diameter is defined as the maximum of the distance between any pair of points in this region. For $\mu\ll1$ this distance is maximized if at least one of the points lies in an infinite-distance region. If the other point lies in the interior, the coefficient $\alpha$ is determined by the exponent of the species scale in this asymptotic regime~\cite{vandeHeisteeg:2023uxj}. Alternatively, the other point may lie in an \textit{inequivalent} infinite-distance region. In this case, we find $\alpha$ to be given by a certain combination of the species scale's exponents in the two infinite-distance regimes which  --- in the case that the geodesic connecting the two points traverses through the interior of the region $\Lambda_s\geq \mu$ --- simply reduces to the sum~\cite{vandeHeisteeg:2023uxj}. The values of $\alpha$ for the theories we considered in this work are summarized in table~\ref{tab:summary}. On the other hand, the coefficient $b$ which is expected to be $\mathcal{O}(1)$ is not as easy to determine and in most cases is negative (cf. table~\ref{tab:summary}). This implies that the asymptotic behavior generically gives an overestimation for the diameter of the moduli space.
\begin{table}[t!]
{{ 
\begin{center}
\renewcommand*{\arraystretch}{1.5}
	\begin{tabular}{|c|c|c|c|}
    \hline
	Example  & desert species scale & diameter coef. $\alpha$ & diameter coef. $b$ \\ \hline 
    \hyperref[ssec:IIA]{10d IIA} & $0.755$ & $8\sqrt{2}$ &$-2.12$ \\ \hline
    \hyperref[ssec:10dIIB]{10d IIB} & $0.756$ &$2\sqrt{2}$& $-0.073$ \\ \hline
    \hyperref[ssec:9dsugra]{9d M-theory on $T^2$} & $0.513$ & $14$ & $-7.978$ \\ \hline
    \hyperref[ssec:8dsugra]{8d M-theory on $T^3$} & $0.504$ & $\sqrt{42}$&$-4.309$  \\ \hline    \hyperref[heteroticE]{10d Heterotic $E_8\times E_8$} & $0.823$ &$8\sqrt{2}$& $-1.14$ \\ \hline
    \hyperref[heteroticO]{10d Heterotic $SO(32)$} & $0.822$ &$4\sqrt{2}$ &$0.14$ \\ \hline
    \hyperref[ssec:6d dP]{6d F-theory on $dP_r$} & $(9-r)^{-1/4}\;{}^*$ & \multirow{3}{*}{$2$}& $\log\left[\frac{2}{9-r}\right]^*$ \\ \cline{1-2}\cline{4-4}
    \hyperref[sec:Hirzebruch surfaces]{6d F-theory on $\mathbb{F}_{n\leq 2}$}  & $\left(2^{-3/4}\right)^*$ && \multirow{2}{*}{$-\log[\sqrt{n}]^*$} \\ \cline{1-2}
    \hyperref[sec:Hirzebruch surfaces]{6d F-theory on $\mathbb{F}_{n> 2}$}  & $\left(\frac{n^{1/4}}{\sqrt{2+n}}\right)^*$ && \\ \hline
    \hyperref[ssec:5d CY3]{5d M-theory on $X_{(2,86)}$}  & $0.490^*$ & $\sqrt{3}$& $-3.903^*$ \\ \hline 
	 \end{tabular}	
\end{center}}}
\caption{\label{tab:summary} Summary of the results for the maximal value of the species scale (the desert point) and the coefficients $\alpha$ and $b$ for the diameter of the field space \eqref{eq:diameter} in the examples discussed in this work (all in Planck units, the asterisks denote that in these cases we cannot determine the overall normalization, and therefore only compare different backgrounds). }
\end{table}

In the maximally supersymmetric case, the coupling of a certain $R^4$ term is protected by supersymmetry and we show that in $d\geq 8$ it correctly captures the dependence of the species scale on \emph{all} moduli. In particular, comparison with the Planck mass in 11d M-theory fixes the overall normalization of the species scale. At the perturbative level, this coupling only receives tree-level and one-loop contributions, but it can also receive further non-perturbative corrections depending on the details of the theory in consideration. In ten-dimensional Type IIA string theory, these non-perturbative corrections are absent.  In this case, the tree-level term correctly captures the behavior of the species scale in the weak-coupling limit whereas the one-loop term dominates at strong-coupling. As we will show by merely comparing the behavior of this one-loop correction with the general expectation for the species scale, one can infer the existence of an eleven-dimensional effective theory of gravity at strong coupling. In other words, without using detailed information about the exact spectrum of states, the one-loop correction to Type IIA string theory knows about the existence of M-theory! 

On the other hand, for Type IIB string theory it is precisely the D(-1)-instanton corrections that render the coefficient of the $R^4$-coupling modular invariant and ensure that the species scale in the strong coupling limit of Type IIB is again an emergent string limit. The species scale of Type IIB is maximized at the value for the axio-dilaton corresponding to the third root of unity where it evaluates roughly to $\frac34 M_{\rm pl}$, similar but slightly higher than Type IIA. Compactifying maximally supersymmetric theories on tori, we have the species scale of these theories can be expressed through generalized Eisenstein series \cite{Green:1997tv, Green:1997di, Kiritsis:1997em, Pioline:1997pu, Obers:1999um, Green:1999pu, Pioline:2010kb, Green:2010wi, Green:2011vz, Bossard:2020xod}. The modular properties of these functions allow us to identify the desert points as certain fixed points of the $U$-duality groups of the maximally supersymmetric theories. The values for the species scale and the dependence of the diameter of the field space as a function of the cut-off scale are summarized in table~\ref{tab:summary}. 

In theories with 16 supercharges, the first non-vanishing higher-derivative terms continue to arise at the eight-derivative level. The $R^4$-coupling in this case is not protected by supersymmetry but, as reviewed in \cite{Green:2016tfs}, the leading perturbative corrections beyond the one-loop level are expected to vanish. The known corrections to the $R^4$-coupling have been calculated in \cite{Green:2016tfs} and we show that, in case these give the exact expression, they reproduce the expected behavior for the species scale. This provides further support to the claim in \cite{Green:2016tfs} for the exactness of their computation. Additionally, in theories with 16 supercharges, there exists another $R^4$ coupling that arises at one-loop and is protected by supersymmetry. Whereas this coupling does not capture the behavior of the species scale in emergent string limits, we show that for the heterotic $SO(32)$ string on $S^1$ (with Wilson lines breaking the gauge group to $SO(16)\times SO(16)$) this one-loop term captures the dependence of the species scale on the radial modulus correctly and serves as a good approximation to the actual species scale. This illustrates that generically (but not always) the higher-derivative corrections that are protected by supersymmetry do capture scaling behavior of the species scale correctly and provide a useful upper bound for the species scale. 

Finally, we consider theories with eight supercharges in five and six dimensions where the species scale can be computed by the same $R^2$-coupling considered in 4d $\cN=2$ theories in \cite{vandeHeisteeg:2022btw,vandeHeisteeg:2023ubh}. Unlike in the 4d case, the coefficient of this coupling in the higher-dimensional theories does not receive quantum corrections and is purely given by the geometry of the compactification manifold. We show that also in five and six dimensions the asymptotic behavior of the species scale is correctly captured by the coefficient of the $R^2$-term from which we infer the behavior of the species scale also in the interior. Unlike in the theories with 16 and 32 supercharges, our methods do not fix the overall normalization of the species scale in terms of the higher-curvature corrections.

In general, we find that in all the examples we consider the $|\nabla \Lambda_s|^2/\Lambda_s^2$ is always bounded by $\frac{1}{d-2}$ except for two cases: 4d $\cN=2$ where the corrections to the coefficient of the $R^2$-coupling in emergent string limits force the slope to approach its asymptotic value from above and in 8d maximal supergravity where the same happens for the slope of the coefficient of the $R^4$-coupling. In both cases, the correction that pushes $|\nabla \Lambda_s|^2/\Lambda_s^2$ above $\frac{1}{d-2}$ as we approach emergent string limits is logarithmic in the moduli and can be traced back to the running of the coupling due to the light states already present in the EFT. We argue that, since the species scale should account for the light, but massive, states beyond the EFT, this logarithmic running of the coupling should not be part of the definition of the species scale. Similarly the behavior of the species scale close to a conifold point is dominated by the contribution of the light hypermultiplet; treating this case carefully, we show that also in this case the slope is bounded by $\frac{1}{d-2}$ from above. Taking all of these considerations into account, we can refine the bound in \eqref{nablaboundintro} as 
\begin{equation}
    \frac{|\nabla \Lambda_s|^2}{\Lambda_s^2}\leq \frac{1}{d-2}\,,
\end{equation}
in Planck units. As one of the main results of this paper, based on our analysis, we propose that in a consistent theory of gravity this bound is always satisfied at any point in the moduli space.

\section{Species Scale from higher-curvature Corrections}\label{sec:basics}
In this section, we review the definition of the species scale in terms of higher-curvature corrections and lay out the general strategy for our example analysis in the following sections. Consider a general theory of Einstein gravity in $d$-dimensions coupled to a massless (or light) scalar field with two-derivative action
\begin{equation}
    S=\frac{M_{\text{pl},d}^{d-2}}{2} \int \dd^d x \sqrt{-g} \left(R -\frac{1}{2} (\partial \phi)^2\right)\,. 
\end{equation}
Here, $M_{\text{pl},d}$ is the $d$-dimensional Planck scale; for most of our analysis we work in $d$-dimensional Planck units and set $M_{\text{pl},d}=1$, unless it is needed for clarity. In general, this effective action receives corrections by higher-derivative terms that encode the effects of quantum gravity at the effective field theory level. These corrections can be parameterized as
\begin{equation}\label{highercurv}
    S_{\rm corr.} = \frac{M_{\text{pl},d}^{d-2}}{2}\int \dd^dx \sqrt{-g} \left(\sum_{n=2}^{\infty} a_n(\phi) \frac{\mathcal{O}_{2n}(R)}{M_{\text{pl},d}^{2n-2}}\right)\,,
\end{equation}
where $\mathcal{O}_{2n}$ are dimension-$2n$ operators formed from contractions of the Riemann tensor $R$. In this parametrization, the coefficients $a_n(\phi)$ encode the information about the UV completion of the effective theory of gravity. If all these coefficients \emph{were} independent of $\phi$ and were of $\mathcal{O}(1)$, the expansion of the effective gravity action in higher-derivative terms would break-down at curvatures of order of the Planck scale. However, in consistent theories of gravity, we typically expect that there are additional light states beyond the massless level. As argued in \cite{Dvali:2007hz}, these effectively cause the scale at which the effective description of gravity breaks down to be lowered below the Planck scale. This new scale is typically referred to as the \emph{species scale}, $\Lambda_s$. Since the number of light states in a theory varies as a function of the vev for the moduli, so does $\Lambda_s=\Lambda_s(\phi)$. From the perspective of the higher-curvature terms in \eqref{highercurv}, a breakdown of the perturbative expansion at scale $\Lambda_s(\phi)$ implies that the coefficients $a_n(\phi)$ satisfy the general bound
\begin{equation}\label{cnhatcn}
    a_n(\phi) \leq \left(\frac{M_{{\rm pl},d}}{\Lambda_s(\phi)}\right)^{2n-2}\; \hat{a}_n\,, 
\end{equation}
where $\hat{a}_n$ is a theory-dependent moduli-independent constant. On general grounds, we expect only a finite amount of fine-tuning among the higher-derivative terms~\cite{Heckman:2019bzm}. This implies that only a few coefficients do not saturate this bound. In other words, the field-dependence of the coefficients of generic higher-curvature terms should capture the behavior of the species scale as a function of moduli --- up to the coefficient $\hat{a}_n$. 


To infer the dependence of the species scale on the scalar fields, we hence need to calculate the coefficients $a_n(\phi)$ of the higher-curvature corrections. In general, this is a difficult task as  typically closed expressions for $a_n(\phi)$ for general $n$ are unknown. Still, in favorable cases certain coefficients can be evaluated explicitly for any $\phi$. This is, e.g., the case in supersymmetric setups where certain higher-curvature corrections are protected by supersymmetry. However, supersymmetry may also prevent certain higher-curvature terms from appearing in the effective action. If this is the case, the parametrization \eqref{cnhatcn} tells us that $\hat{a}_n=0$ for this term and the term cannot be used to infer the dependence of the species scale on $\phi$. To extract non-trivial information about the $\phi$-dependence of the species scale, we therefore need to find a term in the higher-curvature expansion that is protected by supersymmetry but has $\hat{a}_n\neq 0$. Assuming that we have found an operator $\mathcal{O}_{n_0}(R)$ whose coefficient is non-zero and can be evaluated explicitly, we identify the species scale as 
\begin{equation}\label{Lambdascn}
    \frac{\Lambda_s}{M_{\rm pl,d}} = \left(\frac{a_{n_0}(\phi)}{\hat{a}_{n_0}}\right)^{\frac{1}{2-2n_{\tiny 0}}}\,. 
\end{equation}
Strictly speaking, this gives us an upper bound on the species scale since in principle we should consider all terms in the effective action and take the supremum of all possible values. In maximal and half-maximal supergravities, we consider $n_0=4$ where we can fix $\hat{a}_{4}$ by comparison with the Planck scale of 11d M-theory. For theories with 8 supercharges, we consider $n_0=2$. However, as we cannot compute the dependence of $a_2(\phi)$ on all moduli, $\hat{a}_2$ is not fixed here. \\

Once we identify a higher-derivative term whose coefficient we can calculate reliably, we can infer important information about the theory without knowing the exact spectrum of light states but just from the properties of the species scale. 

\paragraph{Asymptotic behavior of $\Lambda_s$.} According to the Distance and Emergent String Conjectures~\cite{Ooguri:2006in,Lee:2019oct}, at infinite distances in moduli space, a dual weakly-coupled description emerges corresponding either to a perturbative string limit or a decompactification to a higher-dimensional theory. The emergence of this dual theory is signalled by a tower of states becoming exponentially light in Planck units. In the perturbative string limit, the species scale reduces to the string scale of the emergent string and in the decompactification limits is identified with the higher-dimensional Planck scale. In both cases, the species scale decays exponentially $\Lambda_s \sim e^{-\gamma \Delta_\phi} M_{\rm pl,d}$ in the field space distance $\Delta_\phi$ where $\phi$ is the field parametrizing the infinite-distance limit. The coefficient $\gamma$ for a decompactification from $d$ to $D$ dimensions and a $d$-dimensional emergent string limit is respectively given by \cite{vandeHeisteeg:2023ubh} (see also \cite{Etheredge:2022opl,Calderon-Infante:2023ler})
\begin{equation}\label{eq:alphas}
    \gamma_{\rm decomp.}^{d\to D} = \sqrt{\frac{D-d}{(D-2)(d-2)}}\,,\qquad \gamma_{\rm string} = \frac{1}{\sqrt{d-2}}\,. 
\end{equation}
By computing the slope of the higher-derivative coefficient in the asymptotic limit, we can therefore infer the nature of the asymptotic limit without detailed knowledge of the spectrum of light states. 

\paragraph{Slope of $\Lambda_s$.} In \cite{vandeHeisteeg:2023ubh} we argued, based on the consistency of the higher-derivative expansion, that the slope of the species scale should be bounded from above as 
\begin{equation}\label{boundslope}
    \frac{|\nabla \Lambda_s|^2}{\Lambda_s^2} \leq \frac{c}{M_{\rm pl}^{d-2}} \,,
\end{equation}
for some $c\sim \mathcal{O}(1)$. Given the identification \eqref{Lambdascn}, this bound can be checked in explicit setups where certain higher-curvature corrections can be computed exactly. This bound should hold anywhere in moduli space including the interior. On the other hand, in asymptotic infinite-distance regimes, the slope of the species scale is expected to become constant, implying that it decays exponentially in the field distance. Based purely on the properties of the coefficient $a_{n_0}(\phi)$ and without prior knowledge of the states in the theory, we can then infer important properties of the theory in the asymptotic region. Comparison with the coefficients \eqref{eq:alphas} suggests that the $\mathcal{O}(1)$ constant is fixed to
\begin{equation}
    c = \frac{1}{d-2}\, .
\end{equation}
We verify this refined bound in the large class of examples considered in this work.

\paragraph{Consistency under dimensional reduction.} As a simple application, we can consider the behavior of the higher-derivative terms under dimensional reduction of the theory. Therefore, let us start with a $D$-dimensional theory and assume we have identified a higher-curvature operator $\mathcal{O}_{n_0}(R)$ whose coefficient, $a^D_{n_0}(\phi)$, can be determined explicitly. Let us therefore focus on the term 
\begin{equation}\label{term1}
    S_{\rm corr.} = \frac{M_{\text{pl},D}^{D-2}}{2} \int \dd^Dx \sqrt{-g_D} \, a^D_{n_0}(\phi)\, \frac{\mathcal{O}_{n_0}(R)}{M_{\text{pl},D}^{2n_0-2}}\,. 
\end{equation}
If we compactify this theory on a $(D-d)$-dimensional manifold with volume $VM_{\text{pl},D}^{D-d}=\mathcal{V}$ such that 
\begin{equation}
    M_{\text{pl},d}^{d-2} = M_{\text{pl},D}^{d-2} \mathcal{V}\,.
\end{equation}
Then, a dimensional reduction of the term in \eqref{term1} yields 
\begin{equation}
    S_{\rm corr.} \rightarrow  \frac{M_{\text{pl},d}^{d-2}}{2} \int \dd^dx \sqrt{-g_d}\,\left( a^D_{n_0}(\phi)\mathcal{V}^\frac{2n_0-2}{d-2}\right) \,\frac{\mathcal{O}_{n_0}(R)}{M_{\text{pl},d}^{2n_0-2}}\,. 
\end{equation}
The terms in the brackets can be identified as a contribution to the higher-derivative coefficient in the lower-dimensional theory 
\begin{equation}
    a^d_{n_0}(\phi, \mathcal{V}) = a^D_{n_0}(\phi)\mathcal{V}^\frac{2n_0-2}{d-2} +\dots\,,
\end{equation}
where the dots indicate corrections that arise in the lower-dimensional theory. However, since $a^D_{n_0}$ is exact, these corrections have to be sub-leading in the limit $\mathcal{V}\rightarrow \infty$. The species scale in the limit $\mathcal{V}\rightarrow \infty$ is therefore given by 
\begin{equation}
    \left(\frac{\Lambda_s}{M_{\text{pl},d}}\right)^{2-2n_0} = a^d_{n_0}(\phi,\mathcal{V})\,. 
\end{equation}
Let us assume that we take a homogeneous decompactification limit such that we can write $\mathcal{V}=e^{(D-d)\sigma}$. The metric for the field space spanned by $\sigma$ is given by (cf. \cite{vandeHeisteeg:2023ubh}) 
\begin{equation}
    \dd s^2= \frac{(D-2)(D-d)}{(d-2)}(\dd \sigma)^2\,,
\end{equation}
such that, as a function of the field space distance $\Delta_\sigma$, the species scale in the limit $\sigma\rightarrow \infty$ scales as 
\begin{equation}
    \frac{\Lambda_s}{M_{\text{pl},d}} \sim e^{-\sqrt{\frac{D-d}{(D-2)(d-2)}}\Delta_\sigma} \,,
\end{equation}
in accordance with our expectation~\eqref{eq:alphas} for a decompactification from $d$ to $D$-dimensions. 

\paragraph{Desert point.} The benefit of defining the species scale via higher-derivative corrections is that it allows us to calculate the species scale in the bulk of the moduli space, i.e., away from the asymptotic regimes. Of particular interest are the points in the moduli space at which $\Lambda_s$ is maximized, i.e., the desert points of the theory~\cite{Long:2021jlv}. Besides the location of these points in the moduli space, the value of $\Lambda_s$ at the desert is of interest as it gives an estimate for the smallest number of light states a theory of gravity can have. Since light states can only decrease the quantum gravity cut-off, we expect that the value of $\Lambda_s$ at the desert point is bounded from above by the Planck scale. We confirm this in the examples below. We also want to point out that the species scale can have saddle points and not just maxima, as happens, e.g., for 10d Type IIB at self-dual coupling, see section \ref{ssec:10dIIB} for more details.

\paragraph{Diameter of Moduli Space.} Let us consider the low-energy EFT at a cut-off scale $\mu$. Since the species scale vanishes asymptotically in infinite-distance limits it eventually drops below $\mu$ and the effective field description breaks down. Therefore, the part of the moduli space where $\Lambda_s\geq \mu$ and the EFT is valid is compact. A useful quantity that we can associate to this compact space is its diameter, $\diam(\mu)$, which is the maximum of the distance between any pair of points on the moduli space.  To get the largest distance in moduli space for a given $\mu$, at least one 
of the points needs to lie on the boundary of the region $\Lambda_s\geq \mu$. This means that for small $\mu\ll M_{\text{pl},d}$, this point needs to lie in one of the asymptotic regions where 
\begin{equation}
    \Lambda_s \sim A e^{-\gamma \Delta_\phi}\,. 
\end{equation}
The diameter for small $\mu$ then schematically takes the form
\begin{equation}
    \diam(\mu) = -\alpha \log\left(\frac{\mu}{M_{\text{pl},d}}\right) + b \,,
\end{equation}
where $b$ accounts for the coefficient $A$ and the size of the interior of the moduli space where the species scale does not feature an exponential behavior. Via \eqref{eq:alphas}, $\alpha$ is (inversely) related to the exponent $\gamma$ of the species scale in infinite-distance limits. If the diameter is set by the distance between two points lying in inequivalent asymptotic regions, the coefficient $\alpha$ will be some combination of the (inverse) exponents of the species scale in the two regions. Whenever the geodesic connecting these regions passes through the interior it will simply be their sum; otherwise, it will depend on the structure of the moduli space, see section \ref{ssec:9dsugra} on 9d M-theory on $T^2$ for an example.

\section{Species Scale and 32 supercharges}\label{sec:32Qs}
The first instance to which we apply our general approach corresponds to maximally supersymmetric cases. Therefore, consider the first few higher-derivative terms in the effective supergravity action in the following schematic form
\begin{equation}\label{general32}
    S_{\rm corr.} = \frac{M_{\text{pl},d}^{d-2}}{2} \int \dd^dx \sqrt{-g} \left(\frac{a_2(\phi)}{M_{\rm pl}^2} R^2 + \frac{a_3(\phi)}{M_{\rm pl}^3} R^3 + \frac{a_4(\phi)}{M_{\rm pl}^4} R^4 +\dots \right)
\end{equation}
Via \eqref{cnhatcn}, we can relate the $a_i(\phi)$ to the species scale. However, in theories with 32 supercharges the $R^2$ and $R^3$ corrections vanish identically implying $\hat{a}_2=\hat{a}_3=0$. Therefore, we cannot use these terms to infer the species scale, but instead we need to go to the operators involving four powers of the Riemann tensor. Theories with 32 supercharges can be obtained from toroidal compactifications of M-theory in eleven dimensions. In eleven-dimensional supergravity, there exists a single independent contraction of four Riemann tensors
\begin{equation}
    \mathcal{J}_0 = (t_8 t_8 -\frac{1}{24} \epsilon_{11} \epsilon_{11})R^4\,, 
\end{equation}
where $t_8$ and $\epsilon_{11}$ are tensors whose exact form can be found, e.g., in Appendix B of \cite{Grimm:2017okk}.
Since M-theory does not have a moduli space, the coefficient of this $R^4$ is just a constant and the species scale is simply given by $M_{\text{pl},11}$. Given \eqref{cnhatcn}, we can hence read off $\hat{a}_4$ as the coefficient of the $R^4$-term in the effective action \cite{Green:1997di}
\begin{equation}\label{11dterm}
    S_{11,R^4} = \frac{M_{{\rm pl},11}^3}{2} \int \dd^{11}x\sqrt{-g}\,\hat{a}_4\,(t_8 t_8 - \frac{1}{24} \epsilon_{11} \epsilon_{11})R^4\,, \qquad \hat{a}_4 = \frac{1}{{18 \cdot 2^{14}\cdot  (2\pi)^{\frac{10}{3}}}}\,. 
\end{equation}
As shown in \cite{Green:1997di}, this term can be obtained by considering four-graviton scattering in eleven dimensions at one loop with the state running in  the loop corresponding to the massless supergraviton in eleven dimensions. Since supersymmetry relates this term to the $C_3\wedge X_8$ term in the effective action, it does not receive corrections beyond one-loop \cite{Green:1997di} and is therefore exact. The non-renormalization of this term can be used to also argue for one-loop exactness of the corresponding terms for toroidal compactifications. In a series of works \cite{Green:1997tv, Green:1997di, Kiritsis:1997em, Pioline:1997pu, Obers:1999um, Green:1999pu, Pioline:2010kb, Green:2010wi, Green:2011vz, Bossard:2020xod}, the coefficient of this term has been determined exactly in toroidal compactifications of M-theory for supergravities in dimensions $d \geq 3$. These coefficients may be expressed through particular (real-analytic) Eisenstein series given by sums over tensions and masses of BPS states in the theory; see appendix~\ref{app:Eisen} for details on the Eisenstein series appearing in these expressions. In the following, we discuss the species scale for maximally supersymmetric gravitational theories in $d=10,9,$ and $8$ of which the moduli spaces and the U-duality groups are summarized in table~\ref{tab:summary}. In all these examples, the species scale is determined by the coefficient of the $t_8t_8 R^4$-term in the respective higher-derivative action. Since this term is present in 11d M-theory, we can use \eqref{11dterm} to fix the coefficient $\hat{a}_4^{(11)}$ appearing in \eqref{cnhatcn}. 

\begin{table}[h!]
\begin{center}
\renewcommand*{\arraystretch}{1.5}
	\begin{tabular}{|c||c|c|c|}
		\hline
	      $d$ & $E_{n+1}(\mathbb{R})$  & $K_{n+1}$ & $E_{n+1}(\mathbb{Z})$ \\ \hline \hline
       10A & $\mathbb{R}^+$ & 1 & 1 
	\\	\hline
        10B & $SL(2,\bbR)$ & $SO(2)$ & $SL(2,\bbZ)$ \\ \hline
	9 & $SL(2,\bbR)\times \bbR^+$ & $SO(2)$ & $SL(2,\bbZ)$\\	\hline
 	8 & $SL(3,\bbR)\times SL(2,R)$ & $SO(3)\times SO(2)$ & $SL(3,\bbZ) \times SL(2,\bbZ)$\\ \hline
	 \end{tabular}	
\end{center}
\caption{\label{table:maxsugras} The symmetry groups in maximal supergravity in $d=10-n$ giving the moduli space of these theories as $E_{n+1}(\mathbb{Z})\backslash E_{n+1}(\mathbb{R})/K_{n+1}$. }
\end{table}

\subsection{10d Type IIA}\label{ssec:IIA}
\begin{figure}[!t]
\begin{subfigure}{0.48\textwidth}
\begin{center}
\includegraphics[width=0.9\textwidth]{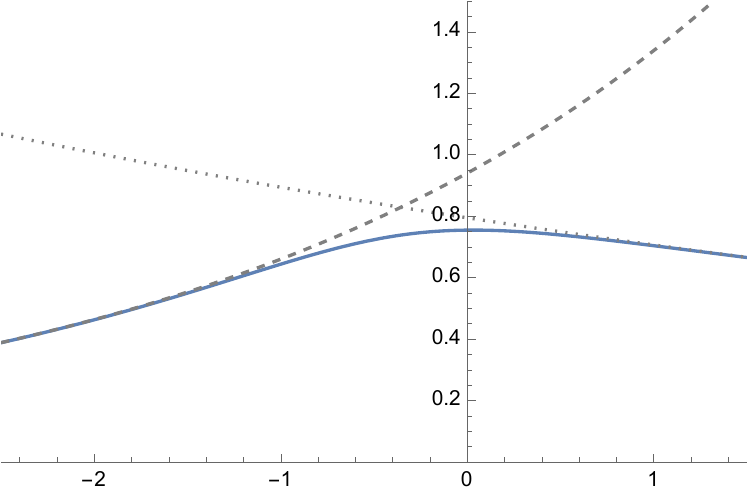}
\end{center}
\begin{picture}(0,0)
\put(112,22){\footnotesize $\Delta_\lambda$}
\put(135,170){\footnotesize $\Lambda_s$}
\end{picture}
\vspace*{-40pt}\caption{Species scale.}
\end{subfigure}
\hspace{2pt}
\begin{subfigure}{0.48\textwidth}
\begin{center}
\includegraphics[width=0.9\textwidth]{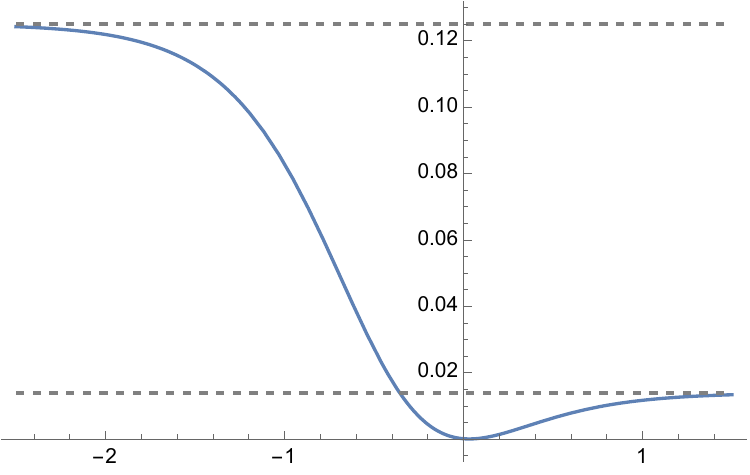}
\end{center}
\begin{picture}(0,0)
\put(112,22){\footnotesize$\Delta_\lambda$}
\put(120,170){\footnotesize$|\Lambda_s'|^2/\Lambda_s^2$}
\put(200,160){\footnotesize $1/8$}
\put(200,58){\footnotesize $1/72$}
\end{picture}
\vspace*{-40pt}\caption{Slope of $\Lambda_s$.\label{fig:IIAgrad}}
\end{subfigure}
\caption{\label{fig:IIAspecies} The species scale \eqref{LambdasIIA} (left) and its slope (right) as functions of $\Delta_\lambda = \lambda/\sqrt{2}$. The limit $\Delta_\lambda \to -\infty $ corresponds to the emergent string limit and $\Delta_\lambda \to \infty $ to the decompactification limit to M-theory. The dashed and dotted curve show the respective contributions of the first and second term in \eqref{LambdasIIA}.}
\end{figure}
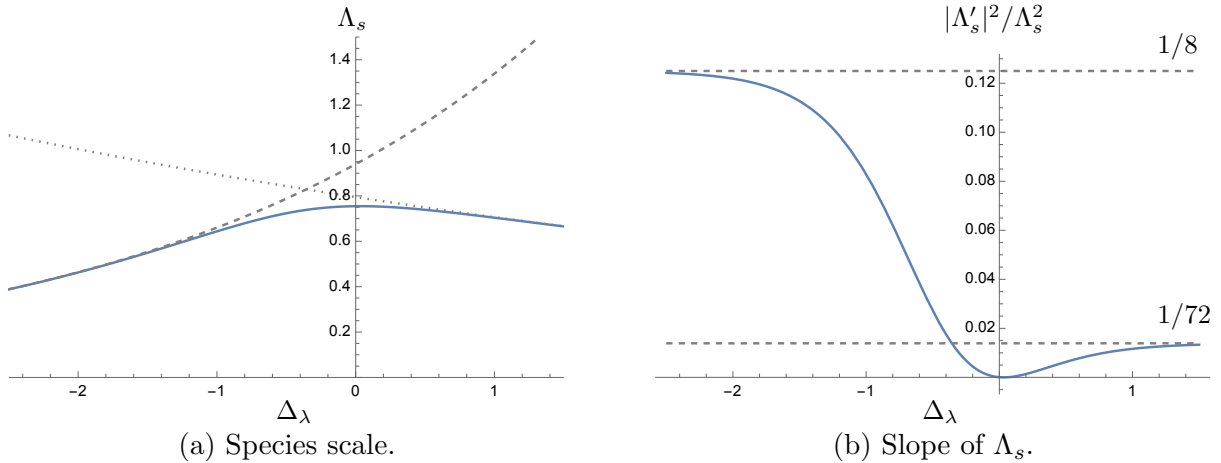

We first consider Type IIA in ten dimensions, which is related to M-theory through a circle compactifications. The compactified theory has a single modulus which corresponds to the Type IIA dilaton $e^{-\lambda}$, or equivalently, the radius of the circle can be expressed as follows
\begin{equation}
    R_{11} M_{\text{pl}, 11} = e^{2/3\,\lambda}\,. 
\end{equation}
The field space is spanned by the dilaton $\lambda$ with metric $\dd s^2 = \frac{1}{2} \dd \lambda^2$. As in 11d M-theory, the $t_8t_8R^4$-term is protected by supersymmetry which ensures that there are no perturbative corrections to its coefficient beyond one-loop. Furthermore, since there are no BPS instantons in 10d Type IIA, there are also no non-perturbative corrections to $a_4(\lambda)$. Therefore, $a_4(\lambda)$ consists of two terms one of which arises as the dimensional reduction of the $t_8t_8 R^4$-term in \eqref{11dterm} on a circle
\begin{equation}\label{eq:IIAfromM}
    S_{10,R^4} \supset \frac{M_{\rm pl,10}^8}{2} \int \dd^{10}x \sqrt{-g}\, \hat{a}_4 \frac{t_8t_8 R^4}{M_{\rm pl,10}^6} \left(\frac{M_{\rm pl,10}}{M_{\rm pl,11}}\right)^{6}\,.
\end{equation}
From our general expression in \eqref{general32}, in the large-radius/strong-coupling limit, we can identify 
\begin{equation}
    a_4^{(10)}(\lambda) = \hat{a}_4  \left(\frac{M_{\text{pl},10}}{\Lambda_s}\right)^{6} \xrightarrow[]{\lambda\rightarrow \infty} \hat{a}_4 \, (2\pi)^{3/4} e^{\lambda/2} \,,
\end{equation}
where we used 
\begin{equation}
    \frac{M_{\text{pl},10}^{8}}{M_{\text{pl},11}^8} = 2\pi R_{11} M_{\text{pl},11}\,.
\end{equation}
As is apparent from the dilaton dependence, this term arises at one-loop level. In fact, both the tree- and one-loop contribution to the $t_8t_8R^4$-term can be computed directly in string theory. These take on the following schematic form 
\begin{equation}
    S_{10,R^4} \sim M_{\text{pl},10}^2 \int \dd^{10}x \sqrt{-g} \left(2\zeta(3) e^{-3\lambda/2} +\frac{2\pi^2}{3} e^{\lambda/2} \right) t_8t_8 R^4 \,.
\end{equation}
This fixes the relative factor between the tree-level and the one-loop term. Comparison with \eqref{eq:IIAfromM} fixes the overall normalization, as in \eqref{general32}, to be
\begin{equation}\label{highderIIA}
\begin{aligned}
     S_{10,R^4} &= \frac{M_{\rm pl,10}^2}{2} \int \dd^{10} x \sqrt{-g}\, a_4^{(10)}(\lambda) \, t_8t_8 R^4\,, \\ 
     \text{with} \qquad a_4^{(10)}(\lambda) &=  \hat{a}_{4} \left(\frac{3\cdot 2^{3/4}\zeta(3)}{\pi^{5/4}} e^{-3\lambda/2} + (2\pi)^{3/4} \, e^{\lambda/2} \right)\,.
\end{aligned}
\end{equation}
According to our general discussion, the species scale in ten-dimensional Type IIA string theory for any value of $\lambda$ is then given by 
\begin{equation}\label{LambdasIIA}
   {\Lambda_s} = \frac{1}{(2\pi)^{1/8}}\left(\frac{3\zeta(3)}{\pi^2} e^{-3\lambda/2} + e^{\lambda/2} \right)^{-1/6}\,. 
\end{equation}
In figure \ref{fig:IIAspecies}, we illustrate how the species scale and its slope vary as a function of $\Delta=\lambda/\sqrt{2}$.  With this preparation, we can now discuss the properties of the species scale.

\paragraph{Slope of $\Lambda_s$.} As expected on general grounds~\cite{vandeHeisteeg:2023ubh}, the slope of the species scale is bounded from above. From figure~\ref{fig:IIAgrad}, we conclude that the maximal value for the slope of the species scale is achieved in the asymptotic weak-coupling limit which determines the $\cO(1)$ constant $c$ in \eqref{boundslope}. This leads to the bound 
\begin{equation}
    \left|\frac{\nabla \Lambda_s}{\Lambda_s} \right|^2\leq \frac{1}{{8}}\,,
\end{equation}
in Planck units. 

\paragraph{Asymptotic behavior.} Next, we study the behavior of the species scale \eqref{LambdasIIA} in the two asymptotic regions. In the weak-coupling limit $\lambda\rightarrow -\infty$, we find that the scaling of the species scale in the canonically normalized field $\Delta_\lambda = -\sqrt{2}\lambda$ on the moduli space is given by
\begin{equation}\label{eq:IIAemergentstring}
    \Lambda_s  =  \frac{\pi^{5/24}}{2^{1/8}(3\zeta(3))^{1/6}}e^{-\Delta_\lambda/(2\sqrt{2})}\,.
\end{equation}
 This agrees with the expected coefficient $1/\sqrt{d-2}$ for an emergent string limit in $d=10$ dimensions. In the strong coupling limit $\lambda\rightarrow \infty$, we find it decays in $\Delta_\lambda = \sqrt{2}\lambda$ as
\begin{equation}\label{eq:IIA11d}
    \Lambda_s =\frac{1}{(2\pi)^{1/8}}e^{-\Delta_\lambda/(6\sqrt{2}) }\,,
\end{equation}
which agrees with the general expectation for the coefficient \eqref{eq:alphas} 
for a decompactification from $d=10$ to $D=11$ dimensions. From the ten-dimensional Type IIA perspective it is in fact rather remarkable that the one-loop contribution to the coefficient of the $t_8 t_8 R^4$ coupling knows about the decompactification to 11d M-theory at strong coupling, as it gives us the right behavior for the species scale in this limit. We therefore could have inferred the existence of this higher-dimensional theory just from studying the higher-derivative terms without prior knowledge about the light spectrum of states in Type IIA.

\paragraph{Desert point.} We next identify the point in the one-dimensional dilaton field space that maximizes the species scale \eqref{LambdasIIA}. We find that this desert point is located at
\begin{equation}
    e^\lambda = \frac{3}{\pi} \sqrt{\zeta(3)} \approx 1.05\, ,
\end{equation}
corresponding to $\lambda\approx 0.05$. This leads to a value of the species scale at the desert point of
\begin{equation}\label{IIAdesert}
    \max(\Lambda_s) = \left(\frac{3^2}{2^{11} \pi \zeta(3)}\right)^{1/24} \approx 0.755\, .
\end{equation}
Notice that the species scale is localized close to but not exactly at $\lambda=0$. This should not come as a surprise since the two infinite-distance limits for $\lambda\rightarrow -\infty$ and $\lambda\rightarrow \infty$ are inequivalent and hence there is no symmetry exchanging the two limits while keeping $\lambda=0$ fixed. Notice further that the maximal value of $\Lambda_s$ is below $M_{\text{pl},10}$ as expected on general grounds.

\paragraph{Diameter.} Lastly, we can compute the diameter of the effective field space available at some cut-off scale $\mu$. The constraint $\Lambda_s\geq \mu$ determines the diameter of the field space to scale as 
\begin{equation}
    \diam(\mu) = -8\sqrt{2} \log[\mu] -\frac{\sqrt{2}}{3}\log\left[ 24 \pi \zeta(3) \right]\, .
\end{equation}
The coefficient of the first term is simply the sum of the separate contributions $\sqrt{72}$ and $\sqrt{8}$ for the decompactification and emergent string limits. The negative shift takes the numerical value
\begin{equation}
    b=-\frac{\sqrt{2}}{3}\log\left[ 24\pi \zeta(3)\right] \approx -2.12\, ,
\end{equation}
which is due to the leading coefficients in \eqref{eq:IIAemergentstring} and \eqref{eq:IIA11d} appearing in the asymptotic scaling of the species scale.

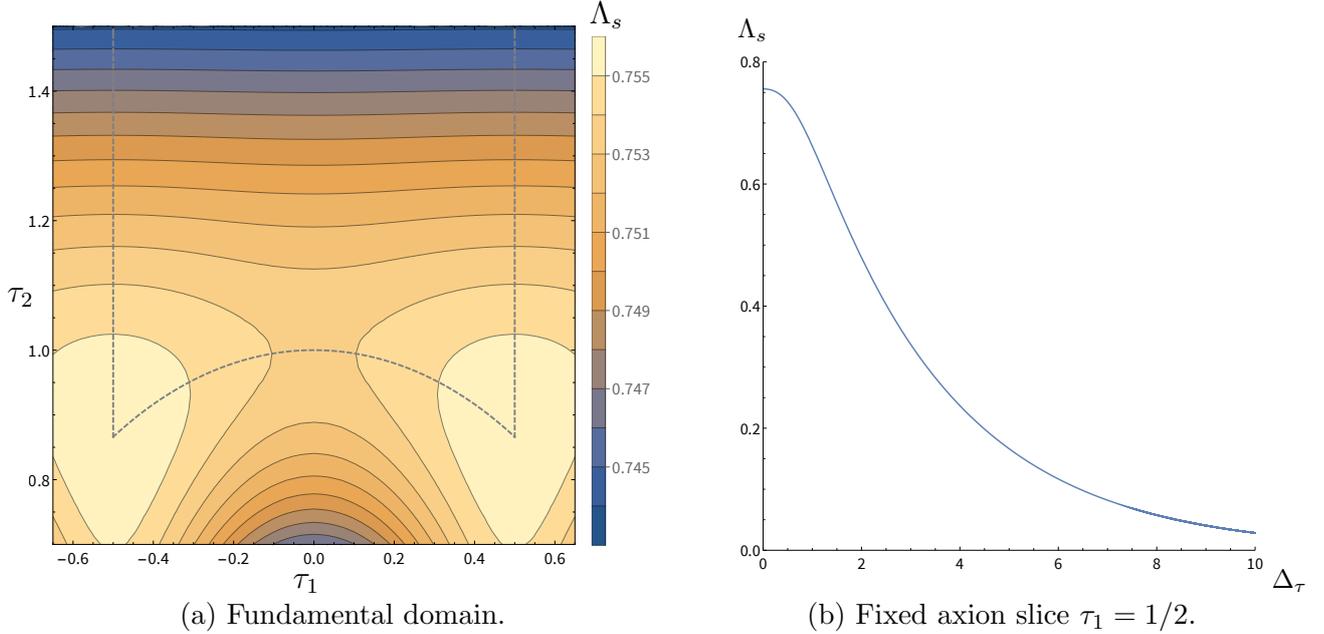
\begin{figure}[!t]
\begin{subfigure}{0.54\textwidth}
\begin{center}
\includegraphics[width=0.9\textwidth]{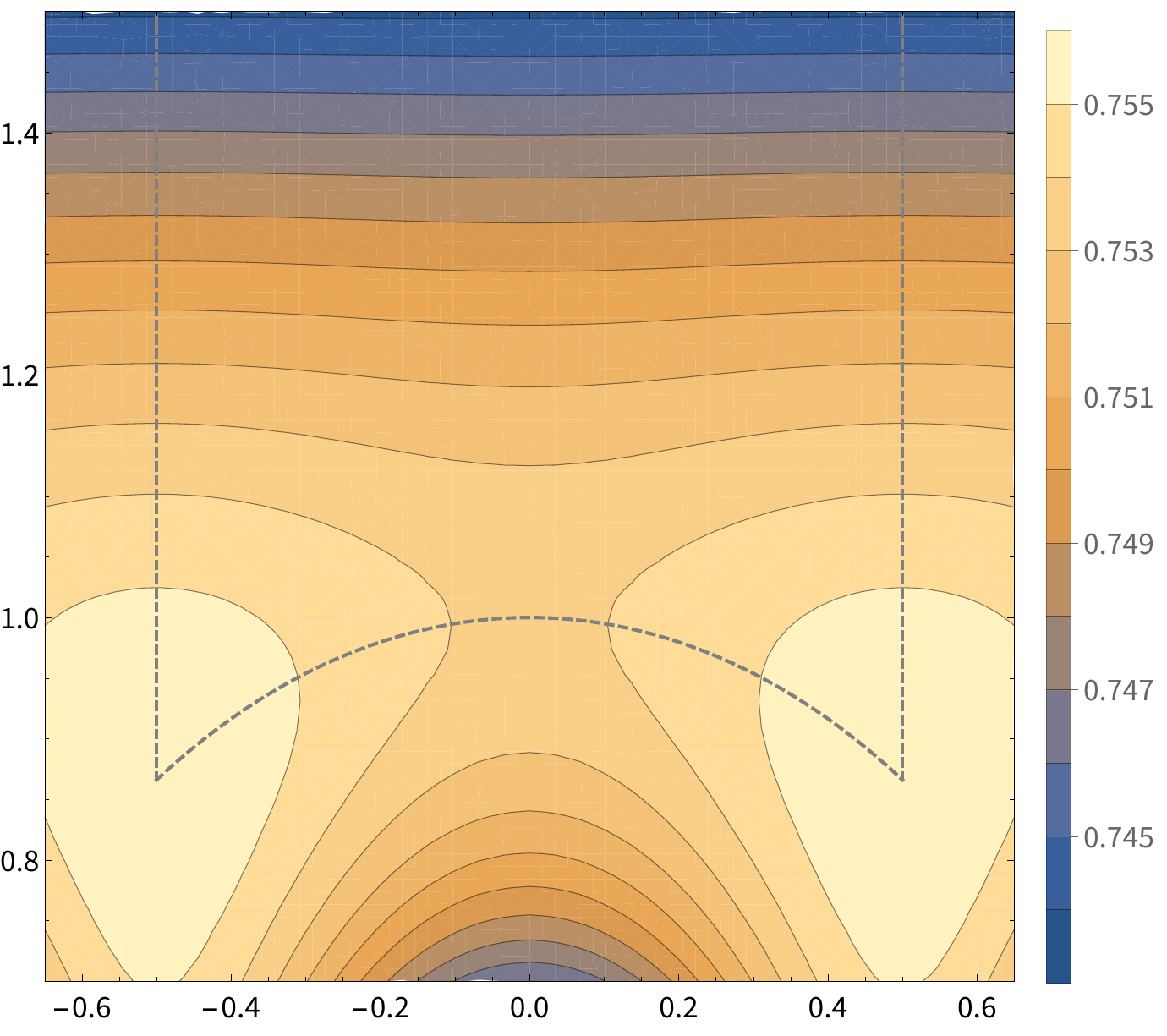}
\end{center}
\begin{picture}(0,0)
\put(112,22){$\tau_1$}
\put(5,130){$\tau_2$}
\put(223,235){$\Lambda_s$}
\end{picture}
\vspace*{-40pt}\caption{Fundamental domain.\label{fig:fund}}
\end{subfigure}
\hspace{2pt}
\begin{subfigure}{0.45\textwidth}
\begin{center}
\includegraphics[width=0.9\textwidth]{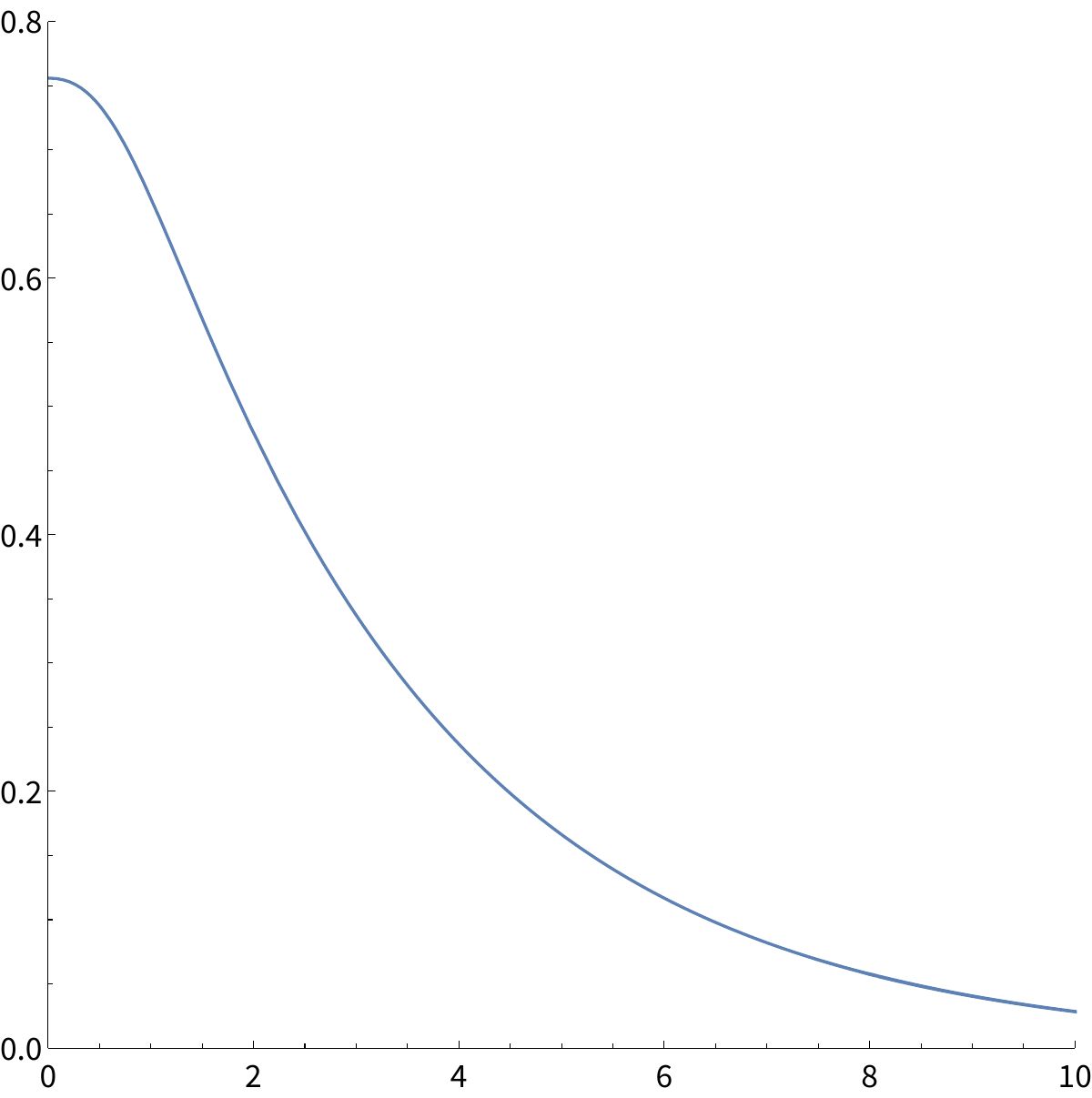}
\end{center}
\begin{picture}(0,0)
\put(210,22){\footnotesize$\Delta_\tau$}
\put(10,230){\footnotesize$\Lambda_s$}
\end{picture}
\vspace*{-40pt}\caption{Fixed axion slice $\tau_1=1/2$.\label{fig:fixedaxion}}
\end{subfigure}
\caption{\label{fig:IIBspecies} Plot of the species scale for ten-dimensional Type IIB. Left: a contour plot of $\Lambda_s$ over the upper half plane, where the fundamental domain bis indicated by a dashed line. Right: $\Lambda_s$ along a  fixed axion slice $\tau_1 = 1/2$ as a function of distance $\Delta_\tau$ in scalar field space.}
\end{figure}

\subsection{10d Type IIB}\label{ssec:10dIIB}
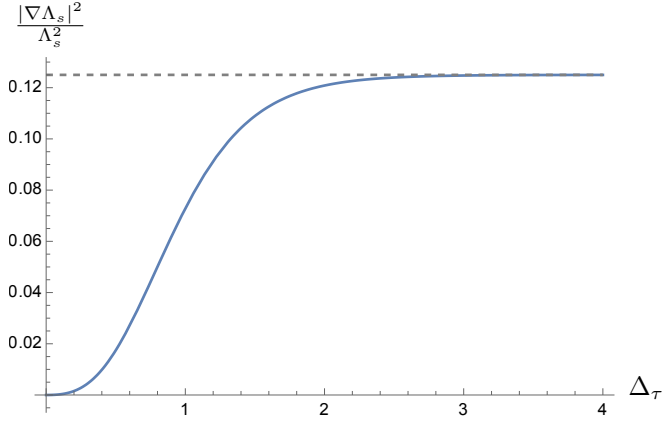
\begin{figure}[!t]
\begin{center}
\includegraphics[width=8 cm]{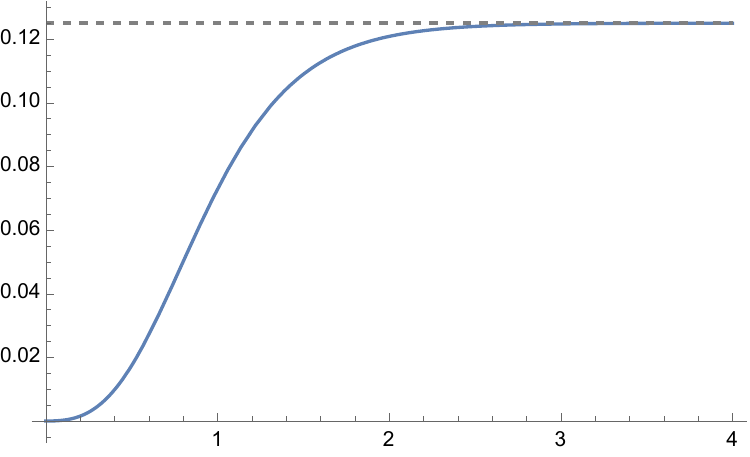}
\end{center}
\begin{picture}(0,0)
\put(130,175){\footnotesize $\frac{|\nabla \Lambda_s|^2}{\Lambda_s^2}$}
\put(360,38){\footnotesize$\Delta_\tau$}
\end{picture}\vspace*{-0.8cm}
\caption{\label{fig:IIBgradient} Slope of the species scale $|\nabla \Lambda_s|^2/\Lambda_s^2$ as a function of the distance $\Delta_\tau = \log[\tau_2]/\sqrt{2}$ in scalar field space, along the constant axion slice $\tau_1=1/2$.}
\end{figure}

We now turn to ten-dimensional Type IIB string theory. In this case, the field space is spanned by the axio-dilaton $\tau=\tau_1+i\tau_2$ which endows the field space with the standard hyperbolic metric
\begin{equation}
    \dd s^2 = \frac{\dd \tau \dd \bar \tau}{2\tau_2^2}\, .
\end{equation}
Again, the relevant term in the effective action corresponds to the $t_8t_8 R^4$ coupling. Thanks to supersymmetry, the dependence of the coefficient $a_4(\tau,\bar{\tau})$ of this term on $\tau$ can be calculated explicitly. As in Type IIA, there do not exist any perturbative contributions to $a_4(\tau, \bar{\tau})$ beyond one-loop. However, there are contributions coming from D($-1$)-instantons. As shown in \cite{Green:1997tv}, the full $\tau$-dependence of the $t_8t_8R^4$ coupling is captured by the $SL(2,\mathbb{Z})$-invariant Eisenstein series $E_{\footnotesize \frac{3}{2}}(\tau, \bar{\tau})$ given by
\begin{equation}\label{eq:IIBeisenstein}
    E_{\frac{3}{2}}(\tau,\bar{\tau}) = \sum_{(p,q)\neq (0,0)} \frac{\tau_2^{3/2}}{|p+q\tau|^3}\, .
\end{equation}
This expression can be understood as summing $M_{p,q}^{-6}$ for every $(p,q)$-string in the BPS spectrum with tension $M_{p,q}$. This Eisenstein series gives the dependence of the species scale on $\tau$; in order to obtain the correct normalization, we consider the weak-coupling limit $\tau_2 \to \infty$ in which the Eisenstein series behaves as
\begin{equation}\label{eq:E32series}
    E_{\frac{3}{2}}(\tau,\bar{\tau}) = 2 \zeta(3) \tau_2^{3/2}+\frac{2\pi^2}{3} \tau_2^{-1/2}+8\pi \sqrt{\tau_2}\sum_{p\neq 0} \sum_{n=1}^\infty \left|\frac{p}{n}\right| K_1(2\pi \tau_2 |p|n) e^{2\pi i p n \tau_1}\, .
\end{equation}
The infinite sum corresponds to the exponentially suppressed contributions from D(-1)-instantons. Recalling that $\tau_2=e^{-\lambda}$ with $\lambda$ the Type IIB dilaton, we recognize the first term as a tree-level contribution and the second term as the one-loop term of which both were also present in Type IIA. Comparison with \eqref{highderIIA} then fixes the normalization of the species scale to 
\begin{equation}\label{eq:speciesIIB}
    \Lambda_s = \left(\frac{3}{2^{1/4}\pi^{5/4}} E_{\frac32}(\tau, \bar\tau)\right)^{-1/6}\, .
\end{equation}
We have depicted the behavior of $\Lambda_s$ over the moduli space in figure \ref{fig:IIBspecies}. We also show the slope of this species scale along a slice in moduli space for fixed $\tau_1=1/2$ in figure \ref{fig:IIBgradient}. Similar to Type IIA, we find that the slope of $\Lambda_s$ is bounded from above by $1/\sqrt{8}$ everywhere in moduli space. 

\paragraph{Asymptotic behavior.} Due to the $SL(2,\mathbb{Z})$-duality, there is only a single kind of infinite-distance limit for ten-dimensional Type IIB string theory corresponding to $\tau_2 \to \infty$. All other infinite-distance limits are related to this one via duality transformations and, due to the $SL(2,\mathbb{Z})$-invariance of the Eisenstein series, the species scale \eqref{eq:speciesIIB} is the same in all infinite-distance limits. From \eqref{eq:speciesIIB}, we then find that the species scale has a power-law behavior $\Lambda_s \sim (\tau_2)^{-1/4}$ in the string coupling. In terms of the field space distance $\Delta_\tau$ given asymptotically by $\Delta_\tau ={\tiny \frac{1}{\sqrt{2}}} \log[\tau_2]$, we find
\begin{equation}\label{eq:IIBemergentstring}
    \Lambda_s  = \frac{\pi^{5/24}}{2^{1/8}(3\zeta(3))^{1/6}} e^{-\Delta_\tau/(2\sqrt{2})},
\end{equation}
which is indeed the expected coefficient $1/\sqrt{d-2}$ for $d=10$. 

\paragraph{Desert point.} We next identify the desert point in the moduli space where the species scale \eqref{eq:speciesIIB} is maximized. To this end, we first notice that due to the SL$(2,\mathbb{Z})$ duality symmetry of the species scale, the extrema of $\Lambda_s$ are located at $i)$ the point $\tau = i$ --- fixed by S-duality --- and $ii)$ the third root of unity $\tau = \rho \equiv -\frac{1}{2}+i\frac{\sqrt{3}}{2}$ --- fixed by the combination of S-duality and axion shift $\tau\rightarrow \tau+1$. We can compute the Eisenstein series numerically over all of moduli space --- as illustrated by figure \ref{fig:IIBspecies} --- confirming that $\tau=\rho$ is the desert point. At these special points we can also compute the Eisenstein series exactly by number theoretical methods; for a detailed analysis we refer to appendix \ref{app:eisensteinSL2}, but let us nevertheless include the values here
\begin{equation}\label{eq:E32values}
    E_{\frac{3}{2}}(i,-i) = 4 \zeta(\tfrac{3}{2})\beta(\tfrac{3}{2}) \approx 9.03\, , \quad E_{\frac{3}{2}}(\rho,\bar{\rho}) = \left(\tfrac{\sqrt{3}}{2}\right)^{1/2} \zeta(\tfrac{3}{2}) \left( \zeta_{\frac{1}{3}}(\tfrac{3}{2}) - \zeta_{\frac{2}{3}}(\tfrac{3}{2})  \right) \approx 8.89\, ,
\end{equation}
where $\beta$ denotes the Dirichlet beta function and $\zeta_a(s)$ the generalized zeta function, see appendix \ref{app:eisensteinSL2} for their precise definitions. The species scale at the desert point is accordingly given by 
\begin{equation}
    \Lambda_s^{\rm max.}=\Lambda_s(\tau=\rho) \approx 0.756\,,
\end{equation}
whereas the species scale at $\tau=i$ is slightly lower 
\begin{equation}
    \Lambda_s(\tau=i) \approx 0.754\,. 
\end{equation}
It is instructive to compare these values with the Type IIA result \eqref{IIAdesert}, which lies in between them. The difference between the IIA and IIB species scales is given solely by the instanton sum in \eqref{eq:E32series}: along the line $\tau_1=0$, this gives a positive contribution to the Eisenstein series, and hence a lower species scale value for IIB at $\tau=i$ compared to IIA; however, by moving along the IIB axion $\tau_1$, we can alter the signs in this instanton sum and achieve a maximal value at $\tau=\rho$ for the species scale. We also want to stress that this example shows there can be points where $\Lambda_s$ has a {\it saddle point}, which here occurs for $\tau=i$, where $d\Lambda_s=0$, but it is neither a minimum nor a maximum. We refer to appendix \ref{app:eisensteinSL2} for the eigenvalues of the Hessian at both of these points.

\paragraph{Diameter.} To determine the diameter of the effective field space set by the bound $\Lambda_s \geq \mu$, we consider a geodesic starting from $\tau=\rho$ along a fixed axion slice $\tau_1=-1/2$ up to the point where $\Lambda_s = \mu$. Note that any other axion value $-1/2 \leq \tau_1 \leq 1/2$ for the endpoint would correspond to an exponential correction, as the length of this segment becomes exponentially small asymptotically. We find that the length of this geodesic and hence the diameter as a function of $\mu$ is given by
\begin{equation}\label{eq:IIBdistance}
    \diam(\mu) = - 2\sqrt{2} \log[\mu] +\frac{1}{6 \sqrt{2}} \log \left[\frac{2^3\pi^5}{3^7 \zeta (3)^4}\right]\, .
\end{equation}
The coefficient corresponds to the expected behavior for an emergent string limit. The shift takes the value
\begin{equation}
    b = \frac{1}{6 \sqrt{2}} \log \left[\frac{2^3\pi^5}{3^7 \zeta (3)^4}\right] \approx -0.073 \, .
\end{equation}
This small shift can be attributed to the contributions $-\log[\sqrt{3}/2]/\sqrt{2}\approx 0.102$ from the short distance between the desert point $\tau_2 = \sqrt{3}/2$ to $\tau_2=1$ and $-\sqrt{2}\log[3 \zeta(3) 2^{3/4}\pi^{-5/4}]/3 \approx -0.175$ coming from the overall coefficient in the scaling of the species scale in \eqref{eq:IIBemergentstring}.

\subsection{M-theory on $T^2$}\label{ssec:9dsugra}
\begin{figure}[!t]
\begin{subfigure}{0.49\textwidth}
\begin{center}
\begin{tikzpicture}
\node[anchor=south west,inner sep=0] at (0,0) {\includegraphics[width=8 cm]{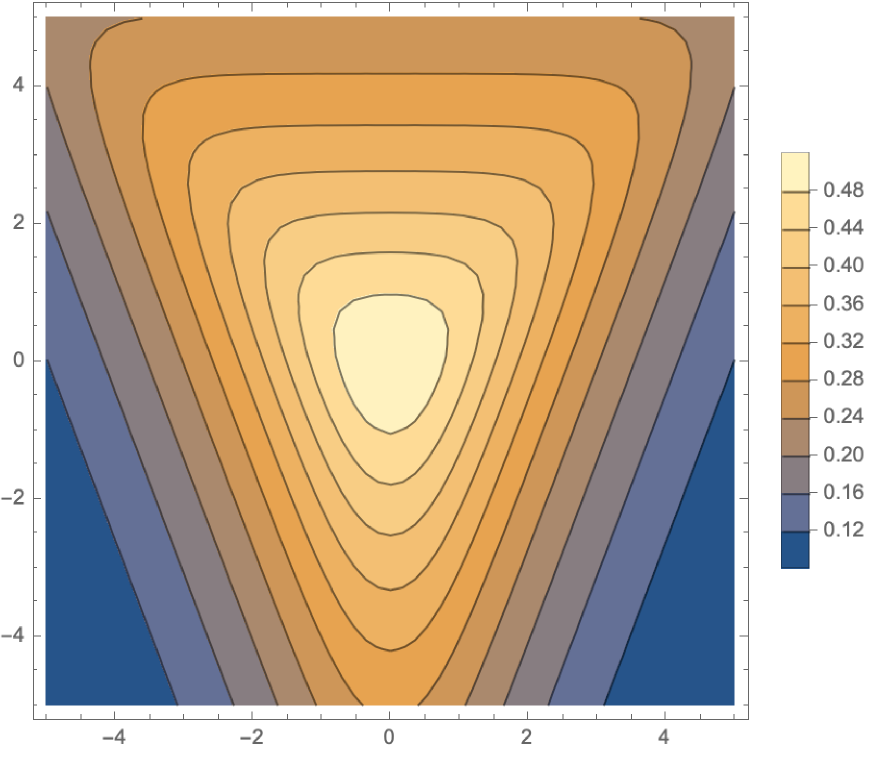}};
\draw[<->] (3.6,1.25) -- node [text width=0.7cm,midway,above ] {1} (3.6,5.7);
\draw[<->] (3.7,1.35) -- node [text width=0.7cm,midway,right ] {2} (5.3,5.4);
\end{tikzpicture}
\end{center}
\begin{picture}(0,0)
\put(200,190){$\Lambda_s$}
\put(-10,130){\footnotesize $\Delta_\cV$}
\put(110,20){\footnotesize $\Delta_{\tau}$}
\end{picture}
\vspace*{-40pt}\caption{Contour plot of $\Lambda_s$.\label{fig:9dspeciescontour}}
\end{subfigure}
\hspace{2pt}
\begin{subfigure}{0.49\textwidth}
\begin{center}
\includegraphics[width=0.95\textwidth]{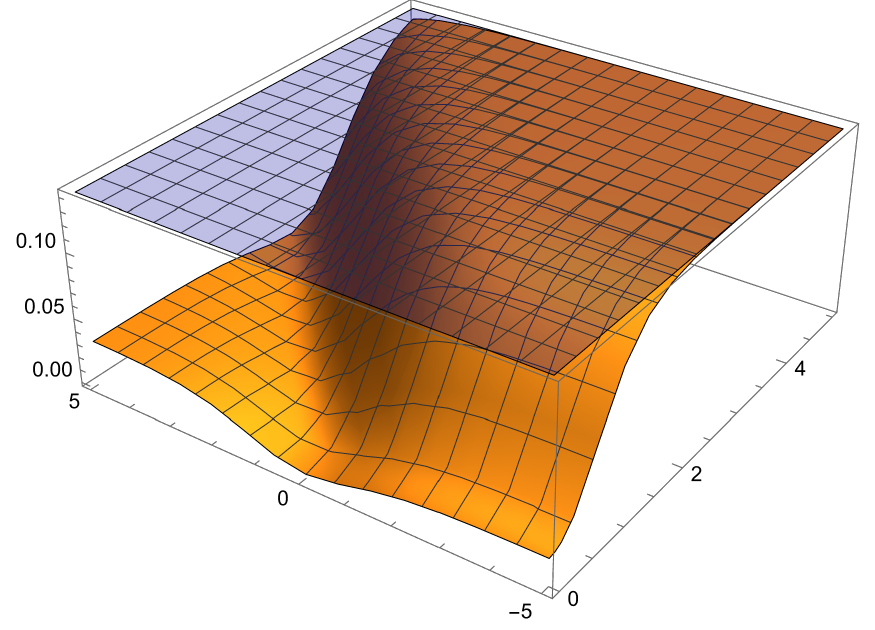}
\end{center}
\begin{picture}(0,0)
\put(73,48){\footnotesize$\Delta_\cV$}
\put(187,62){\footnotesize$\Delta_\tau$}
\put(10,152){\footnotesize$\frac{|\nabla\Lambda_s|^2}{\Lambda_s^2}$}
\end{picture}
\vspace*{-40pt}\caption{Plot of slope $|\nabla\Lambda_s|^2/\Lambda_s^2$.\label{fig:9dslope}}
\end{subfigure}
\caption{\label{fig:9dspecies} Plots of the species scale (left) and slope (right) for M-theory on $T^2$ with constant axion $\tau_1=1/2$. Left: a contour plot of $\Lambda_s$, where arrows 1 and 2 correspond to the geodesic distances \eqref{eq:9ddiam1} and \eqref{eq:9ddiam2} respectively. Right: the slope $|\nabla \Lambda_s|^2/\Lambda_s^2$, bounded from above by $1/7$.}
\end{figure}
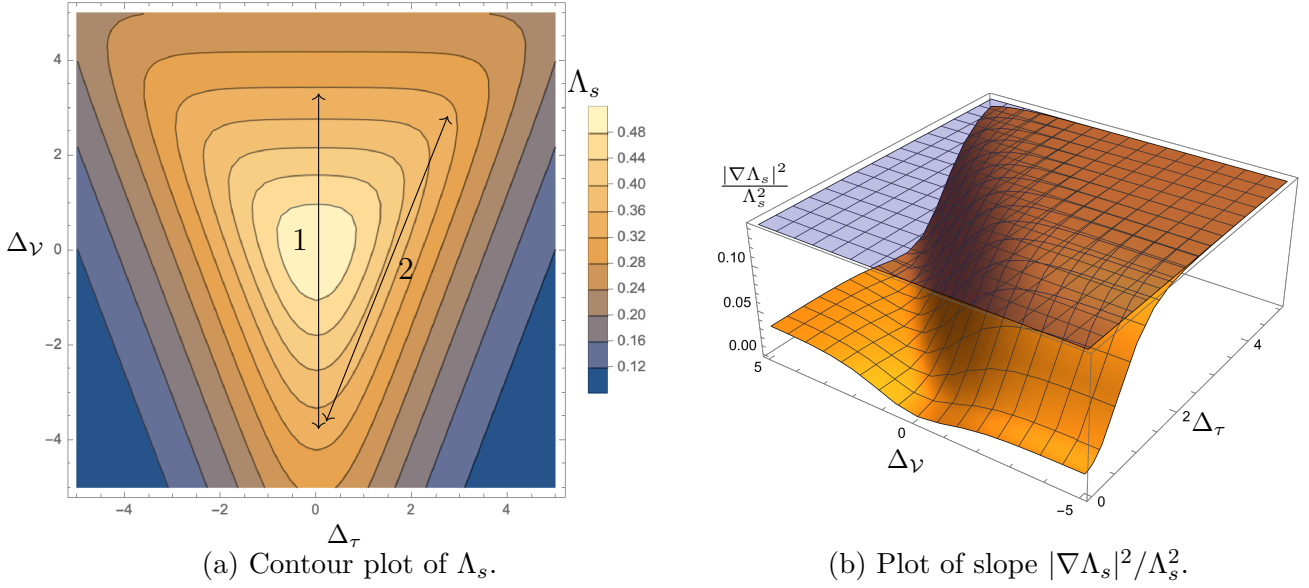

We next consider nine-dimensional supergravity obtained from compactifying M-theory on $T^2$. In this case, the $t_8t_8R^4$-coupling has been computed in \cite{Green:1997di, Green:2010wi} and takes the schematic form
\begin{equation}\label{eq:9dR4}
    S_{R^4}  \sim  M_{\rm pl,9} \int \dd^{9}x  \sqrt{-g}\, \mathcal{V}^{6/7} \left(  \mathcal{V}^{-3/2} E_{\frac{3}{2}}(\tau, \bar \tau) +\frac{2\pi^2}{3}\right)t_8 t_8 R^4\, ,
\end{equation}
where $4\pi^2 \mathcal{V}$ is the volume of the torus in 11d M-theory units and $\tau$ is the complex structure of the torus. Furthermore, the kinetic terms can be determined by the metric\footnote{The coefficient of the $\dd \cV^2$ term agrees with the general expectation for a KK reduction (see e.g.,~\cite{vandeHeisteeg:2023ubh}) from $D=11$ to $d=9$ dimensions.}
\begin{equation}\label{MetricMonT2}
    ds^2 = \frac{\dd\tau \dd{\bar \tau}}{2\tau_2^2} + \frac{9}{14} \frac{\dd \cV^2}{\cV^2}\,. 
\end{equation}
Along trajectories with constant axion $\tau_1$, we can introduce canonically normalized scalar fields
\begin{equation}\label{eq:9dcoordinates}
    \Delta_{\tau} = \log[\tau_2]/\sqrt{2}\, , \qquad \Delta_\cV =\sqrt{14/9} \log[\cV]\, .
\end{equation}
Similar to 10-dimensional Type IIA string theory, we can fix the relation between the species scale and the higher-derivative coefficient $a_4^{(9)}(\tau, \mathcal{V})$ by comparison with 11d M-theory. We therefore realize that in the limit $\mathcal{V}\rightarrow \infty$, the species scale should simply be given by $M_{\text{pl},11}$. In analogy to \eqref{eq:IIAfromM}, we can consider the term 
\begin{equation}\label{9fromM}
    S_{9,R^4} \supset  \frac{M_{\text{pl}, 9}}{2} \int \dd^9 x \sqrt{-g} \hat{a}_4 \frac{t_8t_8 R^4}{M_{\text{pl},9}^6} \left(\frac{M_{\text{pl},9}}{M_{\text{pl},11}}\right)^{6}\, ,
\end{equation}
leading to the identification
\begin{equation}
    a_4^{(9)}(\lambda) = \hat{a}_4  \left(\frac{M_{\text{pl},10}}{\Lambda_s}\right)^{6} \xrightarrow[]{\lambda\rightarrow \infty} \hat{a}_4 \, (4\pi^2 \mathcal{V})^{6/7}\,. 
\end{equation}
This fixes the normalization of the species scale such that, from \eqref{eq:9dR4}, we obtain in Planck units
\begin{equation}\label{eq:9dspecies}
\Lambda_s = \frac{1}{(4\pi^2\cV)^{1/7}}\left(\frac{3}{2\pi^2}\frac{E_{\frac{3}{2}}(\tau,\bar\tau)}{\cV^{3/2}}+1 \right)^{-1/6}\, .
\end{equation}
In figure \ref{fig:9dspeciescontour}, we show a plot of the species scale in terms of $\Delta_\tau$ and $\Delta_{\cV}$. If one wishes, these M-theory coordinates $\tau, \mathcal{V}$ can be mapped to the IIA or IIB dilaton $\lambda_{A/B}$ and circle radius $r_{A/B}$. The precise correspondence between these quantities is given by
\begin{equation}\label{dictionary9d}
    e^{-2D_{A/B}} = \frac{\tau_2^{7/4}}{\mathcal{V}^{3/4}} \,,\quad r_A = \mathcal{V}^{3/4} \tau_2^{1/4}\,,\quad r_B=\frac{1}{r_A}\,,
\end{equation}
where we defined the nine-dimensional dilaton $D_{A/B}$ as
\begin{equation}\label{eq:dilaton9d}
    e^{-2D_{A/B}} = e^{-2\lambda_{A/B}}r_{A/B}\,.
\end{equation}
By substituting these expressions into \eqref{eq:9dR4} and \eqref{MetricMonT2}, one may obtain the $R^4$-term and scalar field metric in the IIA/B coordinates though we choose to work in the M-theory coordinates in this section.

\paragraph{Asymptotic behavior.} Let us now consider how the species scale \eqref{eq:9dspecies} behaves along infinite-distance limits. For theories with maximal supersymmetry in 9d, there are four distinct limits (see also \cite{Calderon-Infante:2023ler}): a 9d emergent string limit, two decompactification limits to 10d Type IIA or IIB supergravity, and a decompactification limit to 11d M-theory. 
\begin{itemize}
    \item We begin with the 9d emergent string limit. In the IIA or IIB coordinates $D_{A/B},r_{A/B}$, this limit corresponds to sending the nine-dimensional dilaton to weak-coupling $D_{A/B} \rightarrow -\infty$ while keeping the radius $r_{A/B}$ constant. Equivalently, using the dictionary \eqref{dictionary9d}, this corresponds to $\tau_2 \sim \cV^{-3} \to \infty$ with $\tau_2 \cV^{3}$ kept fixed. The scaling of $\Lambda_s$ in the distance $\Delta$ along this trajectory is given by
\begin{equation}
    \Lambda_s \sim  e^{-\Delta/\sqrt{7}}\,.
\end{equation}
This scaling with $\Delta$ is obtained straightforwardly in the coordinates $\Delta_\tau,\Delta_\cV$ defined in \eqref{eq:9dcoordinates}, as we then only need to compute Euclidean distances. The coefficient $1/\sqrt{7}$ is consistent with our expectation \eqref{eq:alphas} for the 9d emergent string limit.
\item Next, we consider the two distinct decompactification limits to 10d.  The limit to 10d Type IIA supergravity is obtained by sending $\tau_2 \sim \cV \to \infty$ while keeping $\tau_2/\cV$ fixed; equivalently, this corresponds to scaling the radius $r_A \to \infty$ while keeping the dilaton $\lambda_{A}$ fixed.\footnote{Otherwise we would obtain a decompactification limit super-imposed by a ten-dimensional emergent string limit.} The limit to 10d Type IIB supergravity corresponds to $\mathcal{V}\to 0$ while keeping the dilaton $\tau_2$ fixed. Let us write down the scaling of $\Lambda_s$ in terms of the volume $\cV$ in these limits explicitly
 \begin{equation}\label{eq:9dasymp}
 \begin{aligned}
     \text{IIA}&:  \ \tau_2,\cV \to \infty\, , \ \frac{\tau_2}{\cV} \sim \text{cst}, \quad &\Lambda_s &= \frac{1}{\sqrt[7]{4\pi^2}\sqrt[6]{\frac{3}{\pi^2} \zeta(3) \left(\frac{\tau_2}{\cV}\right)^{3/2}+1}} \cV^{-1/7}\, , \\
     \text{IIB}&: \ \cV \to 0\, , \ \tau_2 = \text{cst},\quad &\Lambda_s &\sim  \frac{1}{\sqrt[7]{4\pi^2}\sqrt[6]{\frac{3}{2\pi^2}E_{\frac{3}{2}}(\tau, \bar \tau)}} \cV^{3/28}\, ,
\end{aligned}
\end{equation}
 where $\tau_2/\cV,\tau$ are kept fixed in the Type IIA and the Type IIB case, respectively. We kept track of the overall factors explicitly, as we will need to tune these later in the computation of the diameter of the field space. Using the Euclidean metric on the coordinates $\Delta_\tau,\Delta_\cV$ defined in \eqref{eq:9dcoordinates}, we find for both trajectories that $\Lambda_s$ scales with the moduli space distance $\Delta$ as
\begin{equation}
    \Lambda_s \sim  e^{-\Delta/\sqrt{56}}\,,
\end{equation}
which is consistent with the expectation of \eqref{eq:alphas}. 
\item Finally we consider the decompactification limit to 11d M-theory. This limit corresponds to decompactifying the $T^2$ by sending $\cV\rightarrow \infty$ while keeping $\tau$ fixed. For the species scale, this gives the scaling of $\Lambda_s$ with the distance as
\begin{equation}
    \Lambda_s = \frac{1}{(2\pi)^{2/7}} \cV^{-1/7} \sim  e^{-\sqrt{2}\Delta/3\sqrt{7}}\,,
\end{equation}
which is also consistent with \eqref{eq:alphas}. Here, we also kept track explicitly of the leading coefficient, as this factor is relevant to the length \eqref{eq:9ddiam1} of the first geodesic considered for the diameter.
\end{itemize}

\paragraph{Species scale polygon.} In figure \ref{fig:9dspeciescontour}, we have provided a plot of constant species scale contours. These contours asymptote to a bilateral triangle for which we briefly elaborate on the physical significance of its corners and sides in relation to the asymptotic limits discussed above. The top left/right corners  correspond to $\cV \sim \tau_2 \to \infty$ and $\cV \sim 1/\tau_2 \to \infty$, respectively, which both lead to a decompactification limit to ten-dimensional Type IIA; the corner at the bottom corresponds to the limit $\cV \to 0$, i.e., the decompactification limit to ten-dimensional Type IIB. In addition, we consider the lines normal to the sides of the triangle passing through the origin: for the top side of the triangle, this corresponds to a decompactification limit to 11d M-theory, while for the left and right side, these yield 9d emergent string limits.

\paragraph{Slope.} The slope of the species scale is depicted in figure~\ref{fig:9dslope}. As indicated in the figure~\ref{fig:9dslope}, the slope is bounded from above, everywhere in the moduli space, by
\begin{equation}
    \frac{|\nabla \Lambda_s|^2}{\Lambda_s^2} \leq \frac{1}{7}\, .
\end{equation}
In this figure, the limit $\Delta_\cV \to \infty$ corresponds to a decompactification to 11d M-theory while the limit $\Delta_\cV \to -\infty$ corresponds to the decompactification to 10d Type IIB.  The valley along $\Delta_\cV,\Delta_\tau \to \infty$ is identified with the decompactification to 10d Type IIA. The plateau corresponds to the 9d emergent string limit. 

\paragraph{Desert point.} Since the field space factorizes between $\tau$ and $\mathcal{V}$, the location of the desert point can be found straightforwardly by first extremizing with respect to the complex structure modulus $\tau$ and subsequently with respect to $\mathcal{V}$. The first step is analogue to our ten-dimensional Type IIB discussion and singles out the third root of unity $\tau=\rho$ as the location of the desert. In turn, extremizing the species scale with respect to the volume $\cV$ yields
\begin{equation}
    \cV = \frac{3^{4/3}}{4 \pi^{4/3}} \big(E_{\frac{3}{2}}(\rho,\bar\rho) \big)^{2/3} \approx 1.009\, ,
\end{equation}
where we have used the numerical value for $E_{\frac{3}{2}}(\rho, \bar \rho)$ computed in \eqref{eq:E32values}. Thus, we find the species scale at the desert point to be 
\begin{equation}\label{eq:desert9d}
    \Lambda_s = \frac{1}{3^{1/42} 7^{1/6} \big(\pi E_{\frac{3}{2}}(\rho,\bar\rho) \big)^{2/21}} \approx 0.513\, ,
\end{equation}
which is substantially lower than the values encountered in ten dimensions for IIA and IIB.

\paragraph{Diameter.} For the diameter of the field space we compare the length of two geodesics: $1)$ a path connecting $\cV=0$ and $\cV = \infty$ with fixed axio-dilaton -- corresponding to a vertical line through the center of the triangle in figure~\ref{fig:9dspecies}; $2)$ a path along the edge of this triangle connecting its bottom and top-right corner, i.e., from $\cV \to 0$ to $\cV \sim \tau_2 \to \infty$. For these endpoints, we need to specify certain order-one constants in the asymptotic behavior in \eqref{eq:9dasymp}: for path $1)$, we need to specify the fixed value for $\tau$, whereas for path $2)$, we need to specify the ratio $\cV/\tau_2$ in the limit $\cV,\tau_2\gg 1$. Below, we work out these two cases in detail.
\begin{itemize}
    \item \emph{Path $1)$ from $\mathcal{V} \to 0$ (10d IIB supergravity) to $\mathcal{V}\to \infty$ (11d supergravity).} To maximize the distance, we must set $\tau=\rho$ since this minimizes $E_{\frac{3}{2}}(\tau,\bar \tau)$, and therefore maximizes the coefficient in \eqref{eq:9dasymp}. We then find the maximal length $\ell_1(\mu)$ of path to be
\begin{equation}\label{eq:9ddiam1}
    \ell_1(\mu) = -\frac{7 \sqrt{7}}{ \sqrt{2}}\log \mu +b_{1}\, .
\end{equation}
The coefficient of the logarithm is the sum of $\sqrt{7\cdot 9/2}$ from decompactification to 11d M-theory and $\sqrt{7 \cdot 8}$ from decompactification to 10d Type IIB. The constant shift of the diameter is given by
\begin{equation}
    b_{1} = -\frac{\sqrt{14}}{3}\log\big[12 \pi E_{\frac{3}{2}}(\rho , \bar\rho)\big]\approx -7.25242\,.
\end{equation}
where we plugged in the numerical value for $E_{\frac{3}{2}}(\rho, \bar \rho)$ from \eqref{eq:E32values}.
    \item \emph{Path $2)$ from $\cV\to 0$ (10d IIB supergravity) to $\cV,\tau_2 \to \infty$ (10d IIA supergravity).} The discussion above tells us that we have to set $\tau=\rho$ in the first limit to 10d Type IIB to maximize the length of path 2). For the second endpoint, we need to determine the ratio $\tau_2/\cV$. The value for this ratio that maximizes the length of path 2) will be determined in the end. Continuing with a generic $\tau_2/\cV$ for now, we find the length $\ell_2(\mu)$ of path 2) to be given by
\begin{equation}\label{eq:9ddiam2}
    \ell_2(\mu) = -14\log \mu + b_{2}(\tau_2/\cV)\, .
\end{equation}    
Note that the coefficient of the logarithm is \textit{not} given by the sum of the contributions coming from the two separate infinite-distance limits, as path $2)$ does not pass through the center of the moduli space. Instead, it is the length of the side of the bilateral triangle in figure \ref{fig:9dspeciescontour} with height $7\sqrt{7/2}$ and width $7\sqrt{2}$. The constant shift $b_2(\tau_2/\cV)$ is given by
\begin{equation}
    b_2\big(\frac{\tau_2}{\cV}\big) = -\tfrac{31}{24}\log[12]+\tfrac{1}{4}\log\big[\frac{\tau_2}{\cV}\big] - \tfrac{7}{6}\log\bigg[ E_{\frac{3}{2}}(\rho,\bar\rho) \big(\pi^2+3\big(\frac{\tau_2}{\cV}\big)^{3/2}\zeta(3)\big)\bigg]+\frac{2\log \pi}{3}\, .
\end{equation}
We can maximize this coefficient as a function of $\tau_2/\cV$ straightforwardly and find that the maximum  is reached for
\begin{equation}
    \frac{\tau_2}{\cV} = \frac{\pi^{4/3}}{3^{4/3} (2\zeta(3))^{2/3}}\, , \quad  b_{2} =  -\tfrac{1}{24}\log\big[ 2^{38}3^{11} \big(7 E_{\frac{3}{2}}(\rho, \bar\rho)\big)^{28} \big(\pi^8 \zeta(3)\big)^{4} \big] \approx -7.97766\, ,
\end{equation}
where we evaluated $E_{\frac{3}{2}}(\rho, \bar\rho)$ numerically using \eqref{eq:E32values}.
\end{itemize}
For small $\mu$, the shortest distance between two points is maximized if they are connected via path 2). Therefore, out of \eqref{eq:9ddiam1} and \eqref{eq:9ddiam2}, the diameter $\diam(\mu)$ as a function of $\mu$ is given by  
\begin{equation}
    \diam(\mu)=\ell_2(\mu) \approx -14 \log \mu -7.97766\,. 
\end{equation}

\subsection{M-theory on $T^3$}\label{ssec:8dsugra}
As a final setup with maximal supersymmetry, we consider eight-dimensional supergravity arising from M-theory compactified on $T^3$, or equivalently, Type IIB on $T^2$ following \cite{Kiritsis:1997em, Green:2010wi}. From table~\ref{table:maxsugras}, we infer that the moduli space of maximal supergravity in 8d is given by 
\begin{equation}
    \cM_{8} = \left(SL(3,\mathbb{Z})\times SL(2,\mathbb{Z})\right)\Big\backslash \frac{SL(3,\mathbb{R}) \times SL(2,\mathbb{R})}{SO(3)\times SO(2)}\,. 
\end{equation}
In particular, the $SL(3,\mathbb{R})$ factor is interesting since its structure differs from the $SL(2,\mathbb{R})$ and $\mathbb{R}_+$ moduli spaces encountered in the previous examples. For definiteness, we consider Type IIB on $T^2$ in the following. The kinetic terms in the eight-dimensional Einstein frame read
\begin{equation}\label{8daction}
    S_8 = \frac{M_{\rm pl,8}^6}{2} \int \dd^8 x \sqrt{-g} \left( R- \frac{1}{6} \frac{\partial \nu^2}{\nu^2}-\frac{1}{2} \frac{\partial U \partial \bar U}{U_2^2} - \frac{1}{2} \frac{\partial \tau \partial \bar \tau}{\tau_2^2} - \nu \frac{|\tau \partial B_N + \partial B_R|^2}{2\tau_2}\right).
\end{equation}
Here, $U=U_1+i U_2$ is the complex structure parameter of the $T^2$, $\tau=\tau_1+i\tau_2$ is the axio-dilaton of ten-dimensional Type IIB string theory, $B_{N}(B_R)$ are the scalars obtained from reducing the NS-NS(R-R) two-form along the $T^2$, and we defined $\nu = 1/(\tau_2 \cV^2)$ where $4\pi^2 \cV$ is the string frame volume of the $T^2$. The complex structure $U$ spans the $SL(2,\mathbb{R})$ component of the moduli space $\cM_8$, whereas the $SL(3,\mathbb{R})$-part is parameterized by $\nu$, $\tau$ and $B_{N,R}$. 

The action~\eqref{8daction} may be brought into a form invariant under the $SL(3,\mathbb{Z})$ U-duality group by introducing~\cite{Hull:1994ys} 
\begin{equation}\label{eq:SL3M}
    M = \nu^{1/3} \begin{pmatrix}
        \frac{1}{\tau_2} & \frac{\tau_1}{\tau_2} & \frac{\Re(B)}{\tau_2} \\
        \frac{\tau_1}{\tau_2} & \frac{|\tau|^2}{\tau_2} & \frac{\Re(\bar \tau B)}{\tau_2} \\
        \frac{\Re(B)}{\tau_2} & \frac{\Re(\bar \tau B)}{\tau_2} & \frac{1}{\nu} + \frac{|B|^2}{\tau_2}
    \end{pmatrix}\, ,
\end{equation}
where $B=B_R+\tau B_N$. In terms of $M$, the action~\eqref{8daction} can then be rewritten as 
\begin{equation}\label{eq:8dkinetic}
    S_8 = \frac{1}{2\kappa_8^2} \int \dd^8 x \sqrt{-g} \left( R- \frac{1}{2} \frac{\partial U \partial \bar U}{U_2^2}+\frac{1}{4} \text{Tr}(\partial M \partial M^{-1}) \right)\, .
\end{equation}
Again, we consider the coefficient of the $t_8t_8R^4$-coupling in the effective action which is given by
\begin{equation}\label{coeff8d}
    \cE^{(2)}(M,U,\bar U) = E^{SL(3)}_{ \frac{3}{2}}(M) + E^{SL(2)}_{1}(U,\bar U)\,.
\end{equation}
Here, the term corresponding to the $SL(3)$ part of the moduli space is defined as
\begin{equation}\label{eq:Esl3}
    E^{SL(3)}_{s} = \sum_{\substack{(m_1,m_2,m_3) \in \bbZ^3 \\ (m_1,m_2,m_3)\neq (0,0,0)}} \nu^{-s/3} \left( \frac{|m_1+m_2 \tau+B m_3|^2}{\tau_2}+\frac{m_3^2}{\nu}\right)^{-s}\,,
\end{equation}
for $s={\tiny \frac32}$. On the other hand, the SL(2)-term is given by
\begin{equation}\label{eq:Esl2}
    E_s^{SL(2)} = \sum_{(p,q)\neq (0,0)} \frac{U_2^s}{|p+q U|^{2s}} \, ,
\end{equation}
for $s=1$. In eight dimensions, the $t_8t_8R^4$-term is conformally invariant and its coefficient in \eqref{coeff8d} is divergent due to the contribution of massless modes to the conformal anomaly. Both terms appearing in \eqref{coeff8d} therefore need to be properly regularized. Evaluating \eqref{eq:Esl2} for $s=1+\epsilon$ and subtracting the pole, one finds 
\begin{equation}\label{hatE1}
    \hat{E}_1^{SL(2)} = -2\pi \log(U_2|\eta(U)|^4)\, .
\end{equation}
up to a constant infrared ambiguity. This is reminiscent of the situation in the vector multiplet sector of Type II string theory compactified on a Calabi--Yau threefold $Y_3$. There, the conformally invariant $R^2$-term also has a coefficient that diverges due to the contributions of massless modes. Regularizing this coefficient yields an expression similar to \eqref{hatE1}\cite{Bershadsky:1993cx,Bershadsky:1993ta} for the case of $Y_3=(K3\times T^2)/\mathbb{Z}_2$.

Similarly, the regularization of $E_{\frac{3}{2}}^{SL(3)}$ has been carried out in detail in \cite{Kiritsis:1997em}. The finite part is given by
\begin{equation}\label{hatESl3}
    \hat{E}_{\frac{3}{2}}^{SL(3)} = 2 \frac{\tau_2^{3/2}}{\nu^{1/2}} \zeta(3) + \frac{2\pi^2}{3} T_2 +4\pi \log \nu^{1/3}+ \mathcal{I}_D + \sum_{(p,q)=1}\mathcal{I}_{p,q} \, .
\end{equation} 
Here $\mathcal{I}_D$ represents the D$(-1)$-instanton contribution given by
\begin{equation}\label{eq:Dinstantons}
    \mathcal{I}_D = 8\pi \sqrt{\frac{\tau_2}{\nu}} \sum_{p\neq0}\sum_{n=1}^\infty\left|\frac{p}{n}\right|K_1(2\pi \tau_2|p|n)e^{2\pi ip n\tau_1}\,,
\end{equation}
with $K_1$ denoting the Bessel function. On the other hand, we have 
\begin{equation}\label{eq:pqstringinstantons}
    \mathcal{I}_{p,q} = -8\pi \text{Re} \log \left[\prod_{n=1}^{\infty} \left(1 - e^{2\pi i n  T_{p,q}}\right)\right]\,,\qquad T_{p,q}= (q B_R - pB_N )+ i|p + q\tau |\cV\,,
\end{equation}
encoding the contributions from Euclidean $[p,q]$-strings wrapping the $T^2$. The effective $t_8t_8 R^4$ coupling is then determined by the regularized Eisenstein series $\hat{E}_1^{SL(2)}$ and $\hat{E}_{\frac{3}{2}}^{SL(3)}$ up to an additive renormalization constant. By studying the full string amplitude this additive constant has been determined in~\cite{Bossard:2023bhv} to be $\frac{22\pi}{3}$ when evaluating the effective coupling at the Planck scale.

\begin{figure}[!t]
\begin{subfigure}{0.49\textwidth}
\begin{center}
\includegraphics[width=8 cm]{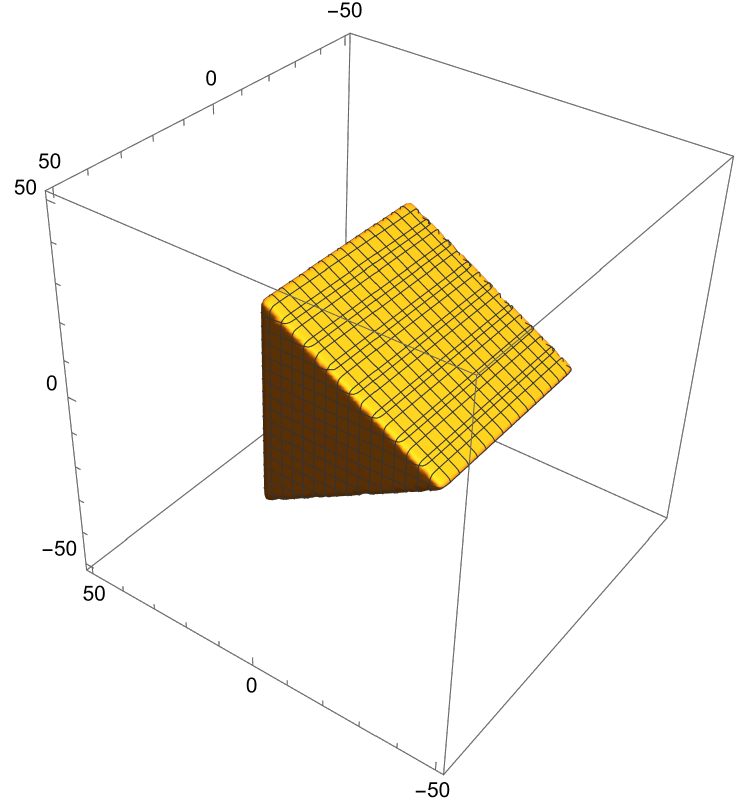}
\end{center}
\begin{picture}(0,0)
\put(55,260){\footnotesize $\Delta_U$}
\put(0,153){\footnotesize $\Delta_\tau$}
\put(70,55){\footnotesize $\Delta_{\nu}$}
\end{picture}\vspace*{-0.0cm}
\vspace*{-40pt}\caption{Plot of the region $\Lambda_s \geq 10^{-4} M_{\rm pl,8}$.\label{fig:8dspeciespolygon}}
\end{subfigure}
\hspace{2pt}
\begin{subfigure}{0.49\textwidth}
\begin{center}
\includegraphics[width=0.95\textwidth]{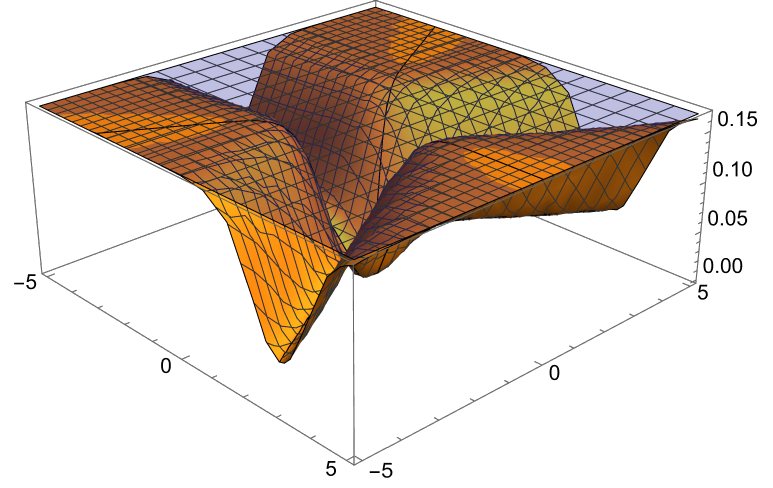}
\end{center}
\begin{picture}(0,0)
\put(45,55){\footnotesize$\Delta_\nu$}
\put(167,52){\footnotesize$\Delta_\tau$}
\put(210,155){\footnotesize$\frac{|\nabla\Lambda_s|^2}{\Lambda_s^2}$}
\end{picture}
\vspace*{-40pt}\caption{Plot of slope $|\nabla\Lambda_s|^2/\Lambda_s^2$.\label{fig:8dslope}}
\end{subfigure}
\caption{\label{fig:8dspecies} Plot of the species scale and its slope of M-theory on $T^3$ as a function of $\Delta_U$, $\Delta_{\tau}$ and $\Delta_\nu$ defined in \eqref{Deltas8d}. We have set all axions to zero $\tau_1=U_1=B_R=B_N=0$. Left: the region where the species scale obeys the lower bound $\Lambda_s \geq 10^{-4} M_{\rm pl,8}$. Right: the slope of the species scale in the $(\Delta_\tau,\Delta_\nu)$-plane together with the plateau at $|\nabla\Lambda_s|^2/\Lambda_s^2=1/6$ for the 8d emergent string limit.} 
\end{figure}
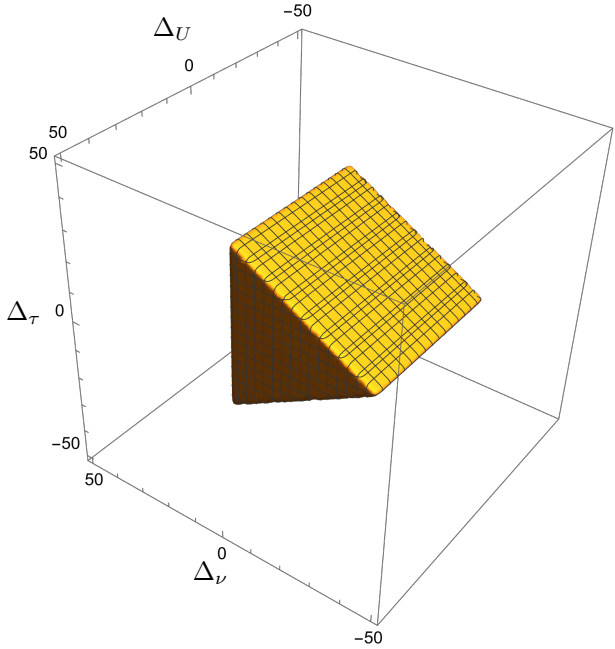
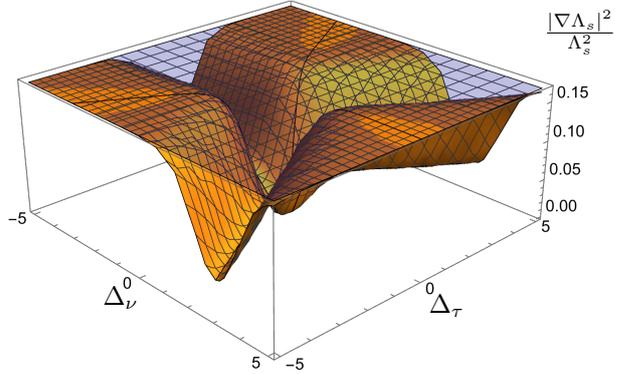

To relate this higher-derivative correction to the species scale, we need to fix its normalization. We therefore realize that in the limit of large-radius and weak-coupling, the species scale should be given by the species scale of ten-dimensional Type IIB string theory discussed in section~\ref{ssec:10dIIB}. In the limit $\cV, \tau_2\rightarrow \infty$, the $t_8t_8R^4$-coupling is dominated by the first term in $\hat{E}_{\frac{3}{2}}^{SL(3)}$ in \eqref{hatESl3} such that 
\begin{equation}\label{E2limit}
    \mathcal{E}^{(2)}\sim 2  \zeta(3) \tau_2^2 \mathcal{V} +\dots \,. 
\end{equation}
Dimensional reduction of ten-dimensional Type IIB string theory on a torus with volume $4\pi^2\mathcal{V}$, we find that in eight-dimension, the coefficient of the $t_8t_8R^4$-term should be given by 
\begin{equation}
    a_4^{(8)}(\tau, \mathcal{V}) \xrightarrow[]{\tau_2,\cV\rightarrow \infty} \hat{a}_4 \left(12 \zeta(3)\, \tau_2^2 \cV +\dots \right)\,. 
\end{equation}
Comparison with \eqref{E2limit} determines the species scale in eight-dimensional maximal supergravity to be 
\begin{equation}\label{eq:species8d}
    \Lambda_s = \frac{1}{4^{1/6}}\left( \hat{E}_{\frac32}^{SL(3)} - 2\pi \log(U_2|\eta(U)|^4)\right)^{-1/6}\, .
\end{equation}
In figure~\ref{fig:8dspecies}, we show the behavior of this species scale $\Lambda_s$ over the moduli space. 

\paragraph{Species scale polygon.} In figure \ref{fig:8dspeciespolygon}, we show the part of (saxionic) field space for which the bound $\Lambda_s \geq \mu$ for $ \mu= 10^{-4}\, M_{\text{pl}, 8}$ is satisfied in terms of the three saxionic coordinates 
\begin{equation}\label{Deltas8d}
    \Delta_\nu = \log[\nu]/\sqrt{6}\,,\quad \Delta_\tau= \log[\tau_2]/\sqrt{2}\,,\quad \Delta_U = \log[U_2]/\sqrt{2}\,.
\end{equation}
The triangular side is parameterized by $\Delta_\tau $ and $\Delta_\nu$ spanning the $SL(3,\mathbb{R})$ factor of the moduli space, while the transverse direction is parameterized by $\Delta_U$ parameterizing $SL(2,\bbR)$ component. The geodescis passing through the corners of the polygon correspond to decompactification limits to 9d, while geodesics normal to rectangular side correspond to emergent string limits in 8d. We will discuss the asymptotic structure of this field space in more detail momentarily.

\paragraph{Slope.} Figure \ref{fig:8dslope} shows the slope $|\nabla \Lambda_s|^2/\Lambda_s^2$ in the $(\Delta_\tau,\Delta_\nu)$-plane of SL$(3,\bbR)$ which can be identified with the triangular side of the polygon in figure~\ref{fig:8dspeciespolygon}. More precisely, the corners of this triangle --- corresponding to the directions of 9d decompactification limits --- are identified as the valleys for the slope, while the normals to the edges of this triangle --- corresponding to 8d emergent string limits --- are identified with the maximal plateaus for the slope. Note that the slope surpasses the emergent string value $1/6$ along these directions; we will explain, in section \ref{sec:refine}, why this behavior is unphysical and that the slope should instead be bounded from above by $1/6$ everywhere. This is achieved by removing zero mode contributions --- logarithmic terms in  $U_2$ and $\nu$ in \eqref{eq:Esl2} and \eqref{hatESl3} --- to the $R^4$-term that should not be included in the species scale.\\ 

\paragraph{Asymptotic behavior.} The parametric behavior of the species scale in the asymptotic limits of the theory has been in analyzed in detail in~\cite{Calderon-Infante:2023ler}. Below we show that our proposal~\eqref{eq:species8d} correctly reproduces the expected parametric behavior of the species scale. 
\begin{itemize}
\item \emph{Emergent string limit.} For this limit, we take the 8d string coupling $g_{s,8} = 1/(V \tau_2^2) = \nu^{1/2}/\tau_2^{3/2}$ to zero while keeping $\tau_2 \nu $ (and $U_2$) constant.  In this limit, the first term in the species scale \eqref{eq:species8d} diverges the quickest. We can express it in terms of the canonically normalized scalars \eqref{Deltas8d} as
\begin{equation}
    \Lambda_s \sim e^{-\frac{1}{2\sqrt{2}}\Delta_\tau-\frac{1}{2\sqrt{6}}\Delta_\nu}\, .
\end{equation}
For the emergent string limit, we then consider the trajectory $(\Delta_\tau,\Delta_\nu) = (\sqrt{3}/2,1/2)\Delta$ with $\Delta$ the geodesic distance, giving us the scaling
\begin{equation}
    \Lambda_s \sim e^{-\Delta/\sqrt{6}}\, ,
\end{equation}
with the expected coefficient $1/\sqrt{d-2}$ in $d=8$.

\item \emph{Decompactification to 11d.} This limit corresponds to taking the large-complex structure limit, $U_2 \to \infty$, for the $T^2$ on which we compactified Type IIB while keeping the volume of $T_2$ and the 10d string coupling $\tau_2$ constant. To see that this limit indeed corresponds to a decompactification to 11d M-theory, let us for simplicity consider a rectangular torus $T^2=S^1_1\times S^1_2$ such that 
\begin{equation}
    U_2 = \frac{r_1}{r_2} \,\qquad T_2 = r_1 r_2\,,
\end{equation}
with $r_{1/2}$ the radii of the respective $S^1$'s. To reach the proposed limit for $U_2$ and $T_2$, we hence need to consider the scaling 
\begin{equation}\label{scalingMtheory}
    r_1 \sim \frac{1}{r_2} \rightarrow \infty\,. 
\end{equation}
The light states in this limit are the KK-modes on $S^1_1$ and the winding modes of both the fundamental string of Type IIB and the D1-brane on $S^1_2$. We therefore have three KK-like towers. Performing a T-duality on $S^1_2$, we hence decompactify to 10d Type IIA. Since we keep the Type IIB string coupling constant, the Buscher rules imply that the limit \eqref{scalingMtheory} corresponds to a strong coupling limit in Type IIA 
\begin{equation}
    g_{\rm IIA} \sim r_1\,,
\end{equation}
such that, indeed, we obtain a decompactification to eleven dimensions. As far as the species scale is concerned, in this limit the last term in \eqref{eq:species8d} dominates. In terms of the canonically normalized scalar $\Delta_U$ defined in \eqref{Deltas8d}, we find the scaling
\begin{equation}
    \Lambda_s \sim e^{-\frac{1}{3\sqrt{2}}\Delta_U} = e^{-\sqrt{\frac{11-8}{(11-2)(8-2)}} \Delta_U}\, ,
\end{equation}
which indeed has the correct exponent \eqref{eq:alphas} for a decompactification from $d=8$ to $D=11$ dimensions.

\item \emph{Decompactification to 10d.} This limit corresponds to keeping the dilaton $\tau_2$ and the complex structure $U_2$ of the torus fixed while sending the volume of the torus $\cV \to \infty$. For this limit, the first two terms in \eqref{eq:species8d} dominate, both leading to the scaling $\Lambda_s \sim \cV^{-1/6}$. From the kinetic terms in \eqref{eq:8dkinetic} we find that the volume scales as $\cV \sim \exp(\sqrt{3/2}\Delta)$ in terms of the moduli space distance $\Delta$ along this trajectory. For the species scale we then find that
\begin{equation}
    \Lambda_s \sim e^{-\frac{1}{2\sqrt{6}}\Delta }\,,
\end{equation}
which agrees with the expected coefficient \eqref{eq:alphas} for a decompactification from $d=9$ to $D=10$ dimensions.

\item \emph{Decompactification to 9d.} Finally, we can take a decompactification limit to 9d by taking the large-volume and large-complex structure limit for the $T^2$ simultaneously, i.e., sending $\cV,U_2\to \infty$ while keeping $\cV/U_2$ fixed. To see that this limit decompactifies to one dimension higher, note that for a rectangular torus  --- $\cV=r_1 r_2$ and $U_2=r_1/r_2$ --- this limit corresponds to taking $r_1 \to \infty$ while keeping $r_2$ fixed. For this limit, the dominant terms in the species scale are the second and third term in \eqref{eq:species8d}. Along this trajectory we find by using the kinetic terms \eqref{eq:8dkinetic} that the moduli $U_2,\cV$ scale with the field space distance $\Delta$ as $U_2\sim \cV \sim e^{\sqrt{6/7}\Delta}$. We then find
\begin{equation}\label{eq:8ddecomp}
    \Lambda_s = \frac{1}{\sqrt[6]{4}\sqrt[6]{E_{\frac{3}{2}}(\tau,\bar \tau)+\frac{2\pi^2}{3}U_2 \sqrt{\nu}}} \nu^{1/12} \sim e^{-\frac{1}{\sqrt{42}}\Delta}\, .
\end{equation}
Here we kept track explicitly of the leading coefficient for the computation of the diameter of the field space later; note in particular that the D-instanton sum in \eqref{eq:Dinstantons} is finite in this limit, and combines with the other $\tau_2$ dependent terms into $E_{\frac{3}{2}}(\tau,\bar \tau)$ given in \eqref{eq:E32series}. The exponent agrees with \eqref{eq:alphas} for a decompactification from $d=8$ to $D=9$ dimensions. 
\end{itemize}

\paragraph{Desert point.} We next determine the point that maximizes the species scale \eqref{eq:species8d}. The extremization of the $SL(2,\bbR)$ moduli space factor yields the third root of unity $U=\rho$, as its dependence is given by the same function as the species scale of $(\text{K3}\times T^2)/\bbZ_2$ studied in \cite{vandeHeisteeg:2022btw}. For the extremization over $SL(3,\bbR)$, we have scanned over duality fixed points of $SL(3,\bbZ)$. We find that the lattice that minimizes $E_{\frac{3}{2}}(M)$ is given by
\begin{equation}\label{eq:8ddesert}
    M_{A_3} =2^{-2/3}\begin{pmatrix}
        2 & -1 & 0 \\
        -1 & 2 & -1 \\
        0 & -1 & 2
    \end{pmatrix}\, , \qquad \tau=-\frac{1}{2}+\frac{i\sqrt{3}}{2}\, , \ \nu=\frac{3\sqrt{3}}{4}\, , \ B_R=-\frac{1}{3}\, , \ B_N = -\frac{2}{3}\, ,
\end{equation}
with $M_{A_3}$ the Cartan matrices of the $A_3$ root lattice, also known as the face-centered cubic (FCC) or hexagonal close-packed (HCC) lattice. The value of the Eisenstein series and species scale is given by
\begin{equation}
        E_{\frac{3}{2}}^{SL(3)} (M_{A_3})= 8.79961\, , \quad \Lambda_s^{\rm max} = \frac{1}{\sqrt[6]{E_{\frac{3}{2}}^{SL(3)} (M_{A_3})+E_{1}^{SL(2)} (\rho, \bar \rho)}} \approx 0.5038
\end{equation}
In appendix~\ref{app:EisensteinSl38d}, we collect the values at other fixed points of $SL(3,\mathbb{Z})$ and show that, indeed, the value of the species scale is smaller at these points. Moreover, we computed the Hessian confirming that $M_{A_3}$ gives a maximum for the species scale. In \cite{Sarnak2006} it was also proven numerically that $M_{A_3}$ indeed give a minimum of the Eisenstein series, confirming our result for the desert point.

\paragraph{Diameter.} We next determine the largest distance between two points inside a finite moduli space region set by $\Lambda_s \geq \mu$. For simplicity, we ignore the axions in the infinite-distance regions, as these only contribute exponential corrections to the diameter. 
\begin{itemize}
    \item Let us first consider the S-duality transformations in $SL(3,\bbZ)\times SL(2,\bbZ)$ that cut out our fundamental domain. The S-duality of $SL(2,\bbZ)$ simply restricts us to the $|U|\geq 1$ regime as usual --- it cuts the polygon in figure \ref{fig:8dspeciespolygon} down the middle of the $\Delta_U$ axis. The S-duality transformations of $SL(3,\bbZ)$ then act on the $(\Delta_\tau,\Delta_\nu)$-plane, corresponding to the triangular-shaped cross section of the remaining half-polygon. As discussed in more detail in appendix~\ref{app:Sl3dualities}, these S-dualities partition this triangle into 6 fundamental regions. We focus on the fundamental regime set by $\tau_2\geq 1$ and $\nu \tau_2 \leq 1$, in which the instanton corrections in \eqref{eq:Dinstantons} and \eqref{eq:pqstringinstantons} are suppressed. In particular, these duality transformations tell us that all corners of the polygon in figure \ref{fig:8dspeciespolygon} are identified, corresponding to the same decompactification limit to $D=9$. This means that the longest distance between two points cannot be given by one of the sides of the polygon; rather, we should consider a geodesic from an interior point --- which we take to be the desert point \eqref{eq:8ddesert} --- to a corner corresponding to the decompactification to $D=9$.
    \item Next we examine this decompactification limit to $D=9$ more closely. We parameterize this limit by sending $U_2 \sim 1/\sqrt{\nu} \to \infty$ while keeping their ratio $U_2\sqrt{\nu}$ and $\tau_2$ fixed. We have to maximize the distance over these fixed parameters. To this end, it is useful to consider the leading behavior of the species scale in this limit given in \eqref{eq:8ddecomp}: we see that $\tau$ appears only in $E_{\frac{3}{2}}(\tau,\bar{\tau})$ in the leading coefficient, which is maximized for $\tau=\rho$. The other parameter $U_2\sqrt{\nu}$ we will keep generic for now, and extremize after the computation of the distance.
    \item With the above preparations in place, let us next compute the distance between the desert point \eqref{eq:8ddesert} and a point along the $D=9$ decompactification limit $U_2 \sim 1/\sqrt{\nu} \to \infty$ with $\tau=\rho$ and $U_2\sqrt{\nu}$ fixed arbitrarily. We then find that the diameter is given by
\begin{equation}
    \diam(\mu) = -\sqrt{42}\log[\mu]+b(U_2 \sqrt{\nu})\, .
\end{equation}
    The coefficient $\sqrt{6 \cdot 7}$ corresponds to a decompactification limit to 9d. The shift is a function of the remaining parameter $U_2\sqrt{\nu}$, given by
\begin{equation}
    b = \frac{-14 \log \left[2 \pi ^2 U_2 \sqrt{\nu}+3 E_{\frac{3}{2}}(\rho,\bar\rho)\right]+6 \log[U_2 \sqrt{\nu}]+17 \log (3)-30 \log (2)}{2 \sqrt{42}}\, .
\end{equation}
Extremizing this coefficient for the fixed parameter $U_2 \sqrt{\nu}$ gives us a maximum at
\begin{equation}
    U_2 \sqrt{\nu} = \frac{9 E_{\frac{3}{2}}(\rho , \bar \rho)}{8\pi^2}\, ,
\end{equation}
with the shift value being
\begin{equation}
    b = -\frac{8 \log (E_{\frac{3}{2}}(\rho, \bar \rho))+12 \log (\pi )+14 \log (7)-15 \log (3)+20 \log (2)}{2 \sqrt{42}} \approx  -4.309\, ,
\end{equation}
which, again, is negative.
\end{itemize}

\section{Species scale and 16 supercharges}\label{sec:16Qs}
After having discussed the cases with maximal supersymmetry in some detail, we now turn to theories with minimal supersymmetry in ten and nine dimensions. To that end, we consider theories arising from Ho\v{r}ava--Witten theory, Type I string theory, and the two heterotic strings. Again, we focus on the higher-derivative terms involving contractions of four Riemann tensors. The situation is therefore similar to the maximally supersymmetric case, but there are some crucial differences: 
\begin{itemize}
    \item Unlike for theories with 32 supercharges, the $t_8t_8 R^4$-interaction in theories with 16 supercharges is not 1/2-BPS. In maximal supergravity, this property ensured that the $t_8t_8R^4$-coupling does not receive any perturbative corrections beyond one-loop level. Such a protection is absent in theories with 16 supercharges even though, as reviewed in \cite{Green:2016tfs}, there is evidence that higher-loop corrections to $t_8t_8R^4$ are also absent in this case. 
    \item Compared to maximal supergravity, there exist two other terms at the eight-derivative level that contribute to the effective action corresponding to $t_8 \tr R^4$- and $t_8(\tr R^2)^2$-couplings. A priori, there is an ambiguity for the coefficients of the individual couplings due to the identity 
    \begin{equation}
          t_8t_8 R^4 -24t_8\tr R^4 + 6 t_8 (\tr R^2)^2 =0 \,.
    \end{equation}
    There exists, however, one combination of these couplings that is related via supersymmetry to the anomaly-cancelling term, i.e., $-12\epsilon_{10} B Y_8^{(\rm gs)}$ which arises at one-loop in the effective heterotic action. Here, the eight-form $Y_8^{(\rm gs)}$ is given, in the absence of a field strength $F$ for the gauge group, by
    \begin{equation}
        Y_8^{(\rm gs)}(R,F=0) = \tr R^4+\frac14 (\tr R^2)^2\,. 
    \end{equation}
    Since it is related to an anomaly, the coefficient of $-12\epsilon_{10} B Y_8^{(\rm gs)}$ does not receive corrections beyond one-loop. Expressing the higher-derivative terms through the superinvariants (cf. \cite{Green:2016tfs})
    \begin{equation}
    \begin{aligned}
        X_1 &= t_8 \tr R^4 - \frac14 \epsilon_{10} B \tr R^4\,,\quad X_2 =t_8 (\tr R^2)^2 - \frac14 \epsilon_{10} B (\tr R^2)^2 \,,\\ 
        \mathcal{I}_2 &= J_0 - 24(X_1 - \frac14 X_2)\,,
    \end{aligned}
    \end{equation}
    where $J_0=t_8t_8R^4 - \frac18 \epsilon_{10}\epsilon_{10} R^4$, one realizes that $-12\epsilon_{10} B Y_8^{(\rm gs)}$ is contained in the combination \cite{Green:2016tfs}
    \begin{equation}
        -\cI_2 +24 X_1 + 18 X_2 = 24\, t_8 \tr R^4 +18\, t_8(\tr R^2)^2 +\frac14 \epsilon_{10} \epsilon_{10} R^4 - 12 \epsilon_{10} B Y_8^{(\rm gs)}\,.
    \end{equation}
    Therefore, the coefficient of the coupling 
    \begin{equation}
       \cI_{\rm anom}= 24\, t_8 \tr R^4 +18\, t_8(\tr R^2)^2 +\frac14 \epsilon_{10} \epsilon_{10} R^4 \,,
    \end{equation}
    is also protected by supersymmetry and does not receive corrections beyond one loop. In addition, there exists a term involving $(\tr R^2)^2$ that already arises at tree-level in heterotic string coming from the $t_8(\tr F^2 - \tr R^2)^2$ which is unrenormalized beyond tree level. 
    \item Given that, compared to the maximally supersymmetric case, we now have three terms appearing at the eight-derivative level. Hence, we need to be more careful when defining the species scale in terms of the eight-derivative terms. Schematically, the effective action at order $R^4$ in $d$-dimension takes the form 
    \begin{equation}\label{16superchargesgeneral}
        S_{R^4,d} = \frac{M_{\rm pl,d}^{d-8}}{2} \int \dd^{d} x \sqrt{-g} \left(a_{4, t_8}(\phi)\, t_8 t_8 R^4 + a_{4,\rm anom}(\phi)\, \mathcal{I}_{\rm anom}+a_{4,\rm tree} (\tr R^2)^2 \right)\,,
    \end{equation}
    where $\phi$ denotes any scalar field in the theory. The species scale is then given by 
    \begin{equation}\label{defspeciesN=1}
        \left(\frac{M_{\text{pl},d}}{\Lambda_s(\phi)}\right)^6 = \max\left\{\frac{a_{4, t_8}(\phi)}{\hat{a}_4}, \frac{a_{4,\rm anom}(\phi)}{\hat{a}_4}, \frac{a_{4,\rm tree}}{\hat{a}_4} \right\} \,,
    \end{equation}
    where, similar to \eqref{cnhatcn}, we divided by a constant $\hat{a}_4$ that sets the overall normalization of the species scale. 
\end{itemize}
In the following, we discuss the species scale in theories with minimal supersymmetry restricting to ten and nine dimensions. 

\subsection{Heterotic $E_8\times E_8$ in 10d}\label{heteroticE}
The eight-derivative terms for the heterotic $E_8\times E_8$ string in ten dimensions have been computed in \cite{Gross:1986mw,Ellis:1987dc,Abe:1988cq} and take the schematic form 
\begin{equation}\label{hete8e8}
    S_{{\rm HE},R^4} \sim M_{\rm pl,10}^2 \int \dd^{10}x \sqrt{-g} \left[\left(\frac{2\zeta(3)}{g_{\rm he}^{3/2}} +\frac{2\pi^2}{3} g_{\rm he}^{1/2} \right)t_8t_8 R^4 -\frac{(\tr R^2)^2}{2g_{\rm he}^{3/2}} +\frac{2\pi^2}{3} g_{\rm he}^{1/2} \cI_{\rm anom} \right]\,,
\end{equation}
where we disregarded terms involving the $B$-field. Here, $g_{\rm he}$ is the string coupling of the heterotic $E_8\times E_8$ string. The above expression fixes the relative factor between the different terms, but does not fix the overall normalization $\hat{a}_4$ for the species scale. As in the Type IIA case discussed in section~\ref{ssec:IIA}, we can determine the relation between the species scale and the coefficients appearing in \eqref{hete8e8} by comparing to the eleven-dimensional M-theory compactified, in this case, on $S^1/\mathbb{Z}_2$, i.e., Ho\v{r}ava--Witten theory. Denoting the radius of the $S^1$ again by $R_{11}$, we can equate
\begin{equation}
    g_{\rm he} = (R_{11} M_{11})^{3/2} \,,
\end{equation}
and repeat the analysis of section~\ref{ssec:IIA} while keeping in mind that the length of the interval $S^1/\mathbb{Z}_2$ is $\pi R_{11}$. In the large-$R_{11}$ limit, the eight-derivative action is dominated by the one-loop terms. In fact, from \eqref{hete8e8}, we find that 
\begin{equation}
    \lim_{g_{\rm he}\rightarrow \infty} a_{4,t_8}(g_{\rm he}) = \lim_{g_{\rm he}\rightarrow \infty} a_{4,\rm anom}(g_{\rm he}) = \hat{a}_4 \pi^{3/4} g_{\rm he}^{1/2} \,. 
\end{equation}
Comparing with the general form of the effective action \eqref{16superchargesgeneral}, we can determine the coefficients to be 
\begin{equation}\label{a4he}
\begin{aligned}
    a_{4,t_8}(g_{\rm he})  &= \hat{a}_4 \left(\frac{3\cdot \zeta(3)}{\pi^{5/4}} g_{\rm he}^{-3/2} + \pi^{3/4} \, g_{\rm he}^{1/2} \right)\,,\\
     a_{4,\rm anom}(g_{\rm he}) &= \hat{a}_4 \, \pi^{3/4} g_{\rm he}^{1/2}\,,\\
     a_{4,\rm tree}(g_{\rm he}) &= \hat{a}_4 \, \frac{3}{4\pi^{5/4}}\, g_{\rm he}^{-3/2}\,. 
\end{aligned}
\end{equation}
From our definition of the species scale in \eqref{defspeciesN=1}, we see that $a_{4,t_8}$ always yields the species scale for any value of the coupling whereas $a_{4,\rm anom}$ is comparable only in the strong-coupling limit. Therefore, the analysis of the asymptotic regimes proceeds completely analogous to the Type IIA case which we, therefore, do not repeat here. Again, the desert point is located at
\begin{equation}
    g_{\rm he} = \frac{3}{\pi}\sqrt{\zeta(3)} \simeq 1.05\,,
\end{equation}
whereas the value of the species scale at the desert is slightly higher
\begin{equation}
    \max(\Lambda_s) \approx 0.823\,. 
\end{equation}
Let us stress again that this result is derived under the assumption that there are indeed no corrections to the $t_8t_8 R^4$-term beyond one-loop. While there is evidence for this from the vanishing of the next-order terms, it is by no means at the same level as the supersymmetric non-renormalization theorems. We note, however, that from the perspective of the species scale, it is at least consistent that higher-loop terms are indeed absent. In fact, our species scale analysis provides further evidence for the absence of such higher-loop corrections to this coupling. Similar to the Type IIA case, we can determine the diameter of the field space, for which $\Lambda_s\geq \mu$ is satisfied, to be 
\begin{equation}
    \diam(\mu) = -8\sqrt{2} \log\left[\mu\right]-\frac{\sqrt{2}}{3} \log[3\pi\zeta(3) ]\,,
\end{equation}
with the constant shift taking the numerical value of
\begin{equation}
    b=-\frac{\sqrt{2}}{3} \log[12\pi\cdot 2^{1/4} \zeta(3) ]\approx -1.14\,. 
\end{equation}

\subsection{Heterotic $SO(32)$ and Type I string}\label{heteroticO}
Let us now turn to the heterotic $SO(32)$ string. At tree- and one-loop level, the eight-derivative contribution to the effective action in the gravity sector takes the same form as \eqref{hete8e8} where we replace the string coupling by $g_{\rm ho}$. Again, following the arguments presented in \cite{Green:2016tfs}, one may expect there to be no higher-loop corrections to the $t_8t_8 R^4$-term in the effective action whereas the term proportional to $\cI_{\rm anom}$ is one-loop exact as a consequence of supersymmetry. However, this cannot be the full answer since the strong-coupling behavior of the heterotic $SO(32)$ string is distinctively different from the strong-coupling behavior of the heterotic $E_8\times E_8$ string: instead of being a decompactification limit, it corresponds to a weak-coupling limit for the Type I string. Hence, the behavior of the species scale cannot be the same in the strong-coupling limits for $SO(32)$ and $E_8\times E_8$. This necessarily implies that the higher-derivative terms need to be different. 

In \cite{Green:2016tfs}, non-perturbative corrections to the $t_8t_8 R^4$-coupling have been computed explicitly, giving rise to the effective eight-derivative action which schematically takes the form 
\begin{equation}\label{hetSo32}
    S_{{\rm HO},R^4} \sim M_{\text{pl},10}^2 \int \dd^{10}x \sqrt{-g} \left[E_{\frac{3}{2}}(ig_{\rm ho}^{-1})t_8t_8 R^4 -\frac{(\tr R^2)^2}{2g_{\rm ho}^{3/2}} +\frac{2\pi^2}{3} g_{\rm ho}^{1/2} \cI_{\rm anom} \right]\,.
\end{equation}
Notice that the coefficient of the $t_8t_8 R^4$-term is similar to the one of Type IIB in ten dimensions discussed in section~\ref{ssec:10dIIB}. The relation between the species scale and the higher-derivative terms can be inferred by recalling that the heterotic $E_8\times E_8$ string on $S^1$ with Wilson lines chosen to break each $E_8$ to $SO(16)$ is T-dual to the heterotic $SO(32)$ string on $S^1$ with Wilson lines breaking $SO(32)$ to $SO(16)\times SO(16)$. Using this duality, we can translate the normalization \eqref{a4he} to the $SO(32)$ string, leading to
\begin{equation}\label{a4h0}
\begin{aligned}
    a_{4,t_8}(g_{\rm ho})  &= \frac{3 \hat{a}_4 }{2 \pi^{5/4}} E_{\frac32}(ig_{\rm ho}^{-1})  \,,\\
     a_{4,{\rm anom}}(g_{\rm ho}) &= \hat{a}_4 \, \pi^{3/4} g_{\rm ho}^{1/2}\,,\\
     a_{4,{\rm tree}}(g_{\rm ho}) &= \, \frac{3 \hat{a}_4}{4\pi^{5/4}}\, g_{\rm ho}^{-3/2}\,. 
\end{aligned}
\end{equation}
Similar to the other heterotic theory, our definition of the species scale in \eqref{defspeciesN=1} singles out the coefficient of the $t_8t_8R^4$-coupling as the species scale defined everywhere in the moduli space. The behavior of the species scale in the asymptotic limits $g_{\rm ho}\rightarrow 0,\infty$ parallels that of the species scale in Type IIB and we correctly reproduce the scaling of the species scale in an emergent string limit in 10d. Therefore, the corrections computed in \cite{Green:2016tfs} are precisely of that form to ensure that the eight-derivative terms correctly capture the species scale, again, providing further evidence for the exactness of the computation of \cite{Green:2016tfs}. For the heterotic $SO(32)$ string, the desert point is located at $g_{\rm ho}=1$ where the species scale is given by 
\begin{equation}
    \max(\Lambda_s)=\Lambda_s(g_{\rm ho}=1) = 0.822\,. 
\end{equation}
We can again determine the diameter of the field space region for which $\Lambda_s\geq \mu$ to be
\begin{equation}
    \diam(\mu) = -4\sqrt{2} \log\left[\mu\right] - \frac{\sqrt{2}}{6}\log\left[\frac{3^4\zeta(3)^4}{\pi^5}\right]\,.
\end{equation}
Compared to Type IIB, we now have two inequivalent limits for $g_{\rm ho}\rightarrow 0$ and $g_{\rm ho}\rightarrow \infty$. Therefore, the prefactor of the $\log \mu$-term differs by a factor of two from \eqref{eq:IIBdistance}. The shift evaluates to 
\begin{equation}
    b= - \frac{\sqrt{2}}{6}\log\left[\frac{3^4\zeta(3)^4}{\pi^5}\right] \approx 0.140\,,
\end{equation}
which, unlike in all previous examples, is positive. 

\subsection{16 supercharges in 9d}
\begin{figure}[!t]
\begin{subfigure}{0.49\textwidth}
\begin{center}
\includegraphics[width=0.95\textwidth]{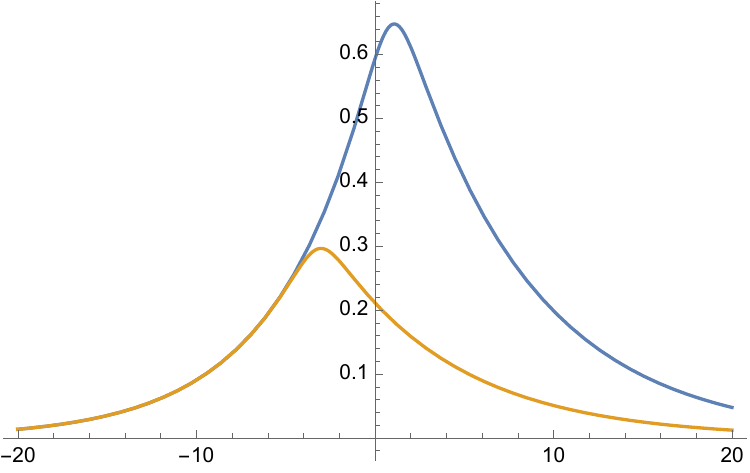}
\end{center}
\begin{picture}(0,0)
\put(95,180){\footnotesize$\left(\frac{a_{4,t_8/{\rm anom}}}{\hat{a}_4}\right)^{-1/6}$}
\put(125,20){\footnotesize$\sigma$}
\put(140,70){\footnotesize{\textcolor{mathorange}{$a_{4,t_8}$}}}
\put(150,115){\footnotesize{\textcolor{dblue}{$a_{4,\rm{anom}}$}}}
\end{picture}\vspace*{-0.8cm}
\caption{Higher-derivative coefficients.}
\end{subfigure}
\begin{subfigure}{0.49\textwidth}
\begin{center}
\includegraphics[width=0.95\textwidth]{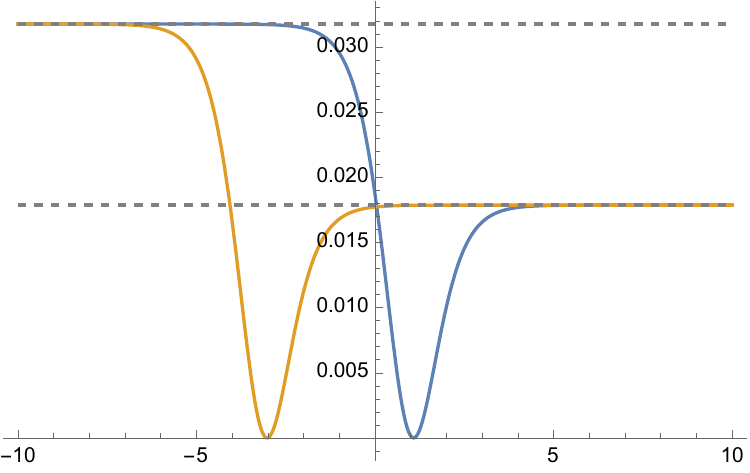}
\end{center}
\begin{picture}(0,0)
\put(80,180){\footnotesize$\frac{1}{36} \frac{|\nabla a_{4,t_8/{\rm anom}}|^2}{a_{4,t_8/{\rm anom}}^2}$}
\put(115,20){\footnotesize$\sigma$}
\put(55,80){\footnotesize{\textcolor{mathorange}{$a_{4,t_8}$}}}
\put(150,80){\footnotesize{\textcolor{dblue}{$a_{4,\rm{anom}}$}}}
\put(0,105){\footnotesize$\frac{1}{56}$}
\put(225,160){\footnotesize$\frac{2}{63}$}
\end{picture}\vspace{-0.8cm}
\caption{Slope of higher-derviative coefficients.}
\end{subfigure}
\caption{\label{fig:9dN=1} Left: The behavior of the terms appearing in the definition of the species scale \eqref{defspeciesN=1} for the heterotic SO(32) string on $S^1$ given in \eqref{a4s9d} plotted as a function of $\sigma = \log r$ for $g_{\rm ho}=1/100$. Right: The slope of the coefficients at eight-derivative level for the SO(32) heterotic string on $S^1$.}
\end{figure}
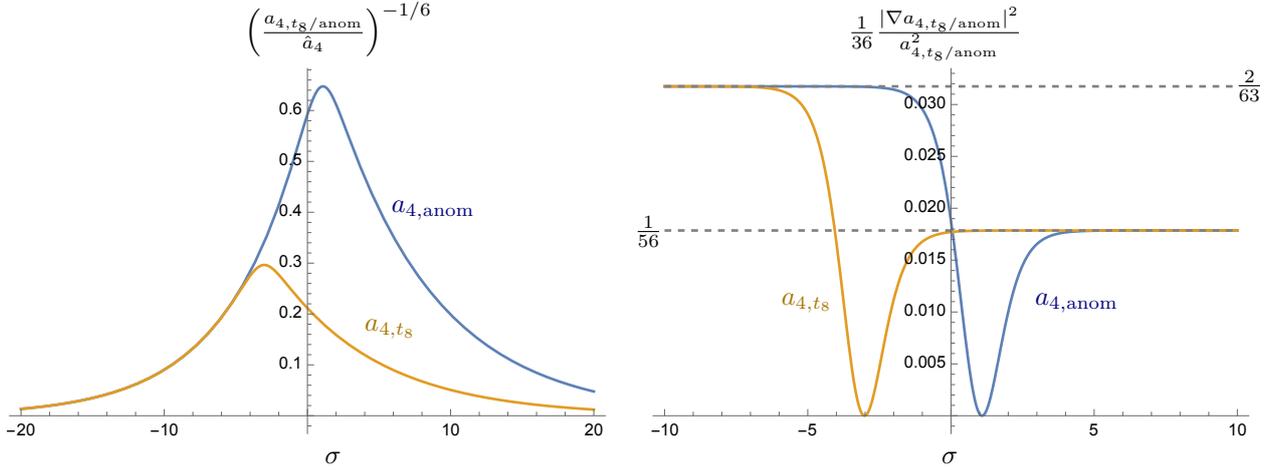

Let us move one dimension down and compactify heterotic string theories on $S^1$.\footnote{For an in-depth analysis of the asymptotic limits in 9d $\cN=1$ theories see \cite{Etheredge:2023odp}.} For simplicity, we choose the Wilson lines in both circle-compactified heterotic theories such that the respective gauge group in each theory is broken to $SO(16)\times SO(16)$. Since in this case, the two heterotic theories are T-dual to each other, the small-radius limit for the $S^1$ in either theory can be well-described. In this setting, the eight-derivative action has been calculated for both heterotic string theories in \cite{Green:2016tfs}. Focusing on the HO theory, the effective 9d action continues to be given by \eqref{16superchargesgeneral} with
\begin{equation}\label{a4s9d}
\begin{aligned}
    a_{4,t_8}(g_{\rm ho}, r)&= \frac{3 \hat{a}_4 }{2 \pi^{5/4}} \left(\frac{2\pi r}{g_{\rm ho}^{1/4}}\right)^{6/7}\left( E_{\frac32}(ig_{\rm ho}^{-1}) + \frac{2\pi^2}{3} \frac{g_{\rm ho}^{1/2}}{r^2}\right)\,,\\
    a_{4,{\rm anom}}(g_{\rm ho},r)&= \hat{a}_4 \, \pi^{3/4} \left(\frac{2\pi r}{g_{\rm ho}^{1/4}}\right)^{6/7}g_{\rm ho}^{1/2}\left(1 + \frac{1}{r^2}\right)\,,\\
    a_{4,{\rm tree}}(g_{\rm ho},r)&= \frac{3\hat{a}_4}{4\pi^{5/4}}  \left(\frac{2\pi r}{g_{\rm ho}^{1/4}}\right)^{6/7} g_{\rm ho}^{-3/2}\,. 
\end{aligned}
\end{equation}
In the following, we are only interested in the dependence that these coefficients have on the radius $r$. We first notice that the functional dependence of $a_{4,t_8}$ and $a_{4,{\rm anom}}$ is similar up to order-one factors. To study the asymptotics of the species scale, we can hence examine either of the two terms. In the limit $r\rightarrow \infty$, the scaling of $\Lambda_s$ in the field space distance $\Delta_r$ is then given by
\begin{equation}
    \Lambda_s \xrightarrow[]{r\rightarrow \infty} e^{-\frac{1}{\sqrt{56}}\Delta_r} \,,
\end{equation}
as expected for a decompactification limit from $d=9$ to $D=10$ dimension. On the other hand for $r\rightarrow 0$ we obtain 
\begin{equation}
    \Lambda_s \xrightarrow[]{r\rightarrow 0\ } e^{-\frac{\sqrt{2}}{3\sqrt{7}}\Delta_r}\,,
\end{equation}
suggesting that this limit is a decompactification limit from $d=9$ to $D=11$ dimensions. And, indeed, recalling that T-duality relates the coupling and radii of the two circle-compactified heterotic string theories via 
\begin{equation}
    g_{\rm he} = \frac{g_{\rm ho}}{r_{\rm ho}}\,,\quad r_{\rm he} = \frac{1}{r_{\rm ho}}\,,
\end{equation}
we observe that the $r_{\rm ho}\rightarrow 0$ limit corresponds to a large-radius strong-coupling limit for the heterotic $E_8\times E_8$ string, i.e., to the decompactification limit to 11d Ho\v{r}ava--Witten theory. Our definition of the species scale in \eqref{defspeciesN=1} singles out the largest coefficient among those in \eqref{a4s9d} to give the species scale. Since 
\begin{equation}
    a_{4, t_8} \geq a_{4,{\rm anom}} \,,
\end{equation}
for any $r$ and fixed $g_{\rm ho}$, the species scale continues to be given by the coefficient of the $t_8t_8 R^4$-coupling. We illustrated this in figure~\ref{fig:9dN=1}. Even though for $r\gtrsim\mathcal{O}(1)$, the coefficient $a_{4,{\rm anom}}$ of the term in the effective action protected by supersymmetry differs by $\mathcal{O}(1)$ factors from the species scale and would for instance predicts a different location of the desert point, it captures the asymptotic scalings of the species scale correctly. Since terms in the effective action, that are protected by supersymmetry, are oftentimes expressed in terms of index-like quantities, one would in general expect that these terms provide an upper bound for the species scale. The example of the heterotic $SO(32)$ string illustrates that the term protected by supersymmetry can indeed be used as a reasonable upper bound on the species scale.

\section{Species scale and 8 supercharges}\label{sec:8Qs}
We now turn to theories with eight supercharges. Of particular interest to us are 6d theories with $\cN=(1,0)$ supersymmetry and 5d $\cN=1$ theories. The case of $\cN=2$ theories in four dimensions has been extensively covered in \cite{vandeHeisteeg:2022btw,vandeHeisteeg:2023ubh,vandeHeisteeg:2023uxj}. In theories with eight supercharges, the moduli space factorizes as 
\begin{equation}
    \cM = \cM_H\times \cM_{V,T}\,,
\end{equation}
where the first factor denotes the hypermultiplet sector which has the same structure in six, five, and four dimensions. The second factor corresponds to the vector multiplet moduli space in five and four dimensions and the tensor multiplet space in six dimension. The structure of this space differs significantly between six, five, and four dimensions. Due to the small amount of supersymmetry, most of the higher-curvature terms are not BPS protected, making their exact computation difficult. Unlike in the theories studied in the previous sections, theories with eight supercharges allow for non-trivial $R^2$-terms implying that $\hat{a}_2\neq 0$ in general. In particular, there exists a four-derivative coupling whose coefficient can be computed explicitly. In 4d $\cN=2$, the coefficient of this term is given by the topological genus-one free energy $F_1$ which has been used in \cite{vandeHeisteeg:2022btw,vandeHeisteeg:2023ubh,vandeHeisteeg:2023uxj} to study the dependence of the species scale on the vector multiplet moduli space. 

In four dimensions, this term is protected from perturbative corrections and can be evaluated explicitly. If we consider Type IIA compactifications on a Calabi--Yau threefold $Y_3$, the contributions to $F_1$ can be split into a classical piece, which is proportional to the second Chern class, and a sum over worldsheet instantons. If we lift this Type IIA setup to a five-dimensional theory with $\cN=1$ supersymmetry corresponding to M-theory compactified on $Y_3$, the contribution from worldsheet instantons vanishes and we are left with the classical piece only. This piece can equivalently be obtained by reducing the $t_8t_8R^4$-term on the Calabi--Yau background. Using~\cite{Grimm:2013gma}
\begin{equation}
    t_8t_8R^4 = 3\cdot 2^8 \tr R^2 (c_2 \wedge J) \,,
\end{equation}
one obtains 
\begin{equation}\label{5dR2term}
    S_{R^2,5} = \frac{M_{{\rm pl},5}}{2} \left(\frac{b_2}{24\, \cV^{1/3}}\int_{Y_3} c_2 \wedge J \right)\int_{\cM^{1,4}} \dd^5 x  \tr R^2\,,\qquad b_2= \frac{1}{2^4 (2\pi)^{10/3}}\,,
\end{equation}
where $J$ is the K\"ahler form on $Y_3$ and $\cV=\frac16 \int J^3$ is the volume of $Y_3$. It is convenient to expand the K\"ahler form as
\begin{equation}
    J = X^I J_I \,,
\end{equation}
where $J_I$ is a basis of 2-forms. The coefficient of the $\tr R^2$-coupling is therefore independent of an overall rescaling of $J$ corresponding to a modulus in a hypermultiplet. As in 4d $\cN=2$, the coefficient of the $\tr\,R^2$-coupling in \eqref{5dR2term} is exact and does not receive any further corrections.

In case $Y_3$ is genus-one fibered, i.e., $Y_3:T^2\rightarrow B_2$ for some K\"ahler surface $B_2$, we can further lift to a six-dimensional theory with $\cN=(1,0)$ supersymmetry corresponding to F-theory on $Y_3$. Given the fibration structure of $Y_3$, it is natural to split the K\"ahler moduli, $X^I$, $I=1, \dots, h^{1,1}(Y_3)$, of $Y_3$ into base moduli, $X_b^a$, $a=1,\dots, h^{1,1}(B_2)$, and fibral moduli, $X_f^\alpha$, $\alpha=h^{1,1}(B_2)+1, \dots, h^{1,1}(Y_3)$. To obtain the six-dimensional limit, one first realizes that F-theory on $Y_3\times S^1$ is dual to M-theory on $Y_3$. The duality identifies~\cite{Grimm:2010ks}
\begin{equation}
    r_{S^1} M_{11} = \frac{1}{\tau}\,,
\end{equation}
where $M_{11}$ is the fundamental M-theory scale and $\tau$ the volume of the generic fiber which is related to the fibral volumes 
\begin{equation}
    \tau= \sum_{\alpha=h^{1,1}(B_2)+1}^{h^{1,1}(Y_3)} c_\alpha v^\alpha_f\,,
\end{equation}
for some $c_\alpha$ that are fixed by the precise geometry of the fibration. The six-dimensional theory is obtained in the $r_{S^1}\rightarrow \infty$ limit corresponding to $\tau\rightarrow 0$. Since the overall volume of $Y_3$ has to remain constant in this limit, the actual F-theory limit corresponds to the scaling 
\begin{equation}
    X_f^\alpha \rightarrow \lambda^{-1}\,,\qquad X_b^a \rightarrow \lambda^{1/2}\,,\qquad \lambda\rightarrow \infty\,. 
\end{equation}
Since the effective action of F-theory is obtained as a scaling limit of M-theory, this in particular includes the $\tr\,R^2$-term in \eqref{5dR2term}. We therefore need to consider the F-theory lift of the coefficient of $\tr R^2$ in \eqref{5dR2term}. Interpreting $c_2(Y_3)$ as a curve class on $Y_3$, we have only its components along the base survive in the F-theory limit. For $Y_3$ a smooth Weierstrass model with zero section $S_0$, we can use the adjunction formula 
\begin{equation}
    c_2(Y_3) = c_2(B_2) + 11 c_1(B_2)^2 + 12 S_0 \wedge c_1(B_2)\,.
\end{equation}
Here, only the last term corresponds to a curve in the base whereas the first two terms do not contribute to $\int c_2 \wedge J$ in the F-theory limit. In principle, there could be additional contributions to $c_2$ surviving in the F-theory limit in case we do not have a smooth Weierstrass model as is, e.g., the case in the presence of a non-Higgsable cluster. In this case, $Y_3$ is singular and we need to perform a (series of) small resolutions to obtain a smooth $\tilde{Y}_3$. Let $[\mathbb{P}^1]$ denote the class of the exceptional curves introduced by the small resolutions. The second Chern class of $\tilde{Y}_3$ is then related to that of $Y_3$ via (see e.g.,~\cite{Anderson:2022bpo})
\begin{equation}
    c_2(\tilde{Y}_3) = c_2(Y_3) - [\mathbb{P}^1]\,. 
\end{equation}
Since in the F-theory limit, the volume of the resolution $\mathbb{P}^1$s vanishes, we find that 
\begin{equation}
    \int_{\tilde{Y}_3} c_2(\tilde{Y_3}) \wedge J \stackrel{\text{F-theory}}{\rightarrow}\int_{Y_3} c_2(Y_3) \wedge J = 12 \int_{B_2} c_1(B_2) \wedge J_b\,,
\end{equation}
such that the $\tr R^2$-term in the 6d F-theory effective action reads
\begin{equation}\label{6dR2term}
    S_{R^2,6} = \frac{M_{\rm pl,6}^2}{2} \left(\frac{b_2}{2\cV_{B_2}^{1/2}}\int_{B_2} c_1(B_2)\wedge J_F \right)\int_{\cM_{1,5}} \dd^6x \tr R^2\,. 
\end{equation}
Here, $J_F$ is the K\"ahler class of the base $B_2$ and $\cV_{B_2}$ is the volume of $B_2$, both measured in Type IIB string units. 

We thus identified a higher-curvature term in supergravity theories with minimal supersymmetry in both six and five dimensions whose coefficient can be calculated explicitly since it is protected by supersymmetry. We can use these terms to study the behavior of the species scale on the scalars in the vector/tensor sector in five and six dimensions. Notice that the protected coefficients are not sensitive to the scalars in the hypermultiplet sector such that they only provide an upper bound for the species scale. Even though the parametric dependence of the coefficients in \eqref{5dR2term} and \eqref{6dR2term} on the scalars in the vector/tensor sector is expected to reflect the scaling of the species scale, recall from section~\ref{sec:16Qs} that in theories with reduced supersymmetry, the actual species scale can differ from the one obtained from terms protected by supersymmetry by $\mathcal{O}(1)$ factors. Therefore, also in the vector/tensor sector, the coefficients in \eqref{5dR2term} and \eqref{6dR2term} provide an upper bound for the species scale. 
 
In the following, we first consider the properties of the species scale as derived from \eqref{6dR2term} in simple examples of six-dimensional F-theory compactifications and then discuss a five-dimensional M-theory example. 

\subsection{Species scale in 6d $\cN=(1,0)$ supergravity}\label{sec:6d}
We start by considering the properties of the species scale in six-dimensional theories with $\cN=(1,0)$ supersymmetry. We focus on F-theory compactifications on elliptically-fibered Calabi--Yau threefolds for which, as described above, the dependence of the species scale on the scalars in the tensor multiplets is captured by 
\begin{equation}\label{speciesFtheory}
    \Lambda_s = \left(\frac{1}{\cV_{B_2}^{1/2}}\int_{B_2} c_1(B_2) \wedge J_f\right)^{-1/2}\,,
\end{equation}
up to the order-one constant $\hat{a}_2$ introduced in \eqref{cnhatcn}. Notice that, unlike in the previous cases, we cannot fix this constant as the relation between the higher-derivative term and the species scale can no longer be read off from eleven-dimensional M-theory. The reason for this is that the coupling in question is independent of the overall volume of the Calabi--Yau threefold such that it is insensitive to the eleven-dimensional decompactifictation limit to M-theory, which is necessary for the matching. We therefore take the definition as in \eqref{speciesFtheory} keeping in mind that it is an upper bound for the species scale up to an $\mathcal{O}(1)$ constant.

\subsubsection{General discussion}
The tensor multiplet moduli space for six-dimensional F-theory has dimension $n_T=h^{1,1}(B_2)-1$ and is embedded in the K\"ahler moduli space of $B_2$ as the hypersurface corresponding to the solution of  
\begin{equation}
    \Omega_{\alpha \beta} X^\alpha X^\beta =1 \,,
\end{equation}
where $\Omega_{\alpha \beta}$ is an $SO(1,n_T)$ invariant inner product. The tensor moduli $j_\alpha$ parameterize this hypersurface. The relevant part of the 6D effective action then reads 
\begin{equation}\label{S6d}
    S_{6d}= \frac{M_{\rm pl,6}^4}{2} \int d^6x \sqrt{-G} \left(R_{6d} - \frac{1}{2} g_{\alpha \beta} \partial_\mu j^\alpha \partial^\mu j^\beta +\dots \right)\,. 
\end{equation}
Here, the field space metric is defined as 
\begin{equation}
    g_{\alpha \beta} = 2 j_\alpha j_\beta -\Omega_{\alpha \beta} \,,
\end{equation}
where we use $\Omega_{\alpha\beta}$ to raise and lower indices. Given that the signature of the tensor branch is $(1,n_T)$, this implies that there is just one kind of infinite-distance limit we can consider in this theory --- an emergent string limit \cite{Lee:2018urn,Lee:2018spm}. In each of these limits, a movable curve with trivial normal bundle shrinks to zero size which is compensated by blowing up other curves in $B_2$ to keep the volume of $B_2$ fixed. Denoting the shrinking curve by $C_0$, the triviality of the normal bundle implies $C_0^2=0$ and the K\"ahler form can be expanded as 
\begin{equation}
    J = X^0 C_0 + J_{\rm rest}\,,
\end{equation}
with the emergent string limits corresponding to $X^0\rightarrow \infty$. These limits fall into two classes depending on whether \cite{Lee:2018spm}
\begin{equation}\label{classificationemstrings}
    \int_{B_2} C_0\cdot c_1(B_2)=2\,\qquad \text{or} \qquad \int_{B_2} C_0\cdot c_1(B_2)=0\,.
\end{equation}
In the former case, a D3-brane on $C_0$ is dual to a heterotic string that becomes weakly-coupled in the limit $X^0\rightarrow \infty$ with tension 
\begin{equation}\label{tensionD3}
    \frac{T}{M_{\rm pl,6}^2} \sim \frac{1}{X^0}\,. 
\end{equation}
On the other hand, the higher-curvature term predicts, via \eqref{speciesFtheory}, a scaling of the species scale
\begin{equation}\label{hetlimits}
    \Lambda_s\rightarrow \left(X^0 \int_{B_2} c_1\wedge C_0+\dots\right)^{-1/2}\sim \frac{1}{\sqrt{X^0}}\,,
\end{equation}
which is consistent with the expectation that in an emergent string limit, the species scale is given by the string scale \eqref{tensionD3}. In terms of the distance, $\Delta$, on moduli space, the species scale scales as 
\begin{equation}
\label{eq:distance in 6d}
    \Lambda_s \sim e^{-\frac12\Delta}\,,
\end{equation}
consistent with \eqref{eq:alphas} for a six-dimensional emergent string limit. Notice that in the second case in \eqref{classificationemstrings}, the protected higher-derivative term does not correctly reproduce the species scale since the leading term in \eqref{hetlimits} vanishes. This is, however, not surprising since in this limit the asymptotically tensionless string corresponds to a Type II string with very mildly broken $\cN=4$ supersymmetry which provides stronger protection to the $R^2$-terms causing a systematic cancellation among the contributions to $a_2$. As a consequence, the $R^2$-term in this limit is not a good approximation to the species scale, and instead higher-curvature corrections should be considered. 

To study the species scale in the interior of the moduli space, we need to specify more details of the geometry of $B_2$. We, therefore, consider in the following a few simple examples to illustrate the properties of the species scale away from asymptotic regimes. 

\subsubsection{F-theory on del Pezzo surfaces}
\label{ssec:6d dP}
As a first class of examples, we study the family of bases corresponding to del Pezzo surfaces $dP_r$, i.e., blow-ups of $\mathbb{P}^2$ in $r$ generic points. A basis of curves on $dP_r$ is given by the hyperplane class $L$, inherited from $\mathbb{P}^2$, and the $r$ exceptional blow-up curves, $E_i$. The intersection pairing in this basis is given by
\begin{equation}\label{eq:dPints}
    L.L=1 \,,\qquad E_i.E_j=-\delta_{ij} \,,\qquad L.E_i=0\,. 
\end{equation}

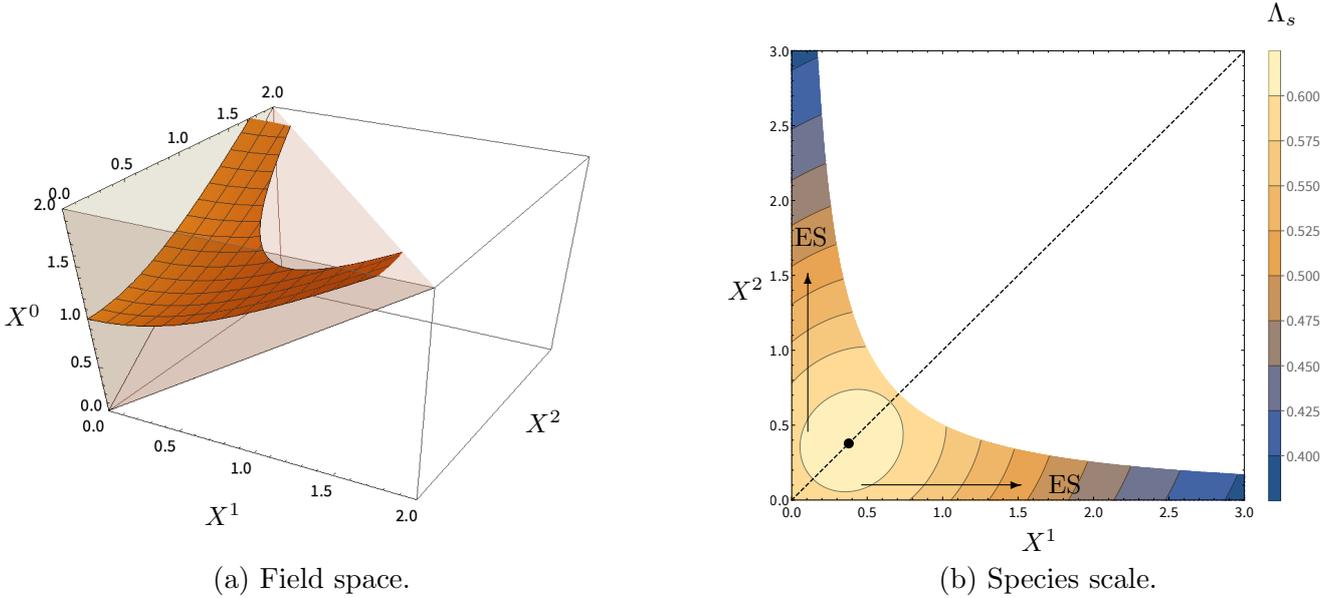
\begin{figure}[!t]
\centering
\vskip 0pt
\begin{subfigure}{0.43\textwidth}
    \centering
    \includegraphics[width=\textwidth]{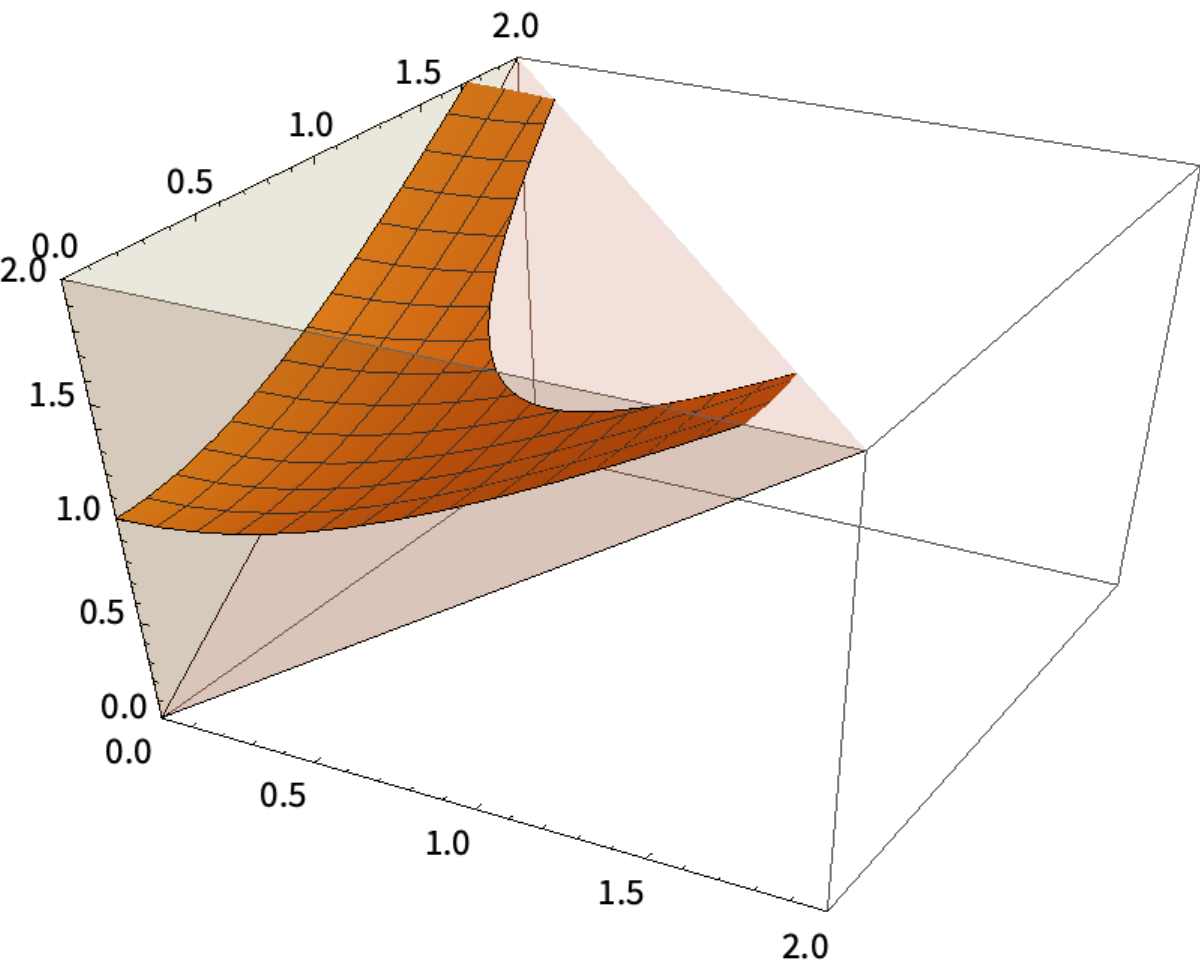}
\begin{picture}(0,0)\vspace*{-1.2cm}
\put(-115,90){\footnotesize $X^0$}
\put(-40,15){\footnotesize $X^1$}
\put(80,50){\footnotesize $X^2$}
\end{picture}\vspace*{-0.8cm}
\caption{Field space.}
\label{fig:dP2 field space}
\end{subfigure}
\hfill
\begin{subfigure}{.43\textwidth}
    \centering
    \includegraphics[width=\textwidth]{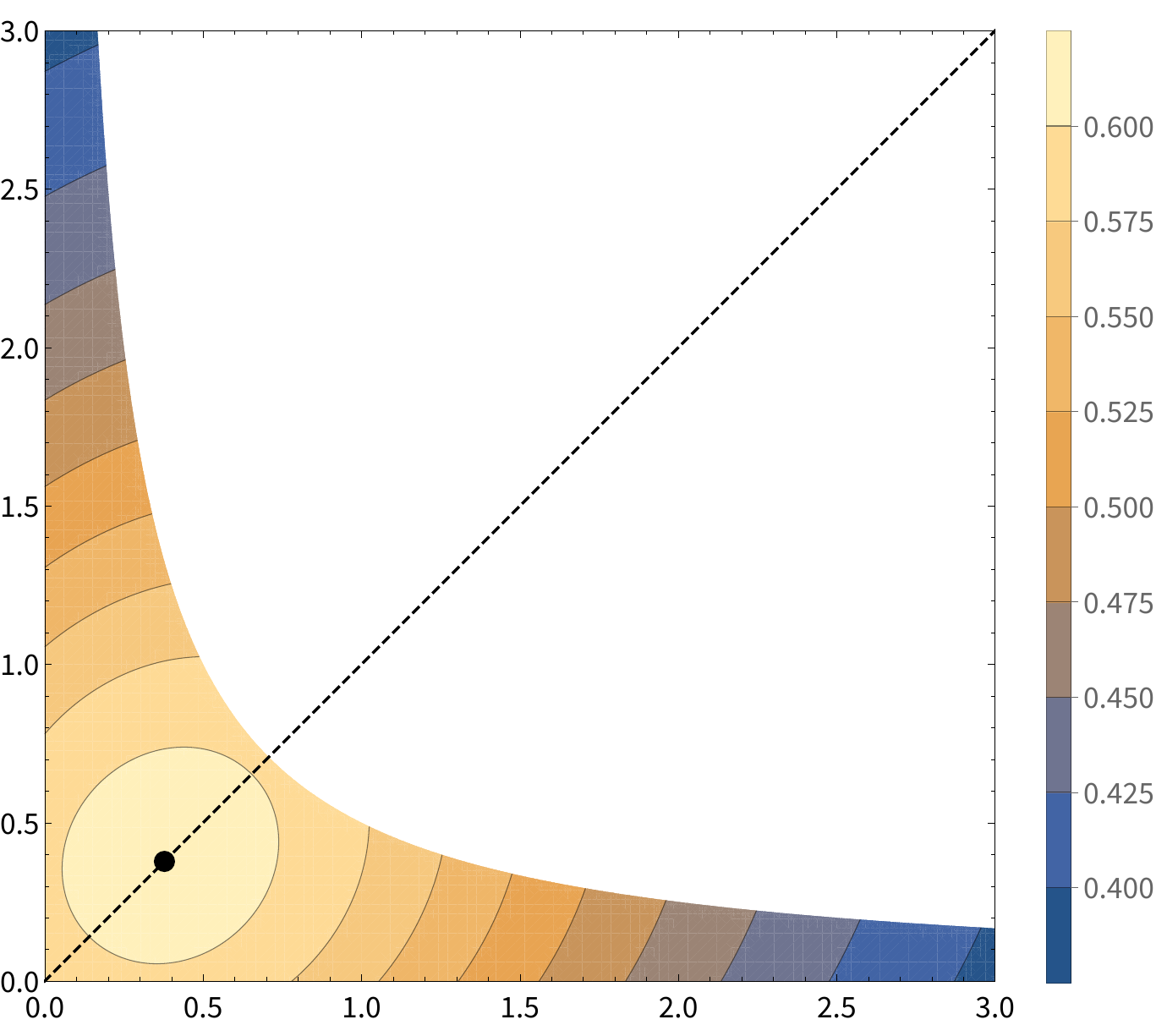}
\begin{picture}(0,0)\vspace*{-1.2cm}
\put(-10,5){\footnotesize $X^1$}
\put(-120,100){\footnotesize $X^2$}
\put(-90,50){\vector(0,1){60}}
\put(-95,120){\footnotesize ES}
\put(-70,30){\vector(1,0){60}}
\put(0,27){\footnotesize ES}
\put(82,205){\footnotesize $\Lambda_s$}
\end{picture}\vspace*{-0.8cm}
\caption{Species scale.}
\label{fig:dP2 species scale}
\end{subfigure}\vspace*{0.3cm}
\caption{\label{fig:dP2} Left: The tensor moduli space of $dP_2$ (solid orange region) corresponding to the constant volume hypersurface $(X^0)^2-(X^1)^2-(X^2)^2=1$ inside the K\"ahler cone defined by $X^1,X^2\geq 0$ and $X^0-X^1-X^2\geq 0$ (opaque orange region). Right: The behavior of $\Lambda_s$ according to (\ref{eq:dPspecies}) along the constant volume hypersurface. The black dot indicates where $\Lambda_s$ is maximized and the dashed line indicates the $\mathbb{Z}_2$ symmetry of the field space.
}
\end{figure}

\paragraph{Field space and metric.}
Let us first characterize the field space and its metric. We do so by expanding the K\"ahler form as
\begin{equation}
    J = X^0 L - \sum_{i=1}^r X^i E_i\, ,
\end{equation}
with coordinates $X^I=(X^0,X^i)$ on the $(r+1)$-dimensional moduli space. We want to restrict to the $r$-dimensional constant volume submanifold. By using the intersection data given in \eqref{eq:dPints}, we find this constraint to be
\begin{equation}
    \mathcal{F}=\left(X^0\right)^2-\sum_{i=1}^r\left(X^i\right)^2=1\, .
\end{equation}
Thus, we parameterize this fixed volume submanifold by
\begin{equation}
\label{eq:dPconstantvolume}
    X^0 = \cosh[x]\, , \quad X^i = u^i\sinh[x] \, , \qquad \sum_{i=1}^r (u^i)^2 = 1\, ,
\end{equation}
where the coordinates $u^i$ are restricted to the sphere $S^{r-1}$, which may for instance be parameterized by the standard spherical coordinates. The pull-back metric on the fixed volume submanifold then reads
\begin{equation}
    \dd s^2 = \dd x^2 + \sinh[x]^2 \dd\Omega_{r-1}^2\, ,
\end{equation}
where $\dd \Omega_{r-1}^2$ denotes the metric on $S^{r-1}$. We still have to supplement this characterization of the field space by the Mori cone constraints on the volumes of the curves. To this end, it is instructive to consider the case of $dP_2$. Its Mori cone is generated by $E_1,E_2,L-E_1-E_2$ (we refer to \cite{Donagi:2004ia,Anderson:2017aux} for an overview of the generators for all del Pezzo surfaces). We parameterize the $S^1$ by $(u^1,u^2)=(\cos \theta, \sin \theta)$, such that a positive volume for $E_{1,2}$ requires $\theta \in [0,\pi/2]$. The volume of $L-E_1-E_2$ reads
\begin{equation}
    \cosh[x]-\sinh[x](\cos \theta + \sin \theta) \geq 0\, .
\end{equation}
The infinite-distance limits correspond to $x\to \infty$, for which the above constraint reduces to $\cos \theta + \sin \theta \leq 1$ which can only be satisfied for $\theta=0,\pi/2$. We therefore find that there are only two possible infinite-distance limits in this field space, corresponding to sending $x \to \infty$ along one of these two directions. A more detailed depiction of this field space has been included in figure \ref{fig:dP2 field space}. For the other del Pezzo surfaces $dP_{r>2}$, one encounters a similar picture where emergent string limits correspond to particular (one-dimensional) rays in the $(r-1)$-sphere along which we send $x\to \infty$; for $dP_{1\leq r \leq 3}$, these rays always lie along the standard axes, while for $dP_{r\geq 4}$, additional rays have to be considered, cf. the emergent string limit for \eqref{eq:dP4string}.

\paragraph{Species scale and asymptotics.} Let us next characterize the physics underlying this field space. We begin with the species scale, which may be determined from the first Chern class of the del Pezzo surface according to \eqref{speciesFtheory}. For $dP_r$, we have $c_1(dP_r) = 3L-\sum_i E_i$, such that we find the species scale to be given by
\begin{equation}\label{eq:dPspecies}
    \Lambda_s = \frac{1}{(3X^0-\sum_i X^i)^{1/2}}\, .
\end{equation}
To study the behavior of the species scale in asymptotic limits, without loss of generality, we can consider the direction $u^i=(1,0,\ldots,0)$ on the sphere $S^{r-1}$. Any other direction in the K\"ahler cone (at constant overall volume) yields the same asymptotics. In the limit $x\rightarrow \infty$, we find
\begin{equation}
\label{eq:dP species asymptotics}
    \Lambda_s  =   e^{-x/2}\, .
\end{equation}
The coefficient $1/\sqrt{4}=1/2$ in the exponent here matches the expectation~\eqref{eq:alphas} for an emergent string limit in $d=6$. Also note that in the normalization in~\eqref{speciesFtheory}, the coefficient of $\Lambda_s$ is always one.\footnote{For this, it is important that we only consider directions inside the constant volume submanifold that also lie in the K\"ahler cone. For example, for $dP_2$ we consider only $u^i=(1,0)$ and $u^i=(0,1)$, but no angles in between these two directions, for which $\Lambda_s$ would have a different leading coefficient. In general, it follows from the fact that for any emergent Type II string limit, the curve wrapped by the D3-brane has degree 2 under $c_1(dP_r)$, see \eqref{classificationemstrings}.}

\paragraph{Tensionless strings and the Weyl group.} The strings that become tensionless along the infinite-distance limits come from D3-branes wrapped on curves with vanishing self-intersection. For $dP_{r=1,2,3}$, these are simply the curves $L-E_i$ with $i=1,\ldots, r$, where for a given curve the corresponding emergent string limit is given by $X^0\sim X^i \to \infty$ (for fixed $i$, and keeping all others constant). For higher del Pezzo surfaces, however, we have to consider a broader spectrum of strings. Starting with $dP_{r\geq 4}$, we have to introduce additional curves of the form
\begin{equation}\label{eq:dP4string}
    C_{ijkl} = 2L-E_i-E_j-E_k-E_l\, ,\quad \text{with}\quad  i<j<k<l\,. 
\end{equation}
These curves also have self-intersection 0 and satisfy $c_1(dP_r).C_{ijkl}=2$. Importantly, these strings are all related under the Weyl group of $E_r$, which acts naturally on the curves. While we refer to e.g.,~\cite{Iqbal:2001ye} for details on this Weyl group, let us summarize some of the main points relevant to our discussion here. In general, a root $\alpha \in H_2(dP_r,\bbZ)$ (that is, $\alpha^2=0$ and $c_1(dP_k).\alpha=0$) defines a transformation that acts as
\begin{equation}\label{eq:weyl}
    C \to C + (C.\alpha)\alpha\, .
\end{equation}
By this transformation rule, we find that simple roots $\alpha_i=E_i-E_{i+1}$ act by exchanging $E_i$ and $E_{i+1}$; this allows us to exchange all curves $L-E_1,\ldots,L-E_r$ with one another. On the other hand, one straightforwardly verifies that $\alpha_k=L-E_1-E_2-E_3$ acts by relating, for instance, $L-E_4$ to $2L-E_1-E_2-E_3-E_4$. The main take-away is that (for $dP_{r<9}$) all emergent strings lie in the same Weyl orbit, and hence just correspond to the same limit in a different duality frame. 

\begin{figure}[!t]
\begin{center}
\includegraphics[width=10 cm]{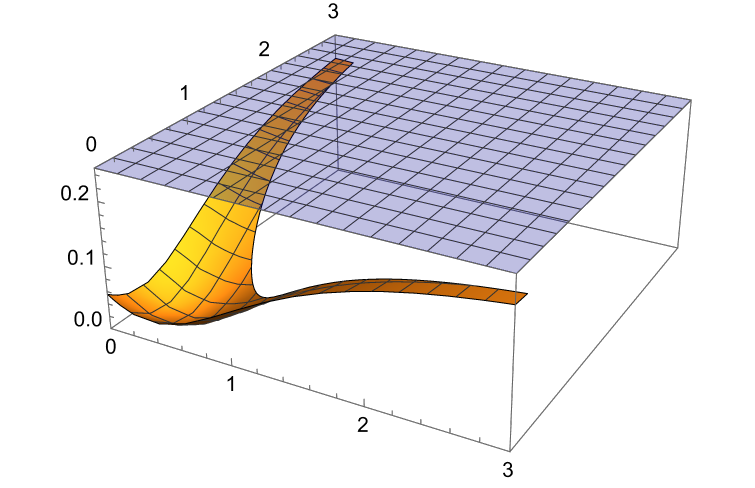}
\end{center}
\begin{picture}(0,0)
\put(90,108){\footnotesize $\frac{|\nabla \Lambda_s|^2}{\Lambda_s^2}$}
\put(190,48){\footnotesize$X^1$}
\put(160,185){\footnotesize$X^2$}
\end{picture}\vspace*{-1cm}
\caption{\label{fig:dPslope} Slope of the species scale $|\nabla \Lambda_s|^2/\Lambda_s^2$ for $dP_2$ in the $(X^1,X^2)$-plane (with $X^0$ fixed by the constant volume constraint \eqref{eq:dPconstantvolume}. The blue plateau indicates the emergent string bound at $1/4$.}
\end{figure}
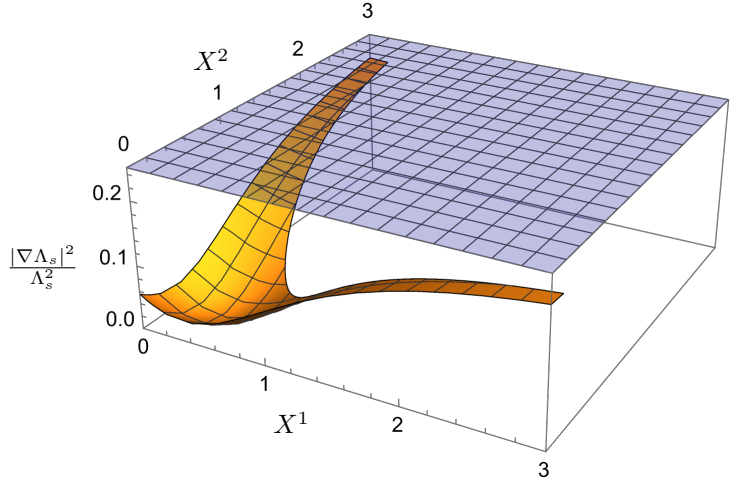

\paragraph{Slope of $\Lambda_s$.} The slope of the species scale $|\nabla \Lambda_s|^2/\Lambda_s^2$ vanishes at the desert point \eqref{eq:dPdesert} in the moduli space and asymptotes to the emergent string value $1/4$. We have provided a plot of this slope in figure \ref{fig:dPslope} for $dP_2$, which illustrates that this slope is indeed bounded from above by $1/4$ everywhere in moduli space. This result extends to all other del Pezzo surfaces $dP_{1\leq r \leq 8}$. For $dP_9$, one has to be more careful, as in addition to the emergent heterotic string limits, there also is a Type II string limit; we will elaborate more on this case later.

\paragraph{Desert point.} Let us next identify the desert point in the moduli space where the species scale is maximized. By symmetry considerations, we can directly restrict to the symmetric locus $X^1=\ldots=X^r$, reducing it to effectively a one-dimensional problem along the constant volume submanifold \eqref{eq:dPconstantvolume}. Extremization of the species scale \eqref{eq:dPspecies} along this symmetric locus then yields the location for the desert point
\begin{equation}\label{eq:dPdesert}
    X^0_{\rm des} = \frac{3}{\sqrt{9-r}}\, , \qquad X^i_{\rm des} = \frac{1}{\sqrt{9-r}}\, ,
\end{equation}
where $\Lambda_s$ takes its maximal value
\begin{equation}
\label{eq:dP desert species}
    \Lambda_s \big|_{\rm des} = (9-r)^{-1/4}\, .
\end{equation}
Note that, again, this analysis does not apply directly to $r=9$ which we consider later on. Other than that, the value of the species scale does increase with $r$ as we switch between the del Pezzo surfaces.

\paragraph{Diameter.} Let us finally compute the diameter of the field space. We consider a geodesic that runs from the desert point \eqref{eq:dPdesert} out towards any of the emergent string limits. As these are related by duality under the Weyl group, we can take simply the one along $u^i=(1,0,\ldots,0)$ and send $x\to \infty$. Using the asymptotic behavior given in \eqref{eq:dP species asymptotics} for the species scale, we find that $\Lambda_s$ crosses the cut-off scale $\mu$ at $x = -2\log[\mu]$. We have carried out the computation of the geodesic distance in appendix \ref{app:delpezzo}. Taking the main result \eqref{eq:geodist} and plugging in the cut-off scale $\mu$, we find
\begin{equation}
\diam(\mu) = -2\log [\mu] + \log \frac{2}{\sqrt{9-r}}\, .
\end{equation}
The coefficient of $\log \mu$ matches with the expectation from \eqref{eq:alphas} for an emergent string limit in 6d, while the constant shift decreases as we increase the index $r$ of the del Pezzo surface $dP_r$.

\paragraph{Rank 9 del Pezzo.} Finally, the del Pezzo $dP_9$ deserves some special attention, as most of our discussion does not apply directly to this case. Namely, our analysis so far breaks down when we consider the following limit
\begin{equation}
    X^0 = \sqrt{1+9\lambda^2}\, , \qquad X^i = \lambda\, ,
\end{equation}
where we send $\lambda \to \infty$. Along this limit, we have a tensionless string obtained from wrapping a D3-brane on the curve given by $c_1(dP_9)=3L-\sum_i E_i$. The difference to the cases before is that this curve has degree zero $d=c_1(dP_9).c_1(dP_9)=0$, whereas all other curves had degree two. As discussed in the beginning of this subsection, the limit therefore corresponds to a Type II emergent string limit. As these Type II strings have only very mildly broken $\mathcal{N}=4$ supersymmetry, there are systematic cancellations that cause the coefficient of the $R^2$-term to vanish rather than diverge asymptotically. For this reason, one should consider other gravitational corrections --- say at order $R^3$ or $R^4$ --- in these limits and, similar to the case of maximal and half-maximal supergravities discussed in section~\ref{sec:32Qs} and \ref{sec:16Qs}, take these as definition for the species scale instead of the coefficient of the $\tr R^2$-coupling.

\subsubsection{F-theory on Hirzebruch surfaces}\label{sec:Hirzebruch surfaces}
For our next class of bases, we consider Hirzebruch surfaces $\mathbb{F}_n$. We can consider these surfaces as base for elliptic fibrations --- and hence as compactification manifold for our 6d F-theory supergravities --- for $n=1,\ldots, 8,$ and $n=12$ (see \cite{Morrison:1996pp}). Our basis of curves on $\mathbb{F}_n$ is given by the generators of the Mori cone, $H,F$ which satisfy
\begin{equation}
    H.H=-n\, , \quad F.H=H.F=1\, , \quad F.F=0\, .
\end{equation}
For convenience, we can expand the K\"ahler form in the dual basis as
\begin{equation}
    J = h \tilde{H} + f \tilde{F}\, .
\end{equation}
This choice of basis is related to $(H,F)$ via the identification $\tilde{F}\sim F$ and $\tilde{H}\sim H+nF$ such that the intersection pairing between $\tilde{H},\tilde{F}$ is given by
\begin{equation}
    \tilde{H}.\tilde{H}=n\, ,\quad \tilde{F}.\tilde{H}=\tilde{H}.\tilde{F}=1\, , \quad \tilde{F}.\tilde{F}=0\, .
\end{equation}
Thus, demanding positive volumes for the Mori cone generators $H,F$ (no tilde) amounts to 
\begin{equation}
\label{eq:Hirzebruch curve volumes}
    h,f\geq 0\, .
\end{equation}
The constant volume submanifold is then given by
\begin{equation}
\label{eq:Hirzebruch volume}
    \mathcal{F} = 2hf+n h^2 = 1\, .
\end{equation}
We parameterize this field space by the canonically normalized scalar $\Delta$ as
\begin{equation}
\label{eq:Hirzebruch parametrize}
    h(\Delta) = e^{-\Delta}\, , \qquad f(\Delta)=\frac{1}{2}(e^{\Delta}-ne^{-\Delta})\, , 
\end{equation}
such that the positivity constraint (\ref{eq:Hirzebruch curve volumes}) is satisfied when $\Delta\geq \log[n]/2$.

Having characterized the field space, we next turn to the species scale. Using $c_1(F_n) = 2\tilde{H}+(2-n)\tilde{F}$ from \eqref{speciesFtheory}, we obtain
\begin{equation}
\label{eq:Hirzebruch species scale}
    \Lambda_s = \frac{1}{\sqrt{2f+(2+n)h}}=\frac{1}{\sqrt{3\cosh{\Delta}-\sinh{\Delta}}}\, .
\end{equation}
Interestingly, the species scale does not depend on the degree $n$ of the Hirzebruch surface $\mathbb{F}_n$.

\begin{figure}[!t]
    \centering
    \vskip 0pt
    \begin{subfigure}{0.43\textwidth}
    \centering
    \includegraphics[width=\linewidth]{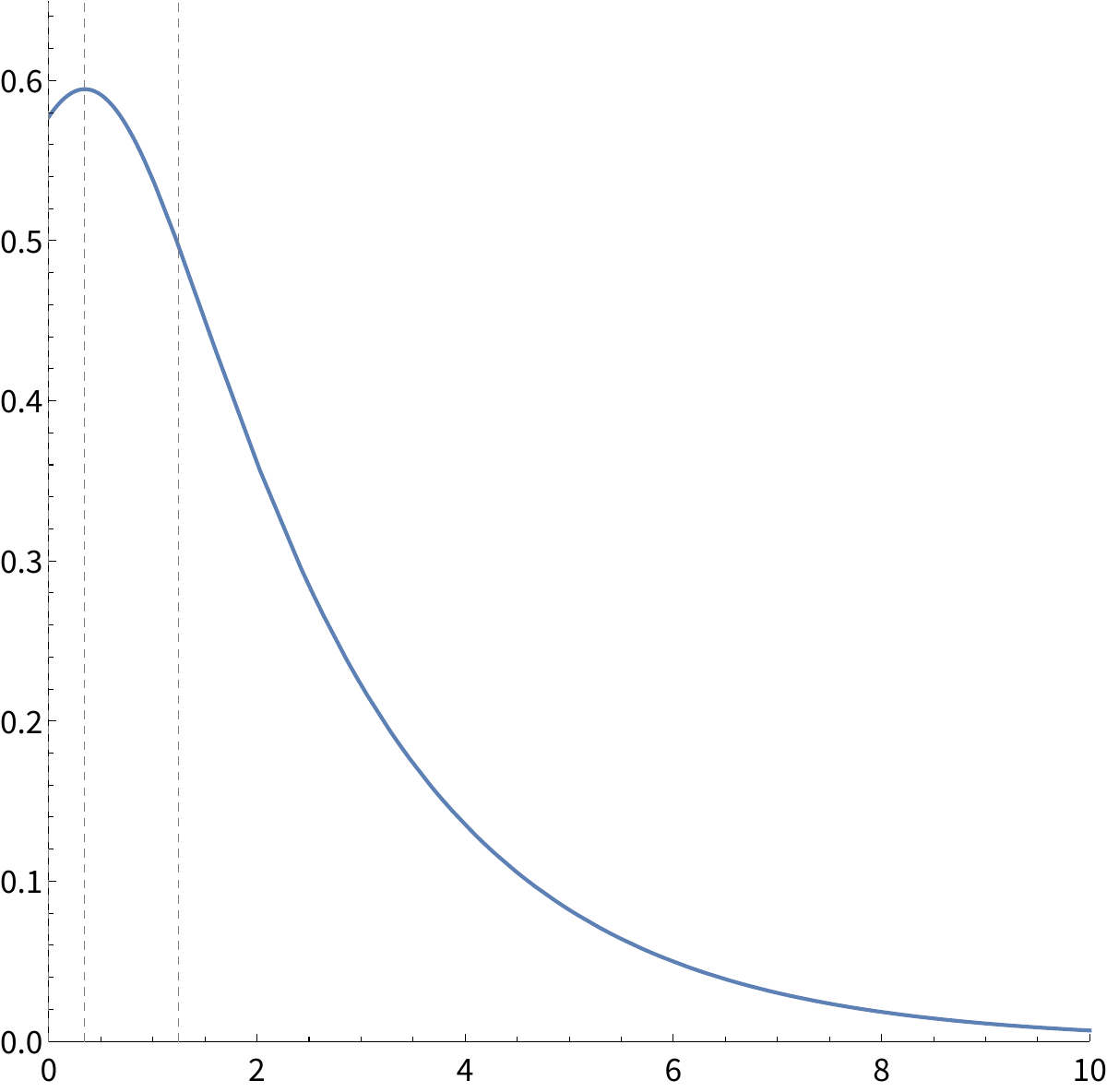}
    \begin{picture}(0,0)\vspace*{-1.2cm}
    \put(0,5){\footnotesize $\Delta$}
    \put(-125,125){\footnotesize $\Lambda_s$}
    \put(-100,225){\scriptsize $\mathbb{F}_1$}
    \put(-90,225){\scriptsize $\mathbb{F}_2$}
    \put(-85,140){\footnotesize $\dots$}
    \put(-75,225){\scriptsize $\mathbb{F}_{12}$}
    \end{picture}\vspace*{-0.8cm}
    \caption{Species scale.}
    \label{fig:Hirzebruch Species Scale} 
    \end{subfigure}
    \hfill
    \begin{subfigure}{0.43\textwidth}
    \centering
    \includegraphics[width=\linewidth]{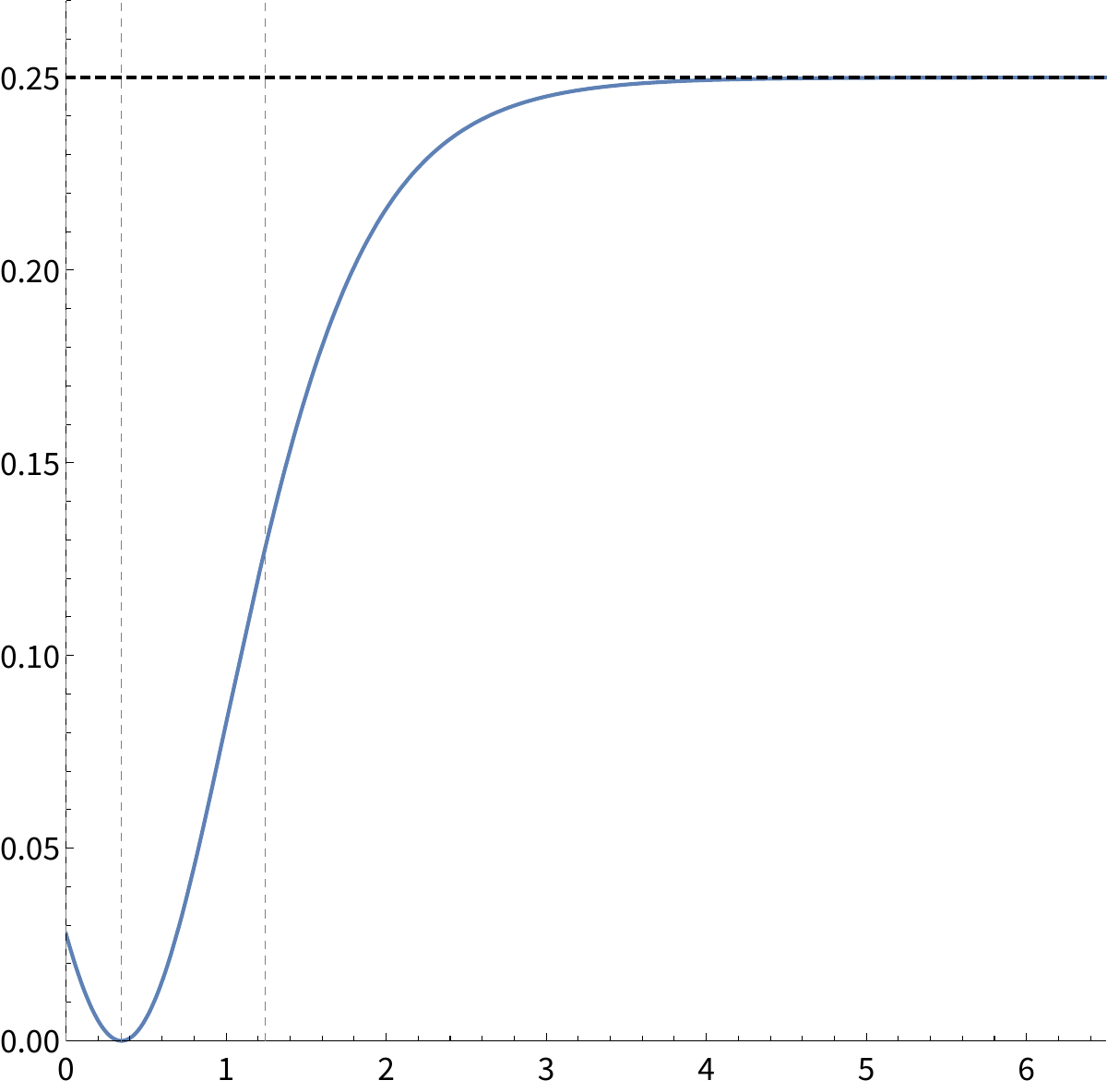}
    \begin{picture}(0,0)\vspace*{-1.2cm}
    \put(0,5){\footnotesize $\Delta$}
    \put(-135,125){\footnotesize $\frac{|\nabla \Lambda_s|^2}{\Lambda_s^2}$}
    \put(-95,225){\scriptsize $\mathbb{F}_1$}
    \put(-85,225){\scriptsize $\mathbb{F}_2$}
    \put(-73,140){\footnotesize $\dots$}
    \put(-60,225){\scriptsize $\mathbb{F}_{12}$}
    \end{picture}\vspace*{-0.8cm}
    \caption{Slope of species scale.}
    \label{fig:Hirzebruch Species Gradient}
    \end{subfigure}
    \vspace*{0.3cm}
    \caption{  
    Species scale (left) and slope $|\nabla \Lambda_s|^2/\Lambda_s^2$ (right) for F-theory base $B_2 = \mathbb{F}_n$ as a function of $\Delta$ defined in (\ref{eq:Hirzebruch parametrize}). The gray vertical dashed lines indicate the range of the boundaries $\Delta=\log[n]/2$ of the tensor branch for $\mathbb{F}_{2\leq n\leq 12}$; the boundary of the field space for $\mathbb{F}_1$ corresponds to the vertical axis.
    }
    \label{fig:Hirzebruch Surfaces}
\end{figure}
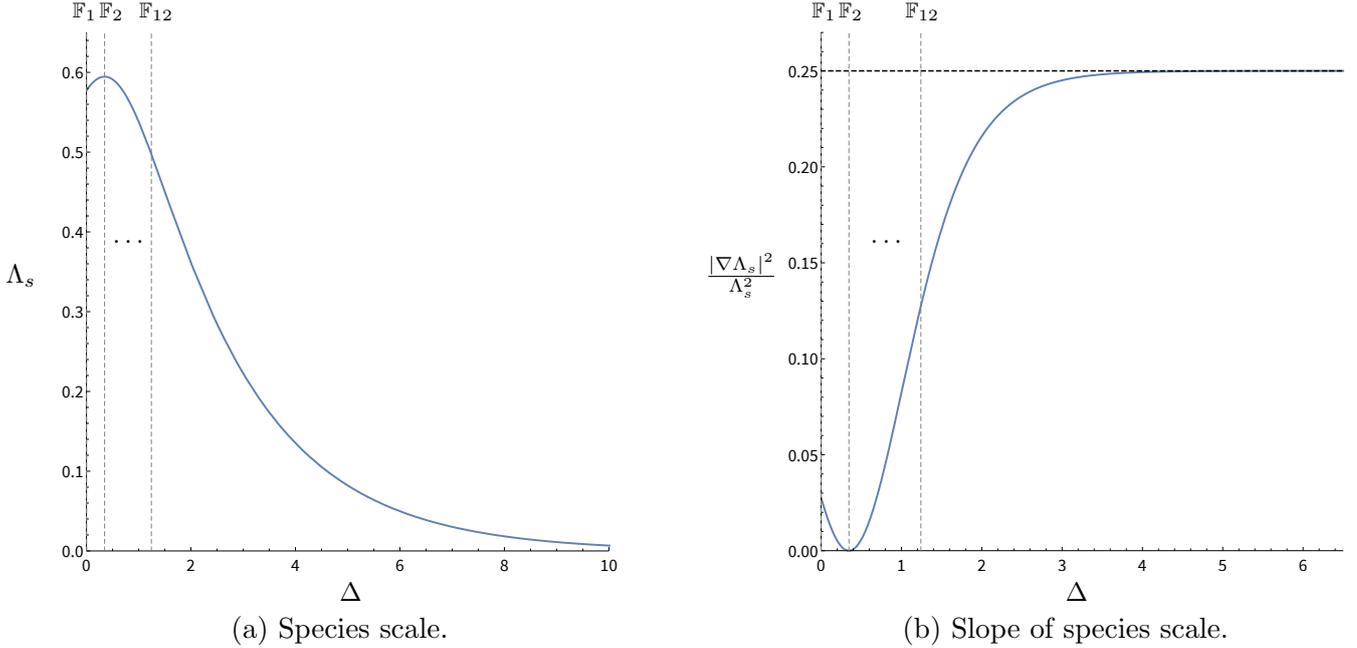

\paragraph{Slope of $\Lambda_s$.} The behavior of the species scale and its slope for all $\mathbb{F}_n$ are plotted in figure~\ref{fig:Hirzebruch Surfaces}. In particular, we observe that as in the $\Delta\to\infty$, the volume of $\tilde{F}$ asymptotes to $f(\Delta)+nh(\Delta)=e^{\Delta}/2\to\infty$. This corresponds to an emergent string limit for which, as we infer from figure~\ref{fig:Hirzebruch Species Gradient}, the slope $|\nabla\Lambda_s|^2/\Lambda_s^2$ approaches $1/4$ from below in this limit --- the correct exponential behavior for the emergent string. However, for each $\mathbb{F}_n$, $\Delta$ must satisfy (\ref{eq:Hirzebruch curve volumes}). Hence for each $\mathbb{F}_n$, $\Lambda_s$ and $|\nabla \Lambda_s|^2/\Lambda_s^2$ plotted in figure~\ref{fig:Hirzebruch Surfaces} should only be considered for $\Delta\geq \log[n]/2$ as indicated by the gray vertical dashed lines in the figures.

\paragraph{Desert point.} We next determine the point where the species scale \eqref{eq:Hirzebruch species scale} is maximized. Extremizing for $\Delta$ gives us a maximum at
\begin{equation}
    \Delta = \tfrac{1}{2}\log[2]\, .
\end{equation}
However, we note that for $n\geq 3$, this point does not lie inside the K\"ahler cone, as we can recall from demanding $f\geq 0$ in \eqref{eq:Hirzebruch parametrize} that $\Delta \geq \log[n]/2$ for $\mathbb{F}_n$; in these cases, the species scale is actually maximized at the boundary of their respective K\"ahler cones. In summary, we find the species scale at the desert point for Hirzebruch surfaces to be
\begin{equation}
\label{eq:Hirzebruch desert}
\Lambda_s\big\vert_{\rm des}=
\begin{cases}
    2^{-3/4}&\text{at }\Delta=\log[2]/2 \text{ for }n=1,2\, ,\\
    \left(\frac{\sqrt{n}}{2+n}\right)^{1/2}&\text{at }\Delta=\log[n]/2\text{ for } 3\leq n\leq 8 \text{ and } n=12\, .
\end{cases}
\end{equation}

\paragraph{Diameter.} Let us now turn to the field range set by some cut-off scale $\mu$. Setting $\Lambda_s=\mu$ and inverting \eqref{eq:Hirzebruch species scale}, we obtain the position in the field space as a function of the cut-off scale $\mu$
\begin{equation}
\label{eq:Hirzebruch distance}
    \Delta = \log \left[\frac{1+\sqrt{1-8\mu^4}}{2\mu^2}\right]\, .
\end{equation}
Therefore, computing the diameter of field space at small $\mu$ for $\mathbb{F}_n$ in the field space metric $\dd s^2 = \dd \Delta^2$ yields the following expression
\begin{equation}
\label{eq:Hirzebruch diameter approx}
    \diam(\mu) =-2\log[\mu] - \frac{1}{2}\log[n] \, .
\end{equation}
The coefficient of $\log[\mu]$ agrees with the emergent string limit. Thus, this is in agreement with \eqref{eq:distance in 6d}.

\subsection{Species scale in 5d $\cN=1$ supergravity}
In this section we consider M-theory compactifications on Calabi--Yau threefolds. We study the vector multiplet moduli spaces arising in these 5d $\mathcal{N}=1$ supergravity theories, where we investigate the shape and size of the effective field space cut out by the species scale. 

\subsubsection{General discussion}
Let us first discuss some general features of vector multiplet moduli spaces for Calabi--Yau threefold compactifications of M-theory. These spaces are spanned by the K\"ahler moduli of the Calabi-Yau manifold, subject to the constraint that the overall volume stays fixed
\begin{equation}
    \mathcal{F}(X) = \frac{1}{6} C_{IJK} X^I X^J X^K = 1 \, ,
\end{equation}
where $X^I$ (with $I=1,\ldots, h^{1,1}$) are the K\"ahler moduli and $C_{IJK}$ are the intersection numbers of the Calabi-Yau manifold. Let us denote the $h^{1,1}-1$ scalar fields that parameterize the $\mathcal{F}=1$ hypersurface by $\phi^i$. The metric on this field space is then obtained as a pull-back as
\begin{equation}
\label{eq:5d metric}
g_{ij} = G_{IJ}(X(\phi)) \ \partial_i X^I(\phi) \partial_j X^J(\phi) \, , \qquad G_{IJ} = -\frac{1}{2}\partial_I \partial_J \log \mathcal{F}(X) \, .
\end{equation}
where the factor of $\tfrac{1}{2}$ accounts for the normalization of the kinetic terms. 

Having characterized the field space, let us next describe the species scale $\Lambda_s$. To this end, let us recall the expression given in (\ref{5dR2term}). 
Thus, the species scale for M-theory compactified on a Calabi--Yau threefold is
\begin{equation}\label{Mtheoryspecies}
   \Lambda_s=\left[\frac{1}{12}c_{2,I}X^I\right]^{-1/2}\, ,
\end{equation}
where $c_{2,I}$ denote the integrated second Chern class numbers. As in the F-theory case discussed previously, also for M-theory on Calabi--Yau threefold, we cannot fix the exact coefficient of the species scale in terms of the higher-derivative coefficient, $\hat{a}_2$ in \eqref{cnhatcn}. In the following, we work with the above expression keeping in mind that we did not fix this exact $\mathcal{O}(1)$ coefficient.\footnote{We keep the factor of $\frac{1}{12}$ such that the overall $\mathcal{O}(1)$ factor in M-theory agrees with that in F-theory discussed in the previous section.} 

Again, we can first study the general behavior of \eqref{Mtheoryspecies} in asymptotic regimes. We therefore recall from \cite{Lee:2019oct,Cota:2022maf} that in the 5d $\cN=1$ vector multiplet moduli space, there exist only three types of infinite-distance limits:
\begin{itemize}
    \item \emph{Limits of Type $T^2$/Decompactification limits to 6d}: For these limits to exist, $Y_3$ needs to allow for a torus fibration $Y_3:T^2\rightarrow B_2$. The infinite-distance limit corresponds to the limit where the fiber shrinks and the base $B_2$ grows homogeneously
    \begin{equation}\label{TypeT2}
    \mathcal{V}_{T^2}^{-1}\sim \mathcal{V}_{B_2}\sim \lambda\rightarrow \infty\,.
    \end{equation} 
    This corresponds to the F-theory limit and we obtain an effectively six-dimensional theory signalled by light KK-modes corresponding to wrapped M2-branes on $T^2$. This leads to a species scale that asymptotically vanishes like 
    \begin{equation}\label{speciesdecom6d}
       \Lambda_s \sim \frac{1}{\lambda^{1/2}}\,.
    \end{equation}
    On the other hand, the scaling \eqref{TypeT2} implies that the K\"ahler moduli corresponding to curves in the base scale like $\lambda^{1/2}$, such that our proposal \eqref{Mtheoryspecies} correctly reproduces the expected scaling for the decompactification limit to 6d. In terms of the field space distance, $\Delta$, it is straight-forward to check that the scaling \eqref{speciesdecom6d} translates to 
    \begin{equation}
       \Lambda_s \sim e^{-\frac{1}{2\sqrt{5}}\Delta}\,,
    \end{equation}
    in agreement with \eqref{eq:alphas}. 
    \item \emph{Limits of Type K3/Emergent heterotic string limits}: In this case, $Y_3$ allows for a K3-fibration $Y_3:K3\rightarrow \mathbb{P}^1_b$ and the infinite-distance limit corresponds to the limit of homogeneously shrinking the K3 fiber while growing the base $\mathbb{P}^1_b$ as 
     \begin{equation}\label{TypeK3}
    \mathcal{V}_{K3}^{-1}\sim \mathcal{V}_{\mathbb{P}^1_b}\sim \lambda\rightarrow \infty\,.
    \end{equation}
    In this limit, an M5-brane wrapping the K3-fiber becomes light. The resulting string in five dimensions is dual to a weakly-coupled heterotic string. Accordingly, the species scale in this limit is set by the tension of this string 
    \begin{equation}\label{speciesemstring5d}
        \Lambda_s \sim \frac{1}{\lambda^{1/2}}\,.
    \end{equation}
    Since the volume of the $\mathbb{P}^1_b$ also diverges as $\lambda$ and $c_2(Y_3)= c_{2,0} [\mathbb{P}^1_b]+\dots $ with $c_{2,0}=24$, our species scale \eqref{Mtheoryspecies} reproduces this scaling in the emergent string limit. In terms of the field space distance, we have 
    \begin{equation}
    \label{eq:5d heterotic emergent limit}
        \Lambda_s \sim e^{-\frac{1}{\sqrt{3}}\Delta}\,,
    \end{equation}
    in accordance with \eqref{eq:alphas}. 
    \item \emph{Limits of Type $T^4$/Emergent Type II string limits}: In this case, $Y_3$ needs to allow for an Abelian fibration $Y_3:T^4\rightarrow \mathbb{P}^1_b$. Similarly to the previous case, the infinite-distance limit corresponds to the fiber shrinking homogeneously and the base becoming large. As a result, the M5-brane wrapping the $T^4$ fiber becomes light leading to a tensionless string in 5d dual to a Type II string. In this case, \eqref{speciesemstring5d} does not serve as a good description of the species scale since for the limits of Type $T^4$, the component of $c_2(Y_3)$ along $\mathbb{P}_1^b$ vanishes, i.e.,~$c_{2,0}=0$. Therefore, \eqref{Mtheoryspecies} does not correctly reproduce the species scale in this limit. As in the analogue limit in 6d F-theory, the reason is again supersymmetry enhancement in the asymptotic limit causing the $R^2$-coupling to be sub-leading due to systematic cancellations occuring in the supersymmetric protected coupling.
\end{itemize}
In the following, we consider a simple example of a Calabi--Yau manifold that possesses a heterotic emergent string limit and use \eqref{Mtheoryspecies} to study the properties of the species scale in the interior of the vector multiplet moduli space.

\subsubsection{Example: Calabi--Yau threefold with $(h^{1,1},h^{2,1})=(2,86)$}
\label{ssec:5d CY3}
We consider the Calabi-Yau threefold given by the intersection of bidegree (4,1) and bidegree (1,1) hypersurfaces inside $\mathbb{P}^4\times \mathbb{P}^1$. This manifold was initially studied in \cite{Greene:1995hu,Greene:1996dh} and in the CICY representation can be written as\begin{equation}
    X_{(2,86)}=\left(
    \begin{array}{c||cc}
        \mathbb{P}^4 & 4&1 \\
        \mathbb{P}^1 & 1&1
    \end{array}
    \right)_{-168}^2.
\end{equation} 
The Hodge numbers for this threefold are $(h^{1,1},h^{2,1})=(2,86)$ such that we refer to it as $X_{(2,86)}$. Following the conventions in \cite{Alim:2021vhs}, the prepotential in phase I is
\begin{equation}
\label{eq:X286 conifold phase I prepotential}
    \mathcal{F}^{(I)}=\frac{5}{6}X^3+2X^2Y\qquad \text{with K\"{a}hler cone } \mathcal{K}_I=\{ X,Y\geq 0\},
\end{equation}
where $X,Y$ correspond to curves in $\mathbb{P}^4$ and $\mathbb{P}^1$. There is a second phase which can be reached via a flop wall located at $Y=0$ where the geometry develops 16 conifold singularities. Passing through the flop wall, we obtain phase II of our moduli space where the prepotential takes on the following form
\begin{equation}
    \mathcal{F}^{(II)}=\frac{5}{6}{X'}^3 +8{X'}^2Y' + 24X'{Y'}^2 + 24{Y'}^3\quad \text{with K\"{a}hler cone }\mathcal{K}_{II}=\{ X',Y'\geq 0\}\, .
\end{equation}
The coordinate change between $\cK_{II}$ and $\cK_I$ is given by $X'=X+4Y$ and $Y'=-Y$. We can then write down the prepotential in $\cK_{II}$ in the coordinates $X,Y$ as
\begin{equation}
\label{eq:X286 conifold phase II prepotential}
    \mathcal{F}^{(II)}=\frac{5}{6}X^3 + 2X^2Y - \frac{16}{6}Y^3 \qquad \text{with K\"{a}hler cone } \mathcal{K}_{II}=\{X\geq -4Y\geq 0\}\, .
\end{equation}
Thus, the extended K\"{a}hler cone is $\mathcal{K}_{\infty}[X_{(2,86)}]=\mathcal{K}_I\cup\mathcal{K}_{II}$. The $\cF=1$ hypersurface in $\cK_I$ can be conveniently parameterized as 
\begin{equation}\label{XofPhi}
    X=e^{-\Delta/\sqrt{3}}\, \qquad \text{and}\qquad Y=-\frac{5}{12}e^{-\Delta/\sqrt{3}}+\frac{1}{2}e^{2\Delta/\sqrt{3}}\, ,
\end{equation}
where $\Delta$ is the canonically normalized scalar in the single vector multiplet moduli space. In terms of $\Delta$, the flop boundary corresponds to 
\begin{equation}\label{phiflop}
    \Delta_{\rm flop}=-\frac{1}{\sqrt{3}}\log \frac65\,.
\end{equation}
There is only one infinite-distance limit which corresponds to $\Delta\rightarrow \infty$. This is an emergent heterotic string limit.\footnote{$\cK_{II}$ is bounded by a flop wall at $Y=0$ and a CFT boundary at $Y=-1/(2\cdot 3^{1/3})$. Therefore, the distance traversed in $\cK_{II}$ is always finite. Furthermore, the finite contribution to $\text{diam}(\mu)$ at small $\mu$ from $\cK_{II}$ is then simply the distance between these two boundaries.} 

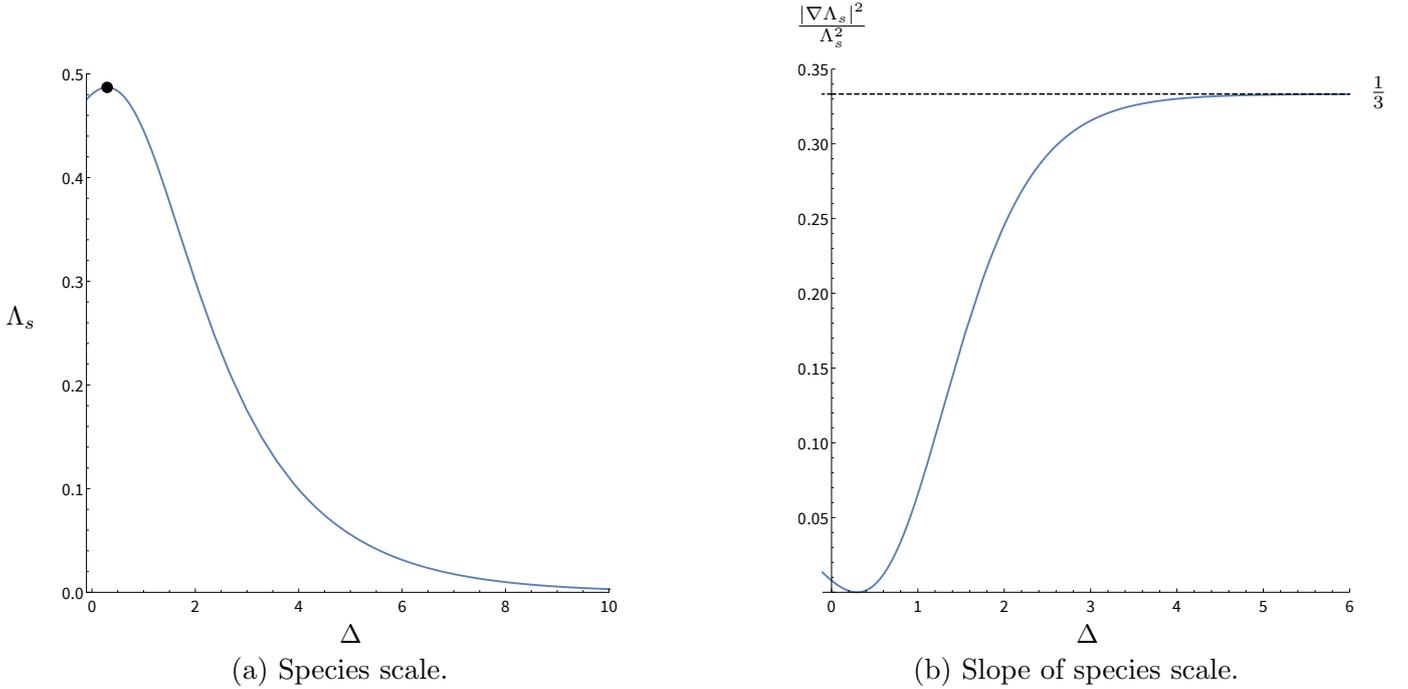
\begin{figure}[!t]
    \centering
    \vskip 0pt
    \begin{subfigure}{0.43\textwidth}
    \centering
    \includegraphics[width=\linewidth]{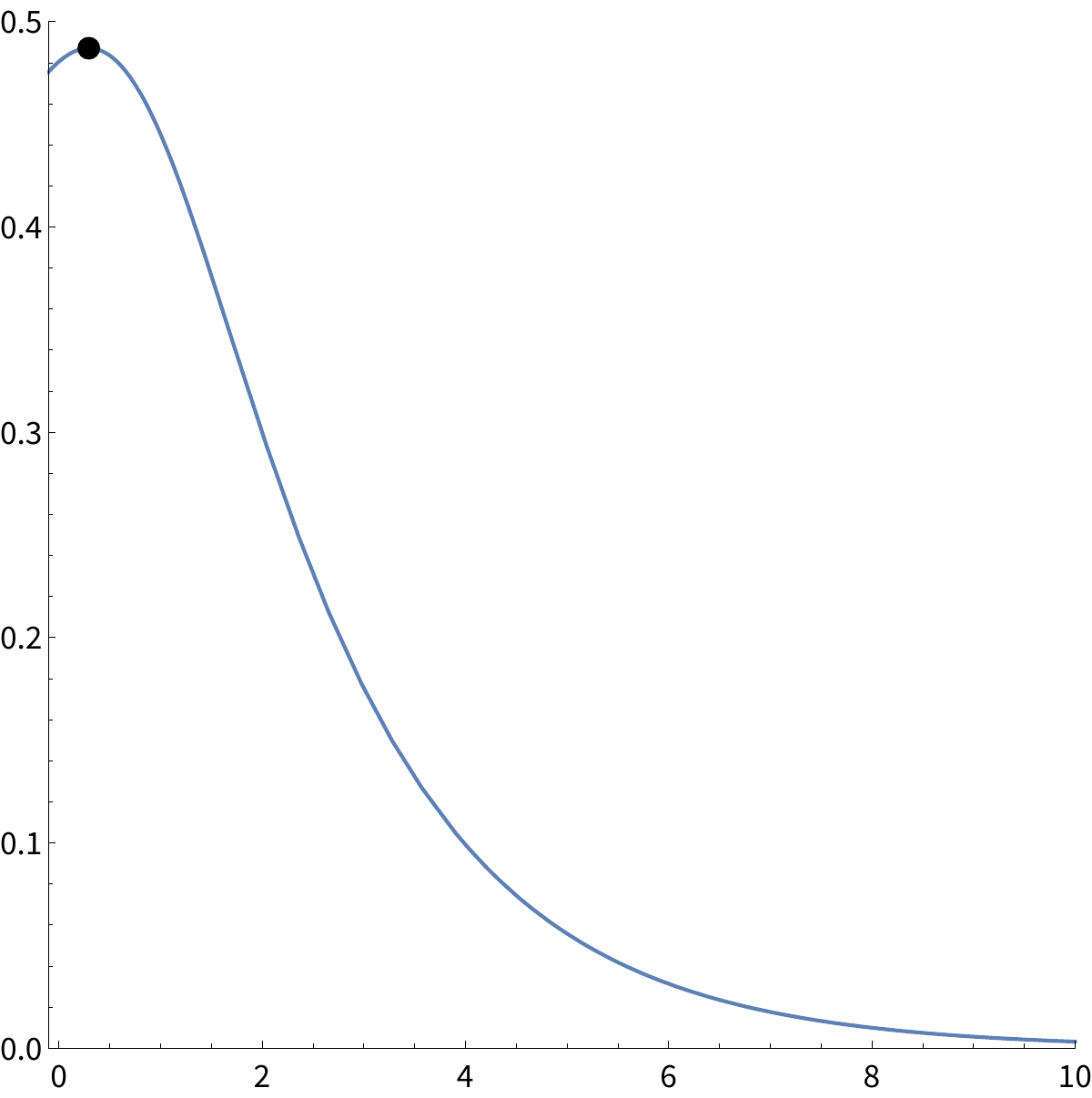}
    \begin{picture}(0,0)\vspace*{-1.2cm}
    \put(0,5){\footnotesize $\Delta$}
    \put(-125,125){\footnotesize $\Lambda_s$}
    \end{picture}\vspace*{-0.8cm}
    \caption{Species scale.} 
    \label{fig:X286 species}
    \end{subfigure}
    \hfill
    \begin{subfigure}{0.43\textwidth}
    \centering
    \includegraphics[width=\linewidth]{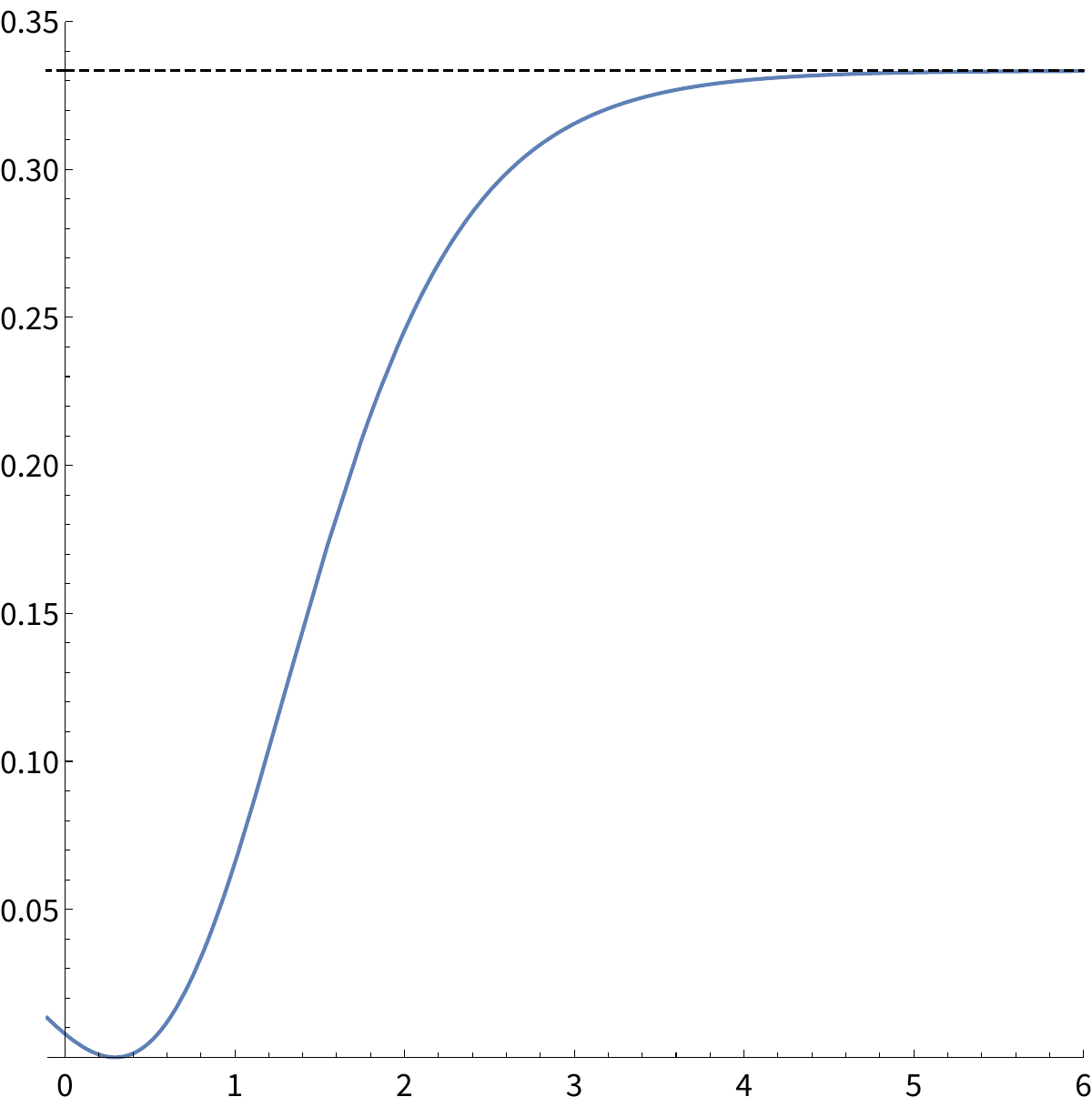}
    \begin{picture}(0,0)\vspace*{-1.2cm}
    \put(0,5){\footnotesize $\Delta$}
    \put(-105,235){\footnotesize $\frac{|\nabla \Lambda_s|^2}{\Lambda_s^2}$}
    \put(110,210){\footnotesize $\frac{1}{3}$}
    \end{picture}\vspace*{-0.8cm}
    \caption{Slope of species scale.}
    \label{fig:X286 species gradient}
    \end{subfigure}\vspace*{0.3cm}
    \caption{ Species scale (left) and slope $|\nabla \Lambda_s|^2/\Lambda_s^2$ (right) for M-theory on $X_{(2,86)}$ in $\cK_I$ as a function of $\Delta$ defined in \eqref{XofPhi}. 
    The black dot on the left indicates the local maximum of the species scale in $\cK_I$.
    }
    \label{fig:X286}
\end{figure}

We are interested in how the species scale varies over this moduli space, which may be computed in terms of the second Chern class via \eqref{Mtheoryspecies}. In general, the second Chern class changes as we pass through a flop wall. For completeness, we provide the integrated second Chern class for this example in the two possible phases
\begin{equation}
    \int_{X_{(2,86)}} c_2^{(I)}\wedge J = 50X+24Y\, , \qquad \int_{X_{(2,86)}} c_2^{(II)} \wedge J= 50X+56Y\, .
\end{equation}
As we are interested in the asymptotic behavior of the species scale and $\cK_{II}$ does not have an infinite-distance limit, let us only write down the species scale \eqref{Mtheoryspecies} in $\cK_I$:
\begin{equation}
\label{eq:X286 species scale}
    \Lambda_s^{(I)}=\left[\frac{1}{12}\left(50 X+24Y\right)\right]^{-1/2} = \left(\frac{10}{3} e^{-\Delta/\sqrt{3}} +e^{2\Delta/\sqrt{3}}\right)^{-1/2}\,.
\end{equation}

\paragraph{Slope of $\Lambda_s$.} In figure~\ref{fig:X286 species}, we show the species scale as a function of the modulus $\Delta$ and in \eqref{fig:X286 species gradient} the slope $|\nabla \Lambda_s|^2/\Lambda_s^2$ is shown. We see that the latter is bounded by $1/3$ everywhere in moduli space with the bound being saturated for $\Delta\rightarrow \infty$ in accordance with our general expectation for a heterotic emergent string limit. 

\paragraph{Desert point.} In phase $\cK_I$, we find an extremum for the species scale as illustrated in figure~\ref{fig:X286}. Its precise location and the value of the species scale at this point is given by
\begin{equation}
\label{eq:5d X286 desert}
    \Lambda_s^{(I)}\left(\Delta=\frac{1}{\sqrt{3}}\log \frac53\right) = \frac{1}{3^{1/6} \cdot 5^{1/3}}\,, 
\end{equation}
which lies below the 5d Planck scale. However, the expression in \eqref{Mtheoryspecies} for $X_{(2,86)}$ is continuous, but not differentiable across the flop wall, and the actual maximum of the species scale is reached in phase $\cK_{II}$ along the CFT boundary at $Y=-\frac14 X=-1/(2\cdot 3^{1/3})$ where we have 
\begin{equation}
    \Lambda_s^{\rm max.} =\frac{1}{\sqrt{2}\cdot 3^{1/3}}\approx 0.490\,,
\end{equation}
which is roughly a factor of two lower than the Planck scale.\footnote{Close to the flop transition, the states becoming massless at this wall in the moduli space should be integrated into the field theory description. In this regime, their effect on the higher-derivative corrections should therefore be subtracted carefully leading to a definition for the species scale that is differentiable across the flop wall. The effect of the light EFT states at the flop wall is the five-dimensional avatar of the logarithmic divergences appearing in 4d that we revisit in section~\ref{sec:refine}.}

\paragraph{Diameter.} Finally, we can evaluate the diameter of the region in moduli space where $\Lambda\geq \mu$ for some cut-off scale $\mu$. 
As before, we set $\Lambda_s=\mu$ and invert (\ref{eq:X286 species scale}) to compute the distance from the flop wall to $\Delta\big\vert_{\Lambda_s(\Delta)=\mu}$ in $\cK_I$. The distance between these two points take the following form in the small $\mu$ expansion
\begin{equation}
    \Delta = \sqrt{3}\log[\mu]-\frac{1}{\sqrt{3}}\log[1440]+\mathcal{O}(\mu^3)\, .
\end{equation}
We also need to consider the bulk contribution which is the finite diameter of $\cK_{II}$. Upon numerical integration, we obtain $\Delta_{\rm bulk,II}\approx 0.294$.\footnote{Here, we only include the numerical value of $\Delta_{\rm bulk,II}$. However, its closed-form expression may be computed and is included in the ancillary Mathematica notebook.} Combining everything together, we obtain the diameter of field space for M-theory compactified on $X_{(2,86)}$ as
\begin{equation}
\label{eq:X286 diameter}
    \text{diam}(\mu) \approx  -\sqrt{3}\log[\mu]-3.903\, .
\end{equation}
The coefficient appearing in front of $\log[\mu]$ agrees with the analysis for a heterotic emergent string limit, i.e.,~(\ref{eq:5d heterotic emergent limit}).

\section{Refined Bound on the Species Scale}\label{sec:refine}
One of the main motivations to study the relation between higher-curvature corrections and the species scale was to extract general patterns for the species scale not specific to certain examples with fixed dimension. Of particular interest is the validity of the bound \eqref{boundslope} on the slope of the species scale.
It was noticed in \cite{vandeHeisteeg:2023ubh}, based on the emergent string conjecture \cite{Lee:2019oct}, that asymptotically there is a bound on the slope of the species scale $\frac{|\nabla \Lambda_s|^2}{\Lambda_s^2} \leq \frac{M_{\rm pl}^{2-d}}{d-2}$.  Motivated by this bound, it was argued, based on the consistency of the higher-derivative expansion of the effective theory of gravity, that a similar upper bound on the slope exists everywhere including the interior of the moduli space. However, this qualitative argument did not allow us to fix the coefficient $c$ appearing in the bound above which was just argued to be $\mathcal{O}(1)$, and counter-examples were found for the original hope that the bound is given by $c={1\over d-2}$. Nevertheless, given the large set of explicit examples in various dimensions studied in this work, we are in a position to revisit the $\mathcal{O}(1)$ factor in the expression~\eqref{boundslope} and propose the originally motivated sharper bound 
\begin{equation}\label{boundrefined}
     \frac{|\nabla \Lambda_s|^2}{\Lambda_s^2} \leq \frac{1}{d-2} \,.
\end{equation}
 In all the examples we studied, this bound is indeed satisfied everywhere in moduli space --- with two notable exceptions that we revisit now. 

As we already mentioned, in \cite{vandeHeisteeg:2023ubh} it was found that the bound \eqref{boundrefined} is naively violated in the vector multiplet sector of 4d $\cN=2$ theories arising from Type II compactifications on Calabi--Yau threefolds $Y_3$. Given the proposal~\cite{vandeHeisteeg:2022btw} of identifying the species scale with the genus-one topological free energy 
\begin{equation}
    \Lambda_s\sim \frac{1}{\sqrt{F_1}}\,,
\end{equation}
it was shown in \cite{vandeHeisteeg:2023ubh} that, with this identification, the constraint \eqref{boundrefined} is already not satisfied in asymptotic emergent string limits. This violation of the proposed bound can be traced back to the behavior of $F_1$ at sub-leading order in the asymptotic regimes. In asymptotic regimes ($t \rightarrow \infty$), the topological genus-one free energy can be schematically written as 
\begin{equation}\label{F1asymp}
    F_1 = \frac{1}{12} c_{2}\,t - \beta \log t + \dots \,, 
\end{equation}
where $c_2$ is related to the integrated second Chern class of $Y_3$ and the coefficient $\beta$ depends on the Hodge numbers of $Y_3$ (see \cite{vandeHeisteeg:2023ubh} for a detailed discussion of $\beta$). If $\beta>0$, it was shown in \cite{vandeHeisteeg:2023ubh} that the slope of $\Lambda_s = F_1^{-1/2}$ approaches its asymptotic value from above. In particular, there exist simple examples of emergent string limits for which $\beta>0$ which violate the bound \eqref{boundrefined}. A similar situation arises already in eight-dimensional maximal supergravity discussed in section~\ref{ssec:8dsugra}. Using the notation introduced in that section, the emergent string limit corresponds to 
\begin{equation}\label{eq:emergentstring8d}
    \tau_2 \sim \lambda \,,\qquad \nu \sim \lambda^{-1} \,,\qquad \lambda \rightarrow \infty\,,
\end{equation}
while keeping $U$ fixed. In this limit, the species scale then scales as 
\begin{equation}
    \Lambda_s =\frac{1}{4^{1/6}} \left(2 \zeta(3) \lambda^2 - \frac{4\pi}{3} \log \lambda +\dots  \right)^{-1/6}\,,
\end{equation}
where the $\dots$ indicate suppressed terms in the limit $\lambda\rightarrow \infty$. We notice that the structure of the term in the brackets is very similar to \eqref{F1asymp} in that the first sub-leading term in the definition of the species scale depends logarithmically on the parameter that is taken to be large in the emergent string limit. 

Therefore, as illustrated in figure~\ref{fig:8dbump}, the bound~\eqref{boundrefined} seems to be violated in the limit \eqref{eq:emergentstring8d}. To understand the origin of the corrections responsible for this violation, we notice that in general the contribution of light modes with mass $m$ to dimension-$n$ operators in $d$-dimension is by dimensional analysis expected to contribute to the effective coupling as $1/m^{n-d}$ and logarithmically if $n=d$. Therefore, the contribution of states describable in the EFT to the couplings diverges if $n\geq d$ as $m\rightarrow 0$. Since the $R^2$-term in four dimensions and the $R^4$-term in eight dimensions have $n=d$, these states lead to logarithmic divergence of the coupling whereas in higher dimensions, they do not lead to a divergence. A similar situation arises close to the conifold where, as shown in \cite{vandeHeisteeg:2023ubh}, the slope of the $F_1$ diverges such that, naively, also the slope of the species scale violates the bound~\eqref{boundrefined}. The divergence in this case is due to the additional hypermultiplet that becomes massless at the conifold and hence should be included in the EFT and not contribute to the moduli-dependence of the species scale. In the following, we first want to discuss how the violation of the bound~\eqref{boundrefined} is avoided in infinite-distance regions and then subsequently discuss how a proper treatment of the conifold is consistent with the bound~\eqref{boundrefined}. 

\subsection{Contribution of EFT states at infinite distance}

In 8d maximal supergravity, the logarithmic dependence of the $R^4$-term on the moduli shows up when regularizing $E_{\footnotesize \frac{3}{2}}^{SL(3)}$ and $E_{\footnotesize 1}^{SL(2)}$, whereas in four dimensions it arises when regulating the IR divergence arising in the definition of 
\begin{equation}
    F_1 = \frac12 \int_{\cF} \frac{\dd^2\tau}{\tau_2} \Tr(-1)^F F_L F_R q^{H_0}\bar{q}^{\bar H_0}\,, 
\end{equation}
due to the zero modes. In fact, in the latter case the term in $F_1$ depending logarithmically on the moduli is fixed by the holomorphic anomaly \cite{Bershadsky:1993ta}. As argued in \cite{Antoniadis:1993ze}, the holomorphic anomaly precisely arises due to the propagation of massless states. Therefore, the logarithmic term in $F_1$ is a consequence of the running of the $R^2$-coupling due to massless modes. Similarly, in eight dimensions, the coefficient $\hat{E}_{\footnotesize\frac32}$ satisfies the Laplacian equation 
\begin{equation}
    \Delta \hat{E}_{\footnotesize\frac{3}{2}} = 4\pi\,,\qquad\text{with}\qquad  \Delta =  4 \tau_2^2 \partial_\tau \bar\partial_{\bar\tau} + \frac{|\partial_{B_N}-\tau \partial_{B_R}|^2}{\nu \tau_2}+3\partial_\nu (\nu^2 \partial_\nu)\, ,
\end{equation}
which can be viewed as the analogue of the holomorphic anomaly equation in four dimensions (see \cite{Berkovits:1994vy,Ooguri:1995cp}) and precisely captures the logarithmic divergence due to massless particles. Since the logarithmic terms parameterizes the running of the higher-curvature couplings due to \emph{massless} particles, it should not affect the scaling of the species scale as a function of moduli which should capture the effect of the massive states. We, therefore, propose that to define the species scale correctly we need to subtract the term corresponding to the running of the coupling due to integrating out states that are already part of the EFT. The exact subtraction scheme is dependent on the exact point in moduli space and chosen duality frame. Notice that since in higher dimensions, the light states do not lead to a divergence of the higher-curvature coupling, we did not need to subtract the moduli-dependent EFT state contribution in, e.g., 10d maximal supergravity.

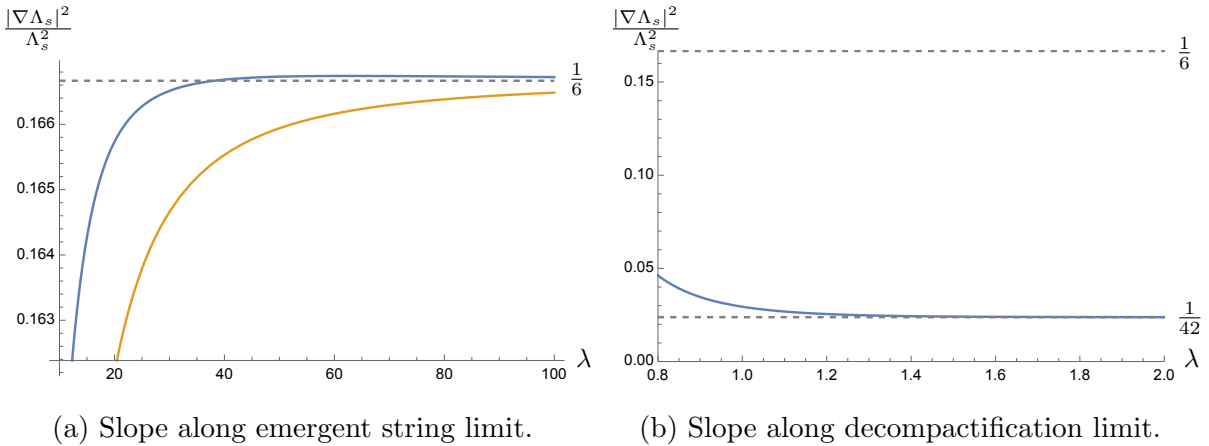
\begin{figure}[!t]
\begin{subfigure}{0.45\textwidth}
\begin{center}
\includegraphics[width=0.95\textwidth]{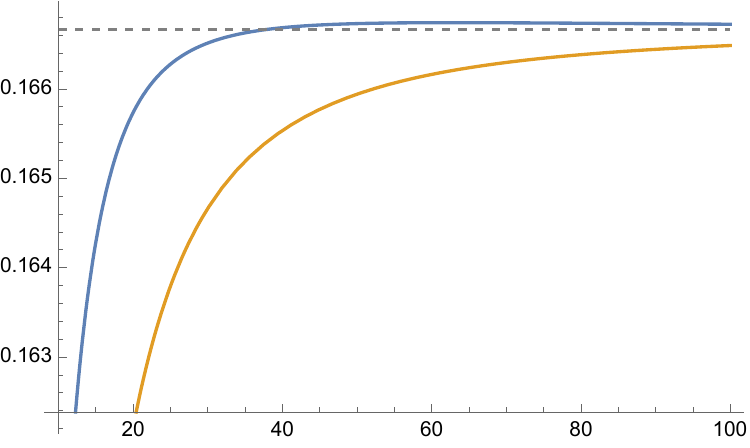}
\end{center}
\begin{picture}(0,0)
\put(215,35){\footnotesize$\lambda$}
\put(0,160){\footnotesize $\frac{|\nabla \Lambda_s|^2}{\Lambda_s^2}$}
\put(212,141){\footnotesize $\frac{1}{6}$}
\end{picture}
\vspace*{-40pt}\caption{Slope along emergent string limit.\label{fig:8dbump}}
\end{subfigure}
\hspace{2pt}
\begin{subfigure}{0.45\textwidth}
\begin{center}
\includegraphics[width=0.95\textwidth]{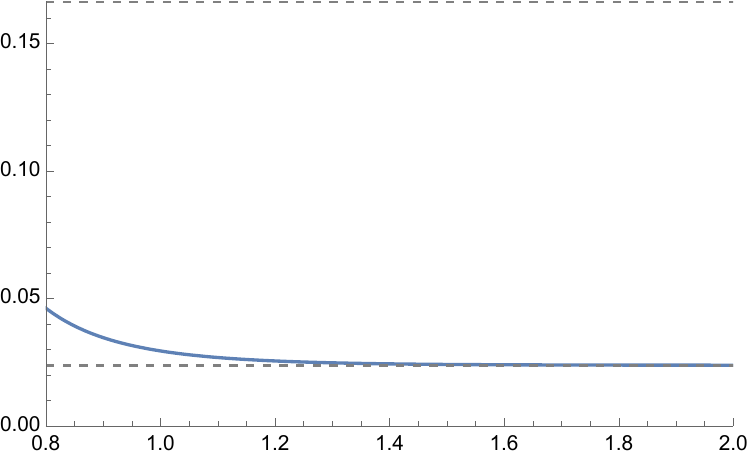}
\end{center}
\begin{picture}(0,0)
\put(215,35){\footnotesize$\lambda$}
\put(0,160){\footnotesize $\frac{|\nabla \Lambda_s|^2}{\Lambda_s^2}$}
\put(212,51){\footnotesize $\frac{1}{42}$}
\put(212,151){\footnotesize $\frac{1}{6}$}
\end{picture}
\vspace*{-40pt}\caption{Slope along decompactification limit.\label{fig:slopeabove}}
\end{subfigure}
\caption{\label{fig:8dgradients} Left: the slope of the species scale 8d M-theory on $T^3$ along the path $(\tau_2,\nu)=(\lambda,1/\lambda)$ (with $U_2=1$ and all axions set to zero): the dashed line denotes the emergent string bound $1/6$, the yellow curve the species scale without logarithmic terms, blue the species scale with logarithmic terms. Right: the slope of species scale along the 9d decompactification limit $(\tau_2,\nu,U_2)=(1,\lambda^{-2},\lambda)$ (with axions set to $(\tau_1,B_R,B_N,U_1)=(\tfrac{1}{2},-\frac{1}{3},-\tfrac{2}{3},\tfrac{1}{2})$), where the blue curve denotes the species scale without logarithmic terms. } 
\end{figure}

In eight dimensions, the subtraction of the moduli-dependent contribution of EFT states amounts to refining our identification of the species scale in \eqref{eq:species8d} to  
\begin{equation}\label{hatlambda8d}
    \hat{\Lambda}_s= \frac{1}{4^{1/6}}\left( \hat{E}_{\frac32}^{SL(3)}-4\pi \log \nu^{1/3} - 2\pi \log(|\eta(U)|^4)\right)^{-1/6}\,.
\end{equation}
Notice that the resulting expression is not U-duality invariant anymore --- neither with respect to $SL(2)$ nor $SL(3)$. From figure~\ref{fig:8dbump}, we see that the slope $|\nabla \hat{\Lambda}_s|^2/\hat{\Lambda}_s^2$ is indeed bounded by $1/6$ in accordance with our bound~\eqref{boundrefined}. 

We can proceed similarly in four dimensions. For simplicity, we focus here on the Enriques Calabi--Yau $Y_3=(K3\times T^2)/\mathbb{Z}_2$. For this manifold, the genus-one topological free energy is given by~\cite{Grimm:2007tm}
\begin{equation}\label{F1K3T2}
    F_1\left(\frac{K3\times T^2}{\mathbb{Z}_2}\right) = -6 \log\left[i(t-\bar{t}) |\eta(t)|^2\right] + F_1^{K3}(s, \bar{s})\,,
\end{equation}
where $t$ is the complexified volume of the torus factor and $s$ collectively stands for the moduli of K3. Here, we are interested in the $\text{Im}\,t\rightarrow \infty$ limit while keeping $s$ constant. This is an emergent string limit and, as shown in \cite{vandeHeisteeg:2023ubh}, the slope of $F_1$ approaches its asymptotic value $1/2$ from above in the limit $\text{Im}\,t\rightarrow \infty$. Since we keep $s$ constant, the details of $F_1^{K3}$ are not important to us, but we can effectively treat it as a positive constant.

\begin{figure}[!t]
\begin{center}
\includegraphics[width=10 cm]{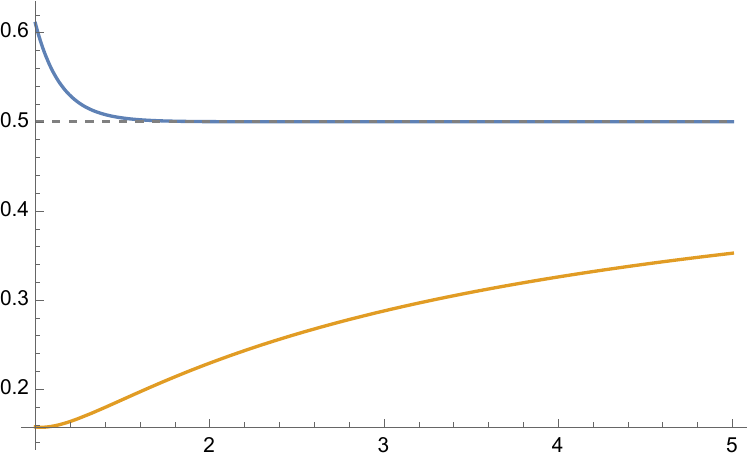}
\end{center}
\begin{picture}(0,0)
\put(386,38){$\tau_2$}
\put(95,210){$|\nabla \Lambda_s|^2/\Lambda_s^2$}
\end{picture}\vspace*{-0.8cm}
\caption{\label{fig:K3T2} Plot of the slope of the species scale for $(K3\times T^2)/\bbZ_2$ along a constant axion slice $\tau_1=1/2$. For blue curve $F_1^{K3}(s, \bar{s})$ is set to zero while for the yellow curve $F_1^{K3}(s, \bar{s})=1$. The dashed line indicates the emergent string bound $1/2$ for the slope.} 
\end{figure}
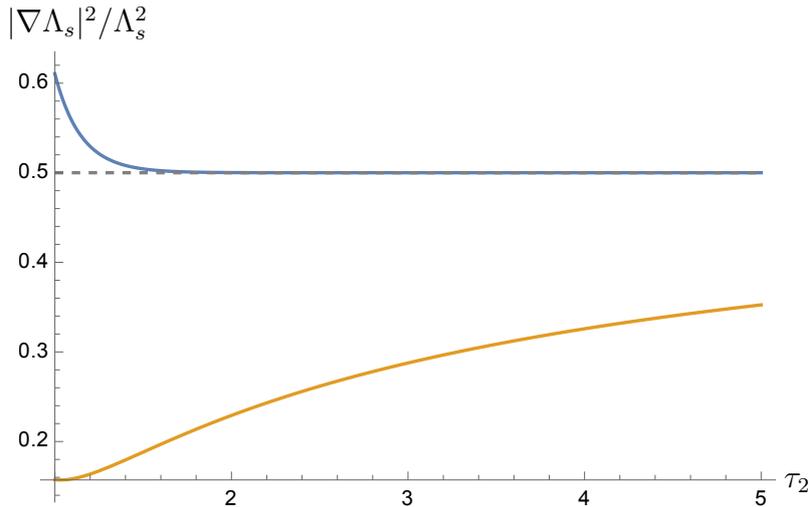

In \eqref{F1K3T2}, the term proportional to $\log[i(t-\bar{t})]$ is precisely the term that at large $\text{Im}\,t$ is fixed by the holomorphic anomaly equation \cite{Bershadsky:1993ta}. According to our general discussion, in the limit $\text{Im}t\rightarrow \infty$, we hence should subtract this term to obtain the correct behavior for the species scale such that we can define 
\begin{equation}
{\hat{\Lambda}_s}= \left(-6 \log |\eta(t)|^2 + \hat{F}_1^{\tiny K3}(s, \bar{s})\right)^{-1/2}\,,
\end{equation}
where $\hat{F}_1^{\tiny K3}$ is obtained by similar subtractions to $F_1^{\footnotesize K3}$ for which the details are, however, not important to our present discussion. The slope of $\hat{\Lambda}_s$ is shown in figure~\ref{fig:K3T2} from which we can see that, indeed, the bound~\eqref{boundrefined} is satisfied for $d=4$. Notice that to satisfy the bound, it is crucial to have the additional $t$-independent $\hat{F}_1^{\tiny K3}$ contributing positively to the species scale. If this term was absent, the bound~\eqref{boundrefined} can be violated as illustrated in figure~\ref{fig:K3T2}. In general, for emergent string limits in dimensions less than 10, we always expect such a constant contribution depending on additional moduli. The fact that their contribution can never be entirely absent reflects the fact that the weakly-coupled strings arising in the emergent string limits still have additional moduli corresponding to the compactification manifold in the dual frame. If the additional contributions to the species scale would vanish in the emergent string limit, this would mean that we could find a genuinely lower-dimensional critical string. The analysis of \cite{Baume:2019sry,Klaewer:2020lfg} illustrates, however, that such limits are expected to be obstructed at the quantum level implying that, indeed, there is always a constant contribution to the species scale in emergent string limits such that the slope $|\nabla \hat{\Lambda}_s|^2/\hat{\Lambda}_s^2$ approaches its asymptotic value from below. 

Notice that this does not mean that, for other type of limits, the species scale necessarily approaches its asymptotic value from below. However, this does not imply a violation of the bound~\eqref{boundrefined} since the asymptotic value for decompacitfication limits is well-below the value for emergent string limits. To illustrate this point, consider again eight-dimensional maximal supergravity with species scale given by \eqref{hatlambda8d}. In this theory, we can consider the decompactification to 9d by taking the limit (see section~\ref{ssec:8dsugra}) $\cV, U_2\rightarrow \infty$. From \eqref{hatESl3} and \eqref{hatlambda8d}, we see that, in this limit, the first subleading term in $\hat{\Lambda}_s$ is given by exponentials such that the sign of the correction to the asymptotic behavior of $\hat{\Lambda}_s$ depends on the value of the axionic coordinates. The slope of the species scale for the choice of axions $(\tau_1,B_R,B_N,U_1)=(\tfrac{1}{2},-\frac{1}{3},-\tfrac{2}{3},\tfrac{1}{2})$ is shown in figure~\ref{fig:slopeabove}. It is clear from there that even though the slope of the species scale approaches its asymptotic value from above, it does not violate the bound~\eqref{boundrefined} anywhere in moduli space.

\subsection{Refined treatment of the conifold}
It was already noticed in \cite{vandeHeisteeg:2022btw} that upon identifying $F_1=\Lambda_s^{-2}$ in 4d $\cN=2$ theories the conifold point requires careful treatment. The reason for this is that close to the conifold, $F_1$ diverges logarithmically 
\begin{equation}
    F_1 = -\frac{1}{12} \log |u|^2 +\dots\,,
\end{equation}
where $u=0\in \mathbb{C}$ corresponds to the conifold locus. This term is generated by the hypermultitplet that becomes massless at $u=0$ and hence needs to be included in the EFT. Similar to the contributions of the massless states to $F_1$ in the asymptotic limits discussed above, the relevant term in $F_1$ generated by the additional hypermultiplet in the EFT depends logarithmically on the conifold modulus $u$. As for the slope $|\nabla F_1|/F_1$, it was shown in \cite{vandeHeisteeg:2023ubh} that for $u\rightarrow 0$ it diverges as 
\begin{equation}
    \frac{|\nabla F_1|}{F_1} \sim \frac{1}{|u|^2(-\log |u|)^3}\,,
\end{equation}
implying that, naively, the bound~\eqref{boundrefined} is violated for small $u$. However, since $F_1$ is calculated in string perturbation theory, it does not include the additional hypermultiplet in the EFT. Therefore, $F_1$ only captures the species scale correctly as long as the mass of the hypermultiplet is above the Type II string scale, i.e.,
\begin{equation}\label{uovergs}
    \frac{u}{g_s} \geq 1\,. 
\end{equation}
In order to ensure the validity of the perturbative string description in the limit $u\rightarrow 0$, we need to co-scale $g_s\sim u$. Since $g_s$ is part of a hypermultiplet, the dependence of the species scale on $g_s$ is not directly captured by $F_1$ and we cannot give a closed form for it at finite $g_s$. However, since the limit $g_s\rightarrow 0$ corresponds to an emergent string limit, we know that $\Lambda_s \sim M_s \sim g_s$ in this limit. We account for this by effectively adding a $u$-independent contribution to $F_1$ 
\begin{equation}
    F_1 = -\frac{1}{12} \log |u|^2 + N(g_s)\,,
\end{equation}
where the additive contribution depends on $g_s$.  In the scaling limit $g_s\sim u$ for which the mass of the conifold state does not drop below the string scale, we can still trust the EFT and check if the conjectured bound on the slope of $\Lambda_s$ is still satisfied. The field space metric for the dilaton in four-dimensions and $u$ for $u,g_s\ll1 $ is given by 
\begin{equation}
   \dd s^2 = 2 g_s^2\,\dd g_s^2 - \log|u|^2 \,\dd u^2\,.
\end{equation}
Using this, we evaluate the slope of the species scale to be 
\begin{equation}\label{slopeconifold}
\begin{aligned}
    \left|\frac{\nabla \Lambda_s}{\Lambda_s} \right|^2 &\sim  \frac{g_s^2}{8} \frac{(\partial_{g_s} N(g_s))^2}{(-\frac{1}{12}\log|u|^2 + N(g_s))^2}-\frac{1}{4 \log|u|^2} \frac{1}{|u|^2(-\log|u|^2 + 12 N(g_s))^2}  \\
    &\hspace{-7pt}\stackrel{g_s=|u|}{\sim } \frac{1}{2} +\frac{1}{12} |u|^2 \log |u|^2 - \frac{1}{576} \frac{|u|^2}{\log |u|^2} + \mathcal{O}(|u|^4)\,.
\end{aligned}
\end{equation}
In the last step, we used $N(g_s)\sim g_s^{-2}$, set $g_s=|u|$ and expanded around $|u|=0$. We note that in the second line the $u$-dependent terms vanish in the limit $|u|\rightarrow 0$. For small $|u|$, the term proportional to $\log|u|^2$ is negative and dominates over the positive contribution from the term proportional to $1/\log|u|^2$. Therefore, in the limit $|u|\sim g_s\rightarrow 0$ the slope $|\nabla \Lambda_s|^2/\Lambda_s^2$ approaches $1/2$ from below and hence satisfies the bound \eqref{boundrefined}.\\

After accounting correctly for the effect of the zero modes on the higher-curvature couplings, we hence see that all our examples satisfy the bound \eqref{boundrefined} at every point in moduli space, providing strong evidence that this bound is satisfied in general. We therefore have a proposed sharp constraint on the species scale and hence the light states in the theory that may be valid at any point in moduli space. 

\section*{Acknowledgements}
It is a pleasure to thank Naomi Gendler for collaboration during the initial stages of this work and John Stout for helpful discussions. We would also like to thank Alek Bedroya and Houri Tarazi for useful comments on an earlier version of this paper. This work is supported by a grant from the Simons Foundation (602883,CV), the DellaPietra Foundation, and by the NSF grant PHY-2013858. We thank the 2023 Summer Workshop of the Simons Center for Geometry and Physics for kind hospitality where this work was initiated.

\appendix

\section{Eisenstein series for $SL(n,\bbZ)$}\label{app:Eisen}
In this appendix we detail the computation of real-analytic Eisenstein series for $SL(n,\bbZ)$. These series show up as coefficients of the $R^4$-corrections in maximal supergravities in 8, 9 and 10 dimensions considered in this work. In general the Eisenstein series for $SL(n,\bbZ)$ is defined as
\begin{equation}\label{eq:defEisenstein}
    E_s^{SL(n)}(M) = \sum_{p \in \mathbb{Z}^n\backslash \{0\}}\frac{1}{|p^T M p |^s}\, .
\end{equation}
where we take $M$ to be any positive-definite, symmetric matrix with determinant $\det M = 1$. In the following we consider this Eisenstein series at $s=1,\tfrac{3}{2}$ for $SL(2,\bbZ)$ and $s=\tfrac{3}{2}$ for $SL(3,\bbZ)$.

\subsection{Eisenstein series for $SL(2,\bbZ)$}\label{app:eisensteinSL2}
In this subsection we consider the Eisenstein series $E_{1}(\tau, \bar \tau) $ and $E_{\frac{3}{2}}(\tau, \bar \tau) $ for $SL(2,\bbZ)$ which we parameterize by the usual coordinate $\tau$ on the upper half-plane. The relation with the matrix $M$ used in \eqref{eq:defEisenstein} is then given by
\begin{equation}
    M = \frac{1}{\tau_2} \begin{pmatrix}
        1 & \tau_1 \\
        \tau_1 & \tau_1^2 + \tau_2^2
    \end{pmatrix}\, .
\end{equation}
There are two special points for this Eisenstein series: $\tau=i$ and $\tau=\rho\equiv \tfrac{1}{2}+i\tfrac{\sqrt{3}}{2}$. In the following we will evaluate the various Eisenstein series and their derivatives at these points. 

\subsubsection*{Eisenstein series $\hat{E}_1(\tau, \bar \tau)$}
Here we consider the Eisenstein series $E_1(\tau, \bar \tau)$. It arises as coefficient of the $R^4$-correction for 8d M-theory on $T^3$ discussed in section \ref{ssec:8dsugra}. As this sum is divergent, we consider the regularized version
\begin{equation}
    \hat{E}_1(\tau , \bar \tau) = - 2\pi \log \tau_2 |\eta^2(\tau)|^2\, ,
\end{equation}
where $\eta(\tau)$ is the Dedekind eta function. This expression may be obtained by expanding around $s=1+\epsilon$ and subtracting the divergent $1/\epsilon$ term with an appropriate constant. We now want to evaluate this Eisenstein series at the special points $\tau=i$ and $\tau=\rho$. As we know the values of the Dedekind eta function at these points, we find that
\begin{equation}
    \hat{E}_1(i,-i) = 2\pi \log \frac{16\pi^3}{\Gamma[\frac{1}{4}]^4}\approx 6.6268\, , \qquad     \hat{E}_1(\rho,\bar\rho) = 2\pi \log \frac{32\pi^4}{\Gamma[\frac{1}{3}]^6} \approx 6.49379\, .
\end{equation}
We next compute the Hessian of the Eisenstein series at these special points, that is, $2\tau_2^2 \partial_i \partial_j \hat{E}_1$ with $i,j=\tau_1,\tau_2$. Here we included the factor of $2\tau_2^2$ to ensure the correct normalization of the kinetic terms for the scalars. For the eigenvalues of the Hessian we find
\begin{equation}
\begin{aligned}
   2\tau_2^2\partial_i \partial_j\hat{E}_1(\tau,\bar \tau) \bigg|_{\tau=i} &= \begin{pmatrix}
     -1.87372 & 0 \\
     0 & 8.1569  
   \end{pmatrix} \, , \\
   \qquad  2\tau_2^2\partial_i \partial_j\hat{E}_1(\tau,\bar\tau)\bigg|_{\tau=\rho} &= \begin{pmatrix}
    4.18879 & 0 \\
    0 & 4.18879
   \end{pmatrix} \, .
\end{aligned}
\end{equation}
We conclude that $\tau=i$ is a saddle point of $\hat{E}_1$ while $\tau=\rho$ is a (global) minimum. The instability of $\tau=i$ is due to a negative eigenvalue along the axionic direction. It is also remarkable that for $\tau=\rho$ the two eigenvalues are precisely equal.

\subsubsection*{Eisenstein series $E_s(\tau, \bar \tau)$}
Here we consider the Eisenstein series $E_s(\tau, \bar \tau)$ for $s>1$. We are interested in the exact computation of these infinite series at $\tau=i,\rho$, which we achieve by number theoretic methods (see \cite{latticesums} for a review). In general, the Eisenstein series may be computed exactly whenever $\tau$ lies in an imaginary quadratic field $\tau \in \mathbb{Q}(i\sqrt{d})$, with $d$ some positive integer, i.e., when the elliptic curve associated to $\tau$ has complex multiplication. We write this value for $\tau$ as
\begin{equation}
    \tau=-\frac{b}{2a}+\frac{i\sqrt{-b^2+4ac}}{2a}\, ,
\end{equation}
where $d=4ac-b^2$. For these values of $\tau$ we can rewrite the Eisenstein series into a so-called Epstein zeta function as
\begin{equation}
    \zeta_Q(s) = \sum_{(p,q)\in \mathbb{Z}^2\backslash 0} \frac{1}{Q(p,q)^s}\, , \qquad Q(p,q) = (p,q) \begin{pmatrix}
        a & b/2 \\
        b/2 & c 
    \end{pmatrix}\begin{pmatrix}
        p \\
        q
    \end{pmatrix} = a p^2 + b pq + c q^2\, .
\end{equation}
where $d$ is now the determinant of this matrix. For the values $\tau=i,\rho$ this yields
\begin{equation}
    E_{s}(i,-i) = \zeta_{Q_i }(s) \, , \qquad     E_{s}(\rho,\bar{\rho}) = \left(\tfrac{\sqrt{3}}{2}\right)^{s} \zeta_{Q_\rho}(s) \, , 
\end{equation}
where we defined
\begin{equation}
    Q_i = \begin{pmatrix}
        1 & 0 \\
        0 & 1
    \end{pmatrix}\, , \qquad Q_\rho = \begin{pmatrix}
        1 & 1/2 \\
        1/2 & 1
    \end{pmatrix}\, .
\end{equation}
The general strategy of evaluating these Epstein zeta functions is to rewrite the lattice sum into the product of two one-dimensional sums. One of the two factors will be $\zeta(s)$, while the other factor depends on the precise bilinear $Q$ that specifies the sum. In the following we carry out these summations for $\tau=i$ and $\tau=\rho$.

\paragraph{Value at $\tau=i$.} Let us first compute the Eisenstein series for $\tau=i$. We can rewrite it in terms of an $L$-function value as
\begin{equation}
    E_{s}(i,-i) = 4 \zeta(s) L(s,\chi_i) = 4\zeta(s) \sum_{n=1}^\infty \frac{\chi_i(n)}{n^s}\, ,
\end{equation}
with the character $\chi_i$ given by
\begin{equation}
    \chi_i(n) = \begin{cases}
        0 &\text{ if $n=0 \mod 2$,} \\
        1 &\text{ if $n=1 \mod 4$,} \\
        -1 &\text{ if $n=3 \mod 4$.}
    \end{cases}
\end{equation}
By writing out the sum over $n$ explicitly one recognizes the Dirichlet beta function $\beta(s)$ as
\begin{equation}\label{eq:Eisensteini}
    E_s(i,-i) = 4 \zeta(s) \beta(s) = 4 \zeta(s) \sum_{n=0}^\infty \frac{(-1)^n}{(2n+1)^s}\, .
\end{equation}

\paragraph{Value at $\tau=\rho$.} We next consider the Eisenstein series at the third root of unity. Similar to $\tau=i$ we can rewrite the lattice sum into an $L$-function value as
\begin{equation}
    E_s(\rho,\bar{\rho}) = 6 \left(\frac{\sqrt{3}}{2}\right)^{s} L(s,\chi_\rho)\, .
\end{equation}
The character $\chi_\rho$ associated to this sum is now given by
\begin{equation}
    \chi_\rho(n) = \begin{cases}
        0 &\text{ if $n=0 \mod 3$,}\\
        1 &\text{ if $n=1 \mod 3$,}\\
        -1 &\text{ if $n=2 \mod 3$.}
    \end{cases}
\end{equation}
This allows us to evaluate the Eisenstein series to be
\begin{equation}\label{eq:Eisensteinrho}
    E_s(\rho) = 6 \left(\tfrac{\sqrt{3}}{2}\right)^{s} 3^{-s}\left[\zeta_{\frac{1}{3}}(s) - \zeta_{\frac{2}{3}}(s) \right]\, ,
\end{equation}
where we defined the generalized Riemann zeta functions
\begin{equation}
    \zeta_a(s) = \sum_{n=0}^\infty \frac{1}{(n+a)^s}\, .
\end{equation}

\subsubsection*{Eisenstein series $E_{\frac{3}{2}}(\tau, \bar \tau)$}
Finally we specialize to $s=\tfrac{3}{2}$, in which case the Eisenstein series computes the $R^4$-correction to 10d Type IIB discussed in section \ref{ssec:10dIIB}. Here we will investigate the stability at $\tau=i,\rho$ more closely. Let us begin by evaluating the Eisenstein series at these points explicitly. By using \eqref{eq:Eisensteini} and \eqref{eq:Eisensteinrho} we find that
\begin{equation}
    E_{\frac{3}{2}}(i,-i) = 4 \zeta(\tfrac{3}{2})\beta(\tfrac{3}{2}) \approx 9.03362\, , \quad E_{\frac{3}{2}}(\rho,\bar{\rho}) = \left(\tfrac{\sqrt{3}}{2}\right)^{1/2} \zeta(\tfrac{3}{2}) \left( \zeta_{\frac{1}{3}}(\tfrac{3}{2}) - \zeta_{\frac{2}{3}}(\tfrac{3}{2})  \right) \approx 8.89275\, .
\end{equation}
We note that these numerical values may also be obtained by taking the expression \eqref{eq:E32series} and including only the first few D(-1)-instanton corrections. We next consider the Hessian $2\tau_2^2 \partial_i \partial_j E_{\frac{3}{2}}(\tau,\bar \tau)$, where the factor $2\tau_2^2$ represents the normalization by the metric and $i,j=\tau_1,\tau_2$. We find that
\begin{equation}
\begin{aligned}
    2\tau_2^2 \partial_i \partial_j E_{\frac{3}{2}}(\tau,\bar \tau)\bigg|_{\tau=i} &= \begin{pmatrix}
        -3.96778 & 0 \\
        0 & 17.5182
    \end{pmatrix}\, , \\
     2\tau_2^2 \partial_i \partial_j E_{\frac{3}{2}}(\tau,\bar \tau)\bigg|_{\tau=\rho} &= \begin{pmatrix}
        6.66956 & 0 \\
        0 & 6.66956
    \end{pmatrix}\, . 
\end{aligned}
\end{equation}
We thus arrive at the same conclusions as for $\hat{E}_1(\tau,\bar\tau)$: there is a global minimum at $\tau=\rho$ while $\tau=i$ is a saddle point (again due to the negative eigenvalue along the $\tau_1$ direction). Also note that the eigenvalues of the Hessian at $\tau=\rho$ are equal as before.

\subsection{Fundamental domain of $SL(3,\bbZ)$}\label{app:Sl3dualities}
In this subsection we characterize the fundamental domain of $SL(3,\bbZ)$. Recall from section \ref{ssec:8dsugra} that this field space is spanned by two saxions $\tau_2,\nu$ and three axions $\tau_1,B_R,B_N$. For simplicity we restrict our attention to vanishing axions $\tau_1=B_R=B_N=0$. Here we will work out the precise S-duality transformations inside $SL(3,\bbZ)$ on the remaining moduli space spanned by $\tau_2,\nu$. 

Let us begin by considering the three analogues of the familiar S-duality of $SL(2,\bbZ)$. These are generated by the matrices
\begin{equation}
    S_1 = \begin{pmatrix}
        0 & -1 & 0 \\
        1 & 0 & 0 \\
        0 & 0 & 1
    \end{pmatrix}\, , \qquad S_2 = \begin{pmatrix}
        1 & 0 & 0 \\
        0 & 0 & -1 \\
        0 & 1 & 0 
    \end{pmatrix}\, , \qquad S_3 = \begin{pmatrix}
        0 & 0 & 1 \\
        0 & 1 & 0 \\
        -1 & 0 & 0 
    \end{pmatrix}\, .
\end{equation}
We can infer the action of these S-duality transformations by acting on the matrix $M$, given in \eqref{eq:SL3M}, that parameterizes our moduli space as $M \to S M S^T$. From this transformation we read off the action on the coordinates $\tau_2,\nu$ (in the case of vanishing axions) as
\begin{equation}
    (\tau_2,\nu) \to (1/\tau_2, \nu)\, , \qquad (\tau_2,\nu) \to (\sqrt{\frac{\tau_2}{\nu}} , \frac{1}{\nu^{1/2}\tau_2^{3/2}})\, , \qquad (\tau_2,\nu) \to (\sqrt{\tau_2\nu}, \frac{\tau_2^{3/2}}{\nu^{1/2}})\, .
\end{equation}
There are two additional S-duality transformations, that may be obtained by combining two out of three operators $S_1,S_2,S_3$ above. Concretely, we take
\begin{equation}
    S_4 = S_1 S_2 = \begin{pmatrix}
        0 & 0 & 1 \\
        1 & 0 & 0 \\
        0 & 1 & 0
    \end{pmatrix}\, , \qquad S_5 = S_1 S_3 = \begin{pmatrix}
        0 & -1 & 0 \\
        0 & 0 & 1 \\
        -1 & 0 & 0
    \end{pmatrix}\, .
\end{equation}
These transformations act on the coordinates as
\begin{equation}
    (\tau_2, \nu) \to (\sqrt{\frac{\nu}{\tau_2}}, \frac{1}{\tau_2^{3/2}\nu^{1/2}})\, , \qquad (\tau_2,\nu) \to (\sqrt{\frac{\nu}{\tau_2}}, \frac{\tau_2^{3/2}}{\nu^{1/2}})\, .
\end{equation}
Altogether this gives us 5 S-duality operators. From the validity of the instanton expansion in \eqref{eq:E32series} we know that we have a fundamental domain given by $\tau_2> 1$ and $\sqrt{\nu \tau_2}<1$. In total there are six such patches in the $(\tau_2,\nu)$-plane that are mapped to each other by the S-duality transformations above. This proves to be useful for us in extending \eqref{eq:E32series} to these other regimes --- for instance in producing figure \ref{fig:8dspecies} --- as we can simply plug this coordinate change into the instanton expansion.

\subsection{Eisenstein series for $SL(3,\bbZ)$}\label{app:EisensteinSl38d}
In this appendix we analyze the Eisenstein series $E_{\frac{3}{2}}(M)$ of $SL(3,\bbZ)$. It arises as the $R^4$-coupling of 8d M-theory on $T^3$ discussed in section \ref{ssec:8dsugra}. As this lattice sum is divergent, we have to regularize it by expanding around $s=\tfrac{3}{2}+\epsilon$ and subtracting the $1/\epsilon$ term with an appropriate constant. The resulting series expansion for $\hat{E}_{\frac{3}{2}}(M)$ has been given explicitly in \eqref{eq:E32series}. In the following we use this series to analyze the special points in moduli space and their stability.

\paragraph{Cubic lattice.} We begin with the standard cubic lattice. It corresponds to the following point in the $SL(3,\bbR)/SO(3)$ moduli space
\begin{equation}
    M_{\mathbb{I}} = \begin{pmatrix}
        1 & 0 & 0 \\
        0 & 1 & 0 \\
        0 & 0 & 1
    \end{pmatrix}\, , \qquad \tau = i\, , \quad \nu = 1\, , \quad B_R =  B_N = 0\, .
\end{equation}
For this point we compute the Eisenstein series and the eigenvalues of its (normalized) Hessian numerically. We find these to be
\begin{equation}
\begin{aligned}
    \hat{E}_{\frac{3}{2}}(M_{\mathbb{I}})&=9.13479\, , \\
    \text{eigvals}(G^{ik}\partial_k \partial_j \hat{E}_{\frac{3}{2}}(M)) \big|_{M=M_{\mathbb{I}}} &= (19.0355,19.0355,-4.31275,-4.31275,-4.31275)\, .
\end{aligned}
\end{equation}
This lattice thus gives a saddle point due to the three (equal) negative eigenvalues.

\paragraph{$A_3$ root lattice.} We next consider the $A_3$ root lattice, also called the face-centered cubic (FCC) or hexagonal close-packing (HCC) lattice. It corresponds to the following point in the $SL(3,\bbR)/SO(3)$ moduli space
\begin{equation}
    M_{A_3} = 2^{-2/3}\begin{pmatrix}
        2 & -1 & 0 \\
        -1 & 2 & -1 \\
        0 & -1 & 2
    \end{pmatrix}\, , \quad \tau = -\frac{1}{2}+\frac{\sqrt{3}i}{2}\, , \ \ \nu = \frac{3\sqrt{3}}{4}\, , \ \ B_R = -\frac{1}{3}\, , \ \ B_N = -\frac{2}{3}\, .
\end{equation}
For this point we compute the Eisenstein series and the eigenvalues of its (normalized) Hessian numerically. We find these to be
\begin{equation}
\begin{aligned}
    \hat{E}_{\frac{3}{2}}(M_{A_3})&=8.79961\, , \\
    \text{eigvals}(G^{ik}\partial_k \partial_j \hat{E}_{\frac{3}{2}}(M)) \big|_{M=M_{A_3}} &= (7.58637, 7.58637,7.58637, 1.18682, 1.18682)\, .
\end{aligned}
\end{equation}
From comparison to other special points we find that the $A_3$ root lattice is the global minimum of $\hat{E}_{\frac32}$. Also note that it has only two distinct eigenvalues.

\paragraph{$A_3^*$ root lattice.} We next consider the $A_3^*$ root lattice; this lattice is dual to the $A_3$ root lattice, and also known as the body-centered cubic (BCC) lattice. It corresponds to the following point in the $SL(3,\bbR)/SO(3)$ moduli space
\begin{equation}
    M_{A_3^*} = 2^{-4/3}\begin{pmatrix}
        3 & -1 & -1 \\
        -1 & 3 & -1 \\
        -1 & -1 & 3
    \end{pmatrix}\, , \qquad \tau = -\frac{1}{3}+\frac{2\sqrt{2}i}{3}\, , \quad \nu = \sqrt{2}\, , \quad B_R = B_N = -\frac{1}{2}\, .
\end{equation}
For this point we compute the Eisenstein series and the eigenvalues of its (normalized) Hessian numerically. We find these to be
\begin{equation}
\begin{aligned}
    \hat{E}_{\frac{3}{2}}(M_{A_3^*})&=8.8031\, , \\
    \text{eigvals}(G^{ik}\partial_k \partial_j \hat{E}_{\frac{3}{2}}(M)) \big|_{M=M_{A_3^*}} &= (7.90579, 7.90579,7.90579 ,0.707685,0.707685)\, .
\end{aligned}
\end{equation}
This lattice thus gives another minimum of $\hat{E}_{\frac32}$, although it is not a global one.

\section{Geodesic distances for del Pezzo surfaces}\label{app:delpezzo}
In this appendix we discuss some details about the computation of geodesic distances in the field space of del Pezzo surfaces. This analysis is of interest to the study of 6d F-theory field spaces in section \ref{sec:6d}. As was already discussed there, the metric on the constant volume submanifold reduces to the generalized hyperbolic metric 
\begin{equation}\label{eq:metricapp}
    \dd s^2 = \dd x^2 + \sinh^2 x \dd \Omega_{r-1}^2\, .
\end{equation}
Recall that geodesics on any sphere $S^{r-1}$ lie along great circles, i.e., circles on the sphere whose origin coincides with that of the sphere. For this metric \eqref{eq:metricapp} this allows us to reduce the problem of finding geodesics by considering an appropriate great circle $S^1 \subset S^{r-1}$. If we parameterize this coordinate by $\theta$, the metric then reduces to
\begin{equation}
    \dd s^2 =  \dd x^2 + \sinh^2 x \dd \theta^2\, .
\end{equation}
Note that this metric is just that of the hyperbolic plane, for which we know that geodesics are either lines of constant $\theta$ or described by
\begin{equation}
    \tanh[x] = \frac{\tanh[x_0]}{\cos( \theta - \theta_0)}\, ,
\end{equation}
here $(x_0,\theta_0)$ is the point closest to the origin. The distance between two points $(x_1,\theta_1),(x_2,\theta_2)$ is given by
\begin{equation}\label{eq:dPdist}
    \Delta = \text{arccosh}[\cosh x_1 \cosh x_2 - \sinh x_1 \sinh x_2 \cos(\theta_2 - \theta_1)] \, .
\end{equation}
We will now be interested in finding geodesics between the desert point and a point along any of the emergent string limits. Concretely, we have the begin and endpoints 
\begin{equation}
    X_1^I=(X_1^0, X_1^i) = (\frac{3}{\sqrt{9-r}}, \frac{1}{\sqrt{9-r}})\, , \qquad     X_2^I = (\cosh x, \sinh x, 0, \ldots, 0)\, ,
\end{equation}
where we chose the second endpoint to lie along the emergent string limit of $H-E_1$. In terms of the coordinates $x$ and $u^i$ on the $(r-1)$-sphere defined by \eqref{eq:dPconstantvolume} we find that these points are mapped to
\begin{equation}
    x_1 = \text{arccosh}[\frac{3}{\sqrt{9-r}}]\, , \ \ u_1^i = \frac{1}{\sqrt{r}}\, , \qquad x_2 = x, \ \ u_2^i=(1,0,\ldots,0)\, .
\end{equation}
Let us now fix a great circle $S^1 \subset S^{r-1}$ that passes through $u_1^i$ and $u_2^i$ such that $u_2^i$ lies at $\theta_2=0$ and $u_1^i$ within $\theta_2 \in [0,\pi/2)$. Then we can compute the angle $\theta_1$ simply from the inner product of the vectors $u_1^i$ and $u_2^i$ to be
\begin{equation}
    \theta_1 = \arccos[\frac{1}{\sqrt{r}}]\, .
\end{equation}
With all of these preparations in place, we are ready to compute the geodesic distance for the diameter of field space. Let us first summarize what are the begin and endpoints in the $(x,\theta)$-coordinates we have established
\begin{equation}
    (x_1,\theta_1) = (\text{arccosh}[\frac{3}{\sqrt{9-r}}],\arccos[\frac{1}{\sqrt{r}}])  \, , \qquad (x_2,\theta_2) = (x, 0)\, .
\end{equation}
Plugging this into the distance formula \eqref{eq:dPdist} we find that
\begin{equation}\label{eq:geodist}
    \Delta = \text{arccosh}\left[\frac{3 \cosh x - \sinh x}{\sqrt{9-r}} \right]\approx x +\log \frac{2}{\sqrt{9-r}}\, ,
\end{equation}
where we approximated for large $x\gg 1$ in the second step.

\bibliography{papers_Max}

\providecommand{\href}[2]{#2}\begingroup\raggedright\begin{thebibliography}{10}

\bibitem{Ooguri:2006in}
H.~Ooguri and C.~Vafa, {\it {On the Geometry of the String Landscape and the
  Swampland}},  {\em Nucl. Phys.} {\bf B766} (2007) 21--33,
  [\href{http://arxiv.org/abs/hep-th/0605264}{{\tt hep-th/0605264}}].

\bibitem{Lee:2019oct}
S.-J. Lee, W.~Lerche, and T.~Weigand, {\it {Emergent strings from infinite
  distance limits}},  {\em JHEP} {\bf 02} (2022) 190,
  [\href{http://arxiv.org/abs/1910.01135}{{\tt arXiv:1910.01135}}].

\bibitem{Etheredge:2022opl}
M.~Etheredge, B.~Heidenreich, S.~Kaya, Y.~Qiu, and T.~Rudelius, {\it
  {Sharpening the Distance Conjecture in diverse dimensions}},  {\em JHEP} {\bf
  12} (2022) 114, [\href{http://arxiv.org/abs/2206.04063}{{\tt
  arXiv:2206.04063}}].

\bibitem{Agmon:2022thq}
N.~B. Agmon, A.~Bedroya, M.~J. Kang, and C.~Vafa, {\it {Lectures on the string
  landscape and the Swampland}},  \href{http://arxiv.org/abs/2212.06187}{{\tt
  arXiv:2212.06187}}.

\bibitem{Dvali:2007hz}
G.~Dvali, {\it {Black Holes and Large N Species Solution to the Hierarchy
  Problem}},  {\em Fortsch. Phys.} {\bf 58} (2010) 528--536,
  [\href{http://arxiv.org/abs/0706.2050}{{\tt arXiv:0706.2050}}].

\bibitem{Dvali:2009ks}
G.~Dvali and D.~Lust, {\it {Evaporation of Microscopic Black Holes in String
  Theory and the Bound on Species}},  {\em Fortsch. Phys.} {\bf 58} (2010)
  505--527, [\href{http://arxiv.org/abs/0912.3167}{{\tt arXiv:0912.3167}}].

\bibitem{Dvali:2010vm}
G.~Dvali and C.~Gomez, {\it {Species and Strings}},
  \href{http://arxiv.org/abs/1004.3744}{{\tt arXiv:1004.3744}}.

\bibitem{Dvali:2012uq}
G.~Dvali, C.~Gomez, and D.~Lust, {\it {Black Hole Quantum Mechanics in the
  Presence of Species}},  {\em Fortsch. Phys.} {\bf 61} (2013) 768--778,
  [\href{http://arxiv.org/abs/1206.2365}{{\tt arXiv:1206.2365}}].

\bibitem{Heidenreich:2018kpg}
B.~Heidenreich, M.~Reece, and T.~Rudelius, {\it {Emergence of Weak Coupling at
  Large Distance in Quantum Gravity}},  {\em Phys. Rev. Lett.} {\bf 121}
  (2018), no.~5 051601, [\href{http://arxiv.org/abs/1802.08698}{{\tt
  arXiv:1802.08698}}].

\bibitem{Grimm:2018ohb}
T.~W. Grimm, E.~Palti, and I.~Valenzuela, {\it {Infinite Distances in Field
  Space and Massless Towers of States}},  {\em JHEP} {\bf 08} (2018) 143,
  [\href{http://arxiv.org/abs/1802.08264}{{\tt arXiv:1802.08264}}].

\bibitem{vandeHeisteeg:2023ubh}
D.~van~de Heisteeg, C.~Vafa, and M.~Wiesner, {\it {Bounds on Species Scale and
  the Distance Conjecture}},  \href{http://arxiv.org/abs/2303.13580}{{\tt
  arXiv:2303.13580}}.

\bibitem{Calderon-Infante:2023ler}
J.~Calder\'on-Infante, A.~Castellano, A.~Herr\'aez, and L.~E. Ib\'a\~nez, {\it
  {Entropy Bounds and the Species Scale Distance Conjecture}},
  \href{http://arxiv.org/abs/2306.16450}{{\tt arXiv:2306.16450}}.

\bibitem{vandeHeisteeg:2022btw}
D.~van~de Heisteeg, C.~Vafa, M.~Wiesner, and D.~H. Wu, {\it {Moduli-dependent
  Species Scale}},  \href{http://arxiv.org/abs/2212.06841}{{\tt
  arXiv:2212.06841}}.

\bibitem{Bershadsky:1993ta}
M.~Bershadsky, S.~Cecotti, H.~Ooguri, and C.~Vafa, {\it {Holomorphic anomalies
  in topological field theories}},  {\em Nucl. Phys. B} {\bf 405} (1993)
  279--304, [\href{http://arxiv.org/abs/hep-th/9302103}{{\tt hep-th/9302103}}].

\bibitem{Cribiori:2022nke}
N.~Cribiori, D.~Lust, and G.~Staudt, {\it {Black hole entropy and
  moduli-dependent species scale}},
  \href{http://arxiv.org/abs/2212.10286}{{\tt arXiv:2212.10286}}.

\bibitem{Cribiori:2023ffn}
N.~Cribiori, D.~Lust, and C.~Montella, {\it {Species Entropy and
  Thermodynamics}},  \href{http://arxiv.org/abs/2305.10489}{{\tt
  arXiv:2305.10489}}.

\bibitem{Cribiori:2023sch}
N.~Cribiori and D.~L\"ust, {\it {A note on modular invariant species scale and
  potentials}},  \href{http://arxiv.org/abs/2306.08673}{{\tt
  arXiv:2306.08673}}.

\bibitem{Long:2021jlv}
C.~Long, M.~Montero, C.~Vafa, and I.~Valenzuela, {\it {The Desert and the
  Swampland}},  \href{http://arxiv.org/abs/2112.11467}{{\tt arXiv:2112.11467}}.

\bibitem{vandeHeisteeg:2023uxj}
D.~van~de Heisteeg, C.~Vafa, M.~Wiesner, and D.~H. Wu, {\it {Bounds on Field
  Range for Slowly Varying Positive Potentials}},
  \href{http://arxiv.org/abs/2305.07701}{{\tt arXiv:2305.07701}}.

\bibitem{Andriot:2023isc}
D.~Andriot, {\it {Bumping into the Species Scale with the Scalar Potential}},
  {\em Fortsch. Phys.} {\bf 71} (2023), no.~10-11 2300139,
  [\href{http://arxiv.org/abs/2305.07480}{{\tt arXiv:2305.07480}}].

\bibitem{Hebecker:2017lxm}
A.~Hebecker, P.~Henkenjohann, and L.~T. Witkowski, {\it {Flat Monodromies and a
  Moduli Space Size Conjecture}},  {\em JHEP} {\bf 12} (2017) 033,
  [\href{http://arxiv.org/abs/1708.06761}{{\tt arXiv:1708.06761}}].

\bibitem{Scalisi:2018eaz}
M.~Scalisi and I.~Valenzuela, {\it {Swampland distance conjecture, inflation
  and $\alpha$-attractors}},  {\em JHEP} {\bf 08} (2019) 160,
  [\href{http://arxiv.org/abs/1812.07558}{{\tt arXiv:1812.07558}}].

\bibitem{Green:1997tv}
M.~B. Green and M.~Gutperle, {\it {Effects of D instantons}},  {\em Nucl. Phys.
  B} {\bf 498} (1997) 195--227,
  [\href{http://arxiv.org/abs/hep-th/9701093}{{\tt hep-th/9701093}}].

\bibitem{Green:1997di}
M.~B. Green and P.~Vanhove, {\it {D instantons, strings and M theory}},  {\em
  Phys. Lett. B} {\bf 408} (1997) 122--134,
  [\href{http://arxiv.org/abs/hep-th/9704145}{{\tt hep-th/9704145}}].

\bibitem{Kiritsis:1997em}
E.~Kiritsis and B.~Pioline, {\it {On R**4 threshold corrections in IIb string
  theory and (p, q) string instantons}},  {\em Nucl. Phys. B} {\bf 508} (1997)
  509--534, [\href{http://arxiv.org/abs/hep-th/9707018}{{\tt hep-th/9707018}}].

\bibitem{Pioline:1997pu}
B.~Pioline and E.~Kiritsis, {\it {U duality and D-brane combinatorics}},  {\em
  Phys. Lett. B} {\bf 418} (1998) 61--69,
  [\href{http://arxiv.org/abs/hep-th/9710078}{{\tt hep-th/9710078}}].

\bibitem{Obers:1999um}
N.~A. Obers and B.~Pioline, {\it {Eisenstein series and string thresholds}},
  {\em Commun. Math. Phys.} {\bf 209} (2000) 275--324,
  [\href{http://arxiv.org/abs/hep-th/9903113}{{\tt hep-th/9903113}}].

\bibitem{Green:1999pu}
M.~B. Green, H.-h. Kwon, and P.~Vanhove, {\it {Two loops in
  eleven-dimensions}},  {\em Phys. Rev. D} {\bf 61} (2000) 104010,
  [\href{http://arxiv.org/abs/hep-th/9910055}{{\tt hep-th/9910055}}].

\bibitem{Pioline:2010kb}
B.~Pioline, {\it {R**4 couplings and automorphic unipotent representations}},
  {\em JHEP} {\bf 03} (2010) 116, [\href{http://arxiv.org/abs/1001.3647}{{\tt
  arXiv:1001.3647}}].

\bibitem{Green:2010wi}
M.~B. Green, J.~G. Russo, and P.~Vanhove, {\it {Automorphic properties of low
  energy string amplitudes in various dimensions}},  {\em Phys. Rev. D} {\bf
  81} (2010) 086008, [\href{http://arxiv.org/abs/1001.2535}{{\tt
  arXiv:1001.2535}}].

\bibitem{Green:2011vz}
M.~B. Green, S.~D. Miller, and P.~Vanhove, {\it {Small representations, string
  instantons, and Fourier modes of Eisenstein series}},  {\em J. Number Theor.}
  {\bf 146} (2015) 187--309, [\href{http://arxiv.org/abs/1111.2983}{{\tt
  arXiv:1111.2983}}].

\bibitem{Bossard:2020xod}
G.~Bossard, A.~Kleinschmidt, and B.~Pioline, {\it {1/8-BPS Couplings and
  Exceptional Automorphic Functions}},  {\em SciPost Phys.} {\bf 8} (2020),
  no.~4 054, [\href{http://arxiv.org/abs/2001.05562}{{\tt arXiv:2001.05562}}].

\bibitem{Green:2016tfs}
M.~B. Green and A.~Rudra, {\it {Type I/heterotic duality and M-theory
  amplitudes}},  {\em JHEP} {\bf 12} (2016) 060,
  [\href{http://arxiv.org/abs/1604.00324}{{\tt arXiv:1604.00324}}].

\bibitem{Heckman:2019bzm}
J.~J. Heckman and C.~Vafa, {\it {Fine Tuning, Sequestering, and the
  Swampland}},  {\em Phys. Lett. B} {\bf 798} (2019) 135004,
  [\href{http://arxiv.org/abs/1905.06342}{{\tt arXiv:1905.06342}}].

\bibitem{Grimm:2017okk}
T.~W. Grimm, K.~Mayer, and M.~Weissenbacher, {\it {Higher derivatives in Type
  II and M-theory on Calabi-Yau threefolds}},  {\em JHEP} {\bf 02} (2018) 127,
  [\href{http://arxiv.org/abs/1702.08404}{{\tt arXiv:1702.08404}}].

\bibitem{Hull:1994ys}
C.~M. Hull and P.~K. Townsend, {\it {Unity of superstring dualities}},  {\em
  Nucl. Phys. B} {\bf 438} (1995) 109--137,
  [\href{http://arxiv.org/abs/hep-th/9410167}{{\tt hep-th/9410167}}].

\bibitem{Bershadsky:1993cx}
M.~Bershadsky, S.~Cecotti, H.~Ooguri, and C.~Vafa, {\it {Kodaira-Spencer theory
  of gravity and exact results for quantum string amplitudes}},  {\em Commun.
  Math. Phys.} {\bf 165} (1994) 311--428,
  [\href{http://arxiv.org/abs/hep-th/9309140}{{\tt hep-th/9309140}}].

\bibitem{Bossard:2023bhv}
G.~Bossard and A.~Loty, {\it {Saturating unitarity bounds at U-duality
  symmetric points}},  {\em JHEP} {\bf 10} (2023) 110,
  [\href{http://arxiv.org/abs/2308.02847}{{\tt arXiv:2308.02847}}].

\bibitem{Sarnak2006}
P.~{Sarnak} and A.~{Str{\"o}mbergsson}, {\it {Minima of Epstein's Zeta function
  and heights of flat tori}},  {\em Inventiones Mathematicae} {\bf 165} (Mar.,
  2006) 115--151.

\bibitem{Gross:1986mw}
D.~J. Gross and J.~H. Sloan, {\it {The Quartic Effective Action for the
  Heterotic String}},  {\em Nucl. Phys. B} {\bf 291} (1987) 41--89.

\bibitem{Ellis:1987dc}
J.~R. Ellis, P.~Jetzer, and L.~Mizrachi, {\it {One Loop String Corrections to
  the Effective Field Theory}},  {\em Nucl. Phys. B} {\bf 303} (1988) 1--35.

\bibitem{Abe:1988cq}
M.~Abe, H.~Kubota, and N.~Sakai, {\it {Loop Corrections to the $E(8)$ X $E(8)$
  Heterotic String Effective Lagrangian}},  {\em Nucl. Phys. B} {\bf 306}
  (1988) 405--424.

\bibitem{Etheredge:2023odp}
M.~Etheredge, B.~Heidenreich, J.~McNamara, T.~Rudelius, I.~Ruiz, and
  I.~Valenzuela, {\it {Running Decompactification, Sliding Towers, and the
  Distance Conjecture}},  \href{http://arxiv.org/abs/2306.16440}{{\tt
  arXiv:2306.16440}}.

\bibitem{Grimm:2013gma}
T.~W. Grimm, R.~Savelli, and M.~Weissenbacher, {\it {On $\alpha'$ corrections
  in N=1 F-theory compactifications}},  {\em Phys. Lett. B} {\bf 725} (2013)
  431--436, [\href{http://arxiv.org/abs/1303.3317}{{\tt arXiv:1303.3317}}].

\bibitem{Grimm:2010ks}
T.~W. Grimm, {\it {The N=1 effective action of F-theory compactifications}},
  {\em Nucl. Phys. B} {\bf 845} (2011) 48--92,
  [\href{http://arxiv.org/abs/1008.4133}{{\tt arXiv:1008.4133}}].

\bibitem{Anderson:2022bpo}
L.~B. Anderson, C.~R. Brodie, and J.~Gray, {\it {Branes and Bundles through
  Conifold Transitions and Dualities in Heterotic String Theory}},
  \href{http://arxiv.org/abs/2211.05804}{{\tt arXiv:2211.05804}}.

\bibitem{Lee:2018urn}
S.-J. Lee, W.~Lerche, and T.~Weigand, {\it {Tensionless Strings and the Weak
  Gravity Conjecture}},  {\em JHEP} {\bf 10} (2018) 164,
  [\href{http://arxiv.org/abs/1808.05958}{{\tt arXiv:1808.05958}}].

\bibitem{Lee:2018spm}
S.-J. Lee, W.~Lerche, and T.~Weigand, {\it {A Stringy Test of the Scalar Weak
  Gravity Conjecture}},  {\em Nucl. Phys.} {\bf B938} (2019) 321--350,
  [\href{http://arxiv.org/abs/1810.05169}{{\tt arXiv:1810.05169}}].

\bibitem{Donagi:2004ia}
R.~Donagi, Y.-H. He, B.~A. Ovrut, and R.~Reinbacher, {\it {The Particle
  spectrum of heterotic compactifications}},  {\em JHEP} {\bf 12} (2004) 054,
  [\href{http://arxiv.org/abs/hep-th/0405014}{{\tt hep-th/0405014}}].

\bibitem{Anderson:2017aux}
L.~B. Anderson, X.~Gao, J.~Gray, and S.-J. Lee, {\it {Fibrations in CICY
  Threefolds}},  {\em JHEP} {\bf 10} (2017) 077,
  [\href{http://arxiv.org/abs/1708.07907}{{\tt arXiv:1708.07907}}].

\bibitem{Iqbal:2001ye}
A.~Iqbal, A.~Neitzke, and C.~Vafa, {\it {A Mysterious duality}},  {\em Adv.
  Theor. Math. Phys.} {\bf 5} (2002) 769--808,
  [\href{http://arxiv.org/abs/hep-th/0111068}{{\tt hep-th/0111068}}].

\bibitem{Morrison:1996pp}
D.~R. Morrison and C.~Vafa, {\it {Compactifications of F theory on Calabi-Yau
  threefolds. 2.}},  {\em Nucl. Phys. B} {\bf 476} (1996) 437--469,
  [\href{http://arxiv.org/abs/hep-th/9603161}{{\tt hep-th/9603161}}].

\bibitem{Cota:2022maf}
C.~F. Cota, A.~Mininno, T.~Weigand, and M.~Wiesner, {\it {The Asymptotic Weak
  Gravity Conjecture in M-theory}},
  \href{http://arxiv.org/abs/2212.09758}{{\tt arXiv:2212.09758}}.

\bibitem{Greene:1995hu}
B.~R. Greene, D.~R. Morrison, and A.~Strominger, {\it {Black hole condensation
  and the unification of string vacua}},  {\em Nucl. Phys. B} {\bf 451} (1995)
  109--120, [\href{http://arxiv.org/abs/hep-th/9504145}{{\tt hep-th/9504145}}].

\bibitem{Greene:1996dh}
B.~R. Greene, D.~R. Morrison, and C.~Vafa, {\it {A Geometric realization of
  confinement}},  {\em Nucl. Phys. B} {\bf 481} (1996) 513--538,
  [\href{http://arxiv.org/abs/hep-th/9608039}{{\tt hep-th/9608039}}].

\bibitem{Alim:2021vhs}
M.~Alim, B.~Heidenreich, and T.~Rudelius, {\it {The Weak Gravity Conjecture and
  BPS Particles}},  {\em Fortsch. Phys.} {\bf 69} (2021), no.~11-12 2100125,
  [\href{http://arxiv.org/abs/2108.08309}{{\tt arXiv:2108.08309}}].

\bibitem{Antoniadis:1993ze}
I.~Antoniadis, E.~Gava, K.~S. Narain, and T.~R. Taylor, {\it {Topological
  amplitudes in string theory}},  {\em Nucl. Phys. B} {\bf 413} (1994)
  162--184, [\href{http://arxiv.org/abs/hep-th/9307158}{{\tt hep-th/9307158}}].

\bibitem{Berkovits:1994vy}
N.~Berkovits and C.~Vafa, {\it {N=4 topological strings}},  {\em Nucl. Phys. B}
  {\bf 433} (1995) 123--180, [\href{http://arxiv.org/abs/hep-th/9407190}{{\tt
  hep-th/9407190}}].

\bibitem{Ooguri:1995cp}
H.~Ooguri and C.~Vafa, {\it {All loop N=2 string amplitudes}},  {\em Nucl.
  Phys. B} {\bf 451} (1995) 121--161,
  [\href{http://arxiv.org/abs/hep-th/9505183}{{\tt hep-th/9505183}}].

\bibitem{Grimm:2007tm}
T.~W. Grimm, A.~Klemm, M.~Marino, and M.~Weiss, {\it {Direct Integration of the
  Topological String}},  {\em JHEP} {\bf 08} (2007) 058,
  [\href{http://arxiv.org/abs/hep-th/0702187}{{\tt hep-th/0702187}}].

\bibitem{Baume:2019sry}
F.~Baume, F.~Marchesano, and M.~Wiesner, {\it {Instanton Corrections and
  Emergent Strings}},  {\em JHEP} {\bf 04} (2020) 174,
  [\href{http://arxiv.org/abs/1912.02218}{{\tt arXiv:1912.02218}}].

\bibitem{Klaewer:2020lfg}
D.~Klaewer, S.-J. Lee, T.~Weigand, and M.~Wiesner, {\it {Quantum corrections in
  4d $N$ = 1 infinite distance limits and the weak gravity conjecture}},  {\em
  JHEP} {\bf 03} (2021) 252, [\href{http://arxiv.org/abs/2011.00024}{{\tt
  arXiv:2011.00024}}].

\bibitem{latticesums}
J.~M. Borwein, M.~L. Glasser, R.~C. McPhedran, J.~G. Wan, and I.~J. Zucker,
  {\em Lattice Sums Then and Now}.
\newblock Encyclopedia of Mathematics and its Applications. Cambridge
  University Press, 2013.

\end{thebibliography}\endgroup
\bibliographystyle{JHEP}

\end{document}